\documentclass{article}
\usepackage[english]{babel}
\usepackage{amsmath}
\usepackage{amsfonts}
\usepackage{amssymb}
\usepackage{amsthm}
\usepackage{mathtools}
\usepackage{natbib}
\usepackage[a4paper,margin=1in]{geometry}
\usepackage{comment}
\usepackage{cancel}
\usepackage{bbm}
\usepackage{enumitem}
\usepackage{graphicx}
\usepackage{caption}
\usepackage{subcaption}
\usepackage{tikz}
\usetikzlibrary{arrows}
\usetikzlibrary{arrows.meta, calc}
\usetikzlibrary{shapes,positioning,decorations.markings}
\usetikzlibrary{decorations.pathreplacing}
\tikzset{> = stealth}
\usetikzlibrary{graphs, graphs.standard}
\usepackage{pgfplots}
\usepackage{xcolor}
\usepackage{authblk}

\usepackage[colorlinks=true, allcolors=blue]{hyperref} %needs to be before cleveref

\newcommand{\R}{\mathbb{R}}
\newcommand{\Q}{\mathbb{Q}}
\newcommand{\Z}{\mathbb{Z}}

\DeclareMathOperator{\Sh}{Sh}
\newcommand{\Acal}{\mathcal{A}}

\newcommand{\Rcal}{\mathcal{R}}

\newcommand{\Xcal}{\mathcal{X}}

\DeclareMathOperator{\Rt}{\mathfrak{R}}
\DeclareMathOperator{\Lf}{\mathfrak{L}}
\DeclareMathOperator{\Pa}{Pa} 			
\DeclareMathOperator{\Ch}{Ch} 		
\DeclareMathOperator{\Anc}{Anc} 		
\DeclareMathOperator{\Desc}{Des}

\newcommand{\sm}{\setminus}						
\newcommand{\ins}{\subseteq} 					
 					
\newcommand{\cmpl}{\mathsf{c}}

\newcommand{\one}{\mathbf{1}}
\newcommand{\two}{\mathbf{2}}

\newcommand{\dcup}{\,\dot{\cup}\,}

\newcommand{\lp}{\left ( }
\newcommand{\rp}{\right ) }

\newcommand{\lB}{\left [ }
\newcommand{\rB}{\right ] }
\newcommand{\lC}{\left \{ }
\newcommand{\rC}{\right \} }
\newcommand{\lI}{\left| }
\newcommand{\rI}{\right| }

\newcommand{\st}{\;\middle |\;}

\newcommand{\tl}{\mathsf{t}}
\newcommand{\hd}{\mathsf{h}}
\newcommand{\tlhd}{\mathsf{th}}
\newcommand{\ul}{\underline}

\newcommand{\bij}{\stackrel{\sim}{\longrightarrow}}

\newcommand*{\tuh}[1][]{\mathrel{\tikz [baseline=-0.25ex,-latex] \draw  (0pt,0.5ex) -- (1.3em,0.5ex) node[midway,above] () {$#1$};}}

\usepackage{cleveref} %must be last

\usepackage{thmtools}

\declaretheorem[name=Theorem,numberwithin=section]{theorem}
\declaretheorem[name=Lemma,sibling=theorem]{lemma}
\declaretheorem[name=Corollary,sibling=theorem]{corollary}
\declaretheorem[name=Proposition,sibling=theorem]{proposition}
\declaretheorem[name=Remark,sibling=theorem]{remark}
\declaretheorem[name=Definition,sibling=theorem]{definition}
\declaretheorem[name=Notation,sibling=theorem]{notation}
\declaretheorem[name=Example,sibling=theorem]{example}

\title{M\"obius transforms and Shapley values for vector-valued functions on weighted directed acyclic multigraphs}

\author[1,2,3]{Patrick Forr\'e}
\author[3,4,5]{Abel Jansma}

\affil[1]{Korteweg-de Vries Institute for Mathematics, University of Amsterdam}
\affil[2]{AI4Science Lab, University of Amsterdam}
\affil[3]{Dutch Institute for Emergent Phenomena}
\affil[4]{Institute for Logic, Language and Computation, University of Amsterdam}
\affil[5]{Institute of Physics, University of Amsterdam}

\date{\small Authors ordered alphabetically by last name.}

\begin{document}
\maketitle

\begin{abstract}
Möbius inversion and Shapley values are two mathematical tools for characterizing and decomposing higher-order structure in complex systems. The former defines higher-order interactions as discrete derivatives over a partial order; the latter provides a principled way to attribute those interactions back to the `atomic' elements of the system. Both have found wide application, from combinatorics and cooperative game theory to machine learning and explainable AI. We generalize both tools simultaneously, in two orthogonal directions: (1)~from real-valued functions to functions valued in any abelian group (in particular, vector-valued functions), and (2)~from partial orders and lattices to \emph{directed acyclic multigraphs} (DAMGs) and weighted versions thereof.

The classical axioms, linearity, efficiency, null player, and symmetry, which uniquely characterize Shapley values on lattices, are insufficient in this more general setting. We resolve this by introducing \emph{projection operators} that recursively re-attribute higher-order synergies down to the roots of the graph, and by proposing two natural axioms: \emph{weak elements} (coalitions with zero synergy can be removed without affecting any attribution) and \emph{flat hierarchy} (on graphs with no intermediate hierarchy, attributions are distributed proportionally to edge counts). Together with linearity, these three axioms uniquely determine the Shapley values via a simple explicit formula (\Cref{cor:shapley-main}), while automatically implying efficiency, null player, symmetry, and a novel \emph{projection} property. The resulting framework recovers all existing lattice-based definitions as special cases, and naturally handles settings, such as games on non-lattice partial orders, which were previously out of reach.

The extension to vector-valued functions and general DAMG-structured hierarchies opens new application areas in machine learning, natural language processing, and explainable AI.
\end{abstract}

\section{Introduction}

Higher-order interactions, the irreducible contributions of groups of elements that cannot be explained by any of their proper sub-groups, are ubiquitous in the natural and social sciences, going by many names: synergy, epistasis, cumulants, and more \cite{tanaka2011multistable,skardal2020higher,schawe2022higher,battiston2020networks,battiston2021physics,rosas2022disentangling,jansma2023higher_phd}. Two mathematical tools stand out for quantifying and decomposing such interactions: \emph{Möbius inversion} defines higher-order interactions as discrete derivatives over a hierarchical structure; \emph{Shapley values} project those interactions back onto the atomic parts of the system in a principled way. Both tools have historically been confined to two restrictions: (i)~the value function must be real- (or ring-)valued, and (ii)~the underlying hierarchy must be a Boolean algebra (occasionally relaxed to lattices). We lift both restrictions simultaneously, extending Möbius inversion and Shapley values to functions valued in any abelian group (in particular, vector-valued functions) on arbitrary directed acyclic multigraphs (DAMGs, see \Cref{def:dagm}).

\begin{figure}[t!]
    \centering
\begin{subfigure}{0.3\textwidth}
$G$ = 
\begin{tikzpicture}[baseline=(current bounding box.center)]
    \node (a) at (0,0) {$a$};
    \node (b) at (1,0) {$b$};
    \node (c) at (2,0) {$c$};
    \node (d) at (0.5,1) {$d$};
    \node (e) at (1.5,1) {$e$};
    \node (f) at (0,2) {$f$};
    \node (g) at (1,2) {$g$};
    \node (h) at (2,2) {$h$};
    \path[->] (a) edge (d);
    \path[->] (b) edge (d);
    \path[->] (b) edge (e);
    \path[->] (c) edge (e);
    \path[->] (d) edge (f);
    \path[->] (d) edge (g);
    \path[->] (e) edge (g);
    \path[->] (e) edge (h);
\end{tikzpicture}
\end{subfigure}
\begin{subfigure}{0.3\textwidth}
$v$ = 
\begin{tikzpicture}[baseline=(current bounding box.center)]
    \node (a) at (0,0) {$1$};
    \node (b) at (1,0) {$2$};
    \node (c) at (2,0) {$3$};
    \node (d) at (0.5,1) {$3$};
    \node (e) at (1.5,1) {$5$};
    \node (f) at (0,2) {$5$};
    \node (g) at (1,2) {$14$};
    \node (h) at (2,2) {$9$};
    \path[->] (a) edge (d);
    \path[->] (b) edge (d);
    \path[->] (b) edge (e);
    \path[->] (c) edge (e);
    \path[->] (d) edge (f);
    \path[->] (d) edge (g);
    \path[->] (e) edge (g);
    \path[->] (e) edge (h);
\end{tikzpicture}
\end{subfigure}
\begin{subfigure}{0.3\textwidth}
$w$ = 
\begin{tikzpicture}[baseline=(current bounding box.center)]
    \node (a) at (0,0) {$1$};
    \node (b) at (1,0) {$2$};
    \node (c) at (2,0) {$3$};
    \node (d) at (0.5,1) {$0$};
    \node (e) at (1.5,1) {$0$};
    \node (f) at (0,2) {$2$};
    \node (g) at (1,2) {$8$};
    \node (h) at (2,2) {$4$};
    \path[->] (a) edge (d);
    \path[->] (b) edge (d);
    \path[->] (b) edge (e);
    \path[->] (c) edge (e);
    \path[->] (d) edge (f);
    \path[->] (d) edge (g);
    \path[->] (e) edge (g);
    \path[->] (e) edge (h);
\end{tikzpicture}
\end{subfigure}

\begin{subfigure}{0.3\textwidth}
$G^{\sm e}$ = 
\begin{tikzpicture}[baseline=(current bounding box.center)]
    \node (a) at (0,0) {$a$};
    \node (b) at (1,0) {$b$};
    \node (c) at (2,0) {$c$};
    \node (d) at (0.5,1) {$d$};
    \node (f) at (0,2) {$f$};
    \node (g) at (1,2) {$g$};
    \node (h) at (2,2) {$h$};
    \path[->] (a) edge (d);
    \path[->] (b) edge (d);
    \path[->] (b) edge (g);
    \path[->] (b) edge (h);
    \path[->] (c) edge (g);
    \path[->] (c) edge (h);
    \path[->] (d) edge (f);
    \path[->] (d) edge (g);
\end{tikzpicture}
\end{subfigure}
\begin{subfigure}{0.3\textwidth}
$v|_{\sm e}$ = 
\begin{tikzpicture}[baseline=(current bounding box.center)]
    \node (a) at (0,0) {$1$};
    \node (b) at (1,0) {$2$};
    \node (c) at (2,0) {$3$};
    \node (d) at (0.5,1) {$3$};
    \node (f) at (0,2) {$5$};
    \node (g) at (1,2) {$14$};
    \node (h) at (2,2) {$9$};
    \path[->] (a) edge (d);
    \path[->] (b) edge (d);
    \path[->] (b) edge (g);
    \path[->] (b) edge (h);
    \path[->] (c) edge (g);
    \path[->] (c) edge (h);
    \path[->] (d) edge (f);
    \path[->] (d) edge (g);
\end{tikzpicture}
\end{subfigure}
\begin{subfigure}{0.3\textwidth}
$w|_{\sm e}$ = 
\begin{tikzpicture}[baseline=(current bounding box.center)]
    \node (a) at (0,0) {$1$};
    \node (b) at (1,0) {$2$};
    \node (c) at (2,0) {$3$};
    \node (d) at (0.5,1) {$0$};
    \node (f) at (0,2) {$2$};
    \node (g) at (1,2) {$8$};
    \node (h) at (2,2) {$4$};
    \path[->] (a) edge (d);
    \path[->] (b) edge (d);
    \path[->] (b) edge (g);
    \path[->] (b) edge (h);
    \path[->] (c) edge (g);
    \path[->] (c) edge (h);
    \path[->] (d) edge (f);
    \path[->] (d) edge (g);
\end{tikzpicture}
\end{subfigure}

\begin{subfigure}{0.3\textwidth}
$G^{\sm\{d,e\}}$ = 
\begin{tikzpicture}[baseline=(current bounding box.center)]
    \node (a) at (0,0) {$a$};
    \node (b) at (1,0) {$b$};
    \node (c) at (2,0) {$c$};
    \node (f) at (0,2) {$f$};
    \node (g) at (1,2) {$g$};
    \node (h) at (2,2) {$h$};
    \path[->] (a) edge (f);
    \path[->] (a) edge (g);
    \path[->] (b) edge (f);
    \path[->] (b) edge [bend left=10] (g);
    \path[->] (b) edge [bend right=10] (g);
    \path[->] (b) edge (h);
    \path[->] (c) edge (g);
    \path[->] (c) edge (h);
\end{tikzpicture}
\end{subfigure}
\begin{subfigure}{0.3\textwidth}
$v|_{\sm\{d,e\}}$ = 
\begin{tikzpicture}[baseline=(current bounding box.center)]
    \node (a) at (0,0) {$1$};
    \node (b) at (1,0) {$2$};
    \node (c) at (2,0) {$3$};
    \node (f) at (0,2) {$5$};
    \node (g) at (1,2) {$14$};
    \node (h) at (2,2) {$9$};
    \path[->] (a) edge (f);
    \path[->] (a) edge (g);
    \path[->] (b) edge (f);
    \path[->] (b) edge [bend left=10] (g);
    \path[->] (b) edge [bend right=10] (g);
    \path[->] (b) edge (h);
    \path[->] (c) edge (g);
    \path[->] (c) edge (h);
\end{tikzpicture}
\end{subfigure}
\begin{subfigure}{0.3\textwidth}
$w|_{\sm\{d,e\}}$ = 
\begin{tikzpicture}[baseline=(current bounding box.center)]
    \node (a) at (0,0) {$1$};
    \node (b) at (1,0) {$2$};
    \node (c) at (2,0) {$3$};
    \node (f) at (0,2) {$2$};
    \node (g) at (1,2) {$8$};
    \node (h) at (2,2) {$4$};
    \path[->] (a) edge (f);
    \path[->] (a) edge (g);
    \path[->] (b) edge (f);
    \path[->] (b) edge [bend left=10] (g);
    \path[->] (b) edge [bend right=10] (g);
    \path[->] (b) edge (h);
    \path[->] (c) edge (g);
    \path[->] (c) edge (h);
\end{tikzpicture}
\end{subfigure}
\caption{
\textbf{Top row:} A DAG $G$ with value function $v$ and synergy function $w = v'_G$ (computed via Möbius inversion). The synergy $w(y)$ at a node $y$ captures the contribution of $y$ that is \emph{not} explained by its ancestors.
\textbf{Middle and bottom rows:} \emph{Projections} $G^{\sm e}$ and $G^{\sm\{d,e\}}$ obtained by removing nodes $e$ and $\{d,e\}$ from $G$ while preserving all remaining hierarchical structure. Crucially, removing an intermediate node introduces parallel edges to keep path counts invariant (hence \emph{multi}graph). This is illustrated by the double edge from $b$ to $g$ in $G^{\sm\{d,e\}}$.
\textbf{Key reasoning for Shapley values:} Since $w(d)=w(e)=0$, nodes $d$ and $e$ are \emph{weak elements} (\Cref{def:weak-element}): their synergy is zero and they carry no information. The Shapley values should therefore be the same across all three graphs. In the fully projected graph $G^{\sm\{d,e\}}$, which is \emph{hierarchically flat} (only roots and leaves), the only symmetric choice is to split $w(g)=8$ uniformly among the 4 directed edges pointing to $g$, giving each edge weight $\frac{1}{4}$. Applying this logic to all nodes yields e.g.\ $\Sh^G_b(v) = w(b)+\frac{1}{2}w(f)+\frac{2}{4}w(g)+\frac{1}{2}w(h)=2+1+4+2=9$.
This reasoning generalizes to all DAMGs and leads to the unique explicit formula in \Cref{eq:shapley-value-def-intro} via \Cref{thm:shapley-main-intro}.
}
\label{fig:damg-projections}
\end{figure}

\paragraph{Guided tour via Figure~\ref{fig:damg-projections}}

\Cref{fig:damg-projections} illustrates the key ideas with a concrete example; we encourage the reader to keep it in view throughout this introduction.

The top row shows a directed acyclic graph (DAG) $G=(V, E)$ of a set $V$ of eight nodes and a set $E$ of 8 edges. In classical game-theory language, the nodes represent ``coalitions''. The roots of the graph, the vertices with no incoming directed edges or \emph{parents}, correspond to individual players, the ``atomic'' parts of the system:
\begin{align}
        \Rt(G) &:= \lC r \in V \st \Pa^G(r) = \emptyset  \rC.
\end{align}

The value function $v$ assigns an observable quantity to each node. Applying \emph{Möbius inversion} to $v$ yields the \emph{synergy function} $w = v'_G$, which isolates at each node the contribution not explained by its ancestors:
\begin{align}
    v_G'(x) &:= \displaystyle v(x) - \sum_{y \in \Anc^G(x) \sm \lC x\rC} v'_G(y).
\end{align} 
Nodes with zero synergy we call \emph{weak elements}. A natural axiom for any attribution rule is that weak elements can be removed without changing any player's Shapley value. Removing a node while preserving all directed-path counts requires introducing parallel edges (hence the \emph{multi}graph), as shown in the middle and bottom rows of \Cref{fig:damg-projections}. After projecting out all intermediate nodes, the resulting graph $G^{\sm\{d,e\}}$ is \emph{hierarchically flat}: only roots and leaves of the graph remain, with every root directly connected to every leaf it can reach.

On this hierarchically flat graph, the attribution problem has a unique symmetric solution: each synergy $w(y)$ at a leaf $y$ must be split proportionally to the number of directed edges from each root to $y$. 

This reasoning extends to all \emph{directed acyclic multigraphs} (DAMGs), leading to the following general closed-form formula:
\begin{align}
        \boxed{\Sh_r^G(v) := \sum_{y \in V} \frac{\pi^G(r,y)}{\pi^G(y)} \cdot v'_G(y),} \label{eq:shapley-value-def-intro}
    \end{align}
where $\pi^G(r,y)$ denotes the number of directed paths from root node $r \in \Rt(G)$ to node $y \in V$ in $G$, and where $\pi^G(y):= \sum_{r \in \Rt(G)}\pi^G(r,y)$ is the number of all rooted directed paths in $G$ ending at $y$. Shapley value attributions are thus the ``fairest'' way to project down synergies to the roots of $G$

On the classical power-set structure (Boolean algebra), the number of paths from a singleton $\{i\}$ to a coalition $U\ni i$ equals $1$ for every $U$ of which $i$ is a member, so path-proportional splitting reduces to the familiar equal-split formula:
\begin{align}
    \Sh_i(v) &= \sum_{U \ni i} \frac{w(U)}{|U|}. \label{eq:classic_shapley}
\end{align}

The main contribution of this paper is to show that the Shapley values defined by \Cref{eq:shapley-value-def-intro} have several intuitive properties, and, conversely, are \emph{uniquely} defined by those properties. Formally we state these results (for the simplified case of $\Q$-vector spaces $A$) as follows:

\begin{theorem}[Main theorem: Shapley values on DAMGs, simplified]
\label{thm:shapley-main-intro}
The Shapley value $\Sh$ assignment, given by \Cref{eq:shapley-value-def-intro}, satisfies the following properties for every DAMG $G=(V,E)$ and $\Q$-vector space $A$:
    \begin{enumerate}[series=intro]
        \item \emph{$A$-Linearity}: For every finite index set $I$, elements $a_i \in A$ and value functions $v_i: V \to \Q$, for $i \in I$, and root $r \in \Rt(G)$ we have: \label{enu:intro-A-linearity}
        \begin{align}
            \Sh^G_r\lp  \sum_{i \in I} v_i \cdot a_i\rp&=  \sum_{i \in I} \Sh^G_r(v_i) \cdot a_i \in A.
        \end{align}
        \item \emph{Weak elements}: For every value function $v: V \to A$ and subset $W \ins V \sm \Rt(G)$ of weak elements\footnote{A \emph{weak element} for $(G,v)$ is a node $x \in V$ with $v'_G(x)=0$.} for $(G,v)$ and every root $r \in \Rt(G)$ we have: \label{enu:intro-weak-elements}
        \begin{align}
            \Sh^G_r(v) &=\Sh^{G^{\sm W}}_r(v|_{V\sm W}),
        \end{align}
        where $v|_{V\sm W}$ is the restriction of $v$ to $V \sm W$, and where $G^{\sm W}$ is the DAMG where the nodes from $W$ are removed, while all directed paths are preserved.
        \item \emph{Flat hierarchy}: For every hierarchically flat\footnote{A \emph{hierarchically flat} DAMG $G=(V,E)$ is a DAMG, where each node $x \in V$ either has no outgoing edges or no incoming edges (or none); so it is a DAMG, where no intermediate nodes exist.} DAMG $G$, every $y \in V$, unanimity value function\footnote{The \emph{unanimity value function} is defined in \Cref{rem:unanimity-game} via:  $\zeta_y(x):=1$ if $x \in \Desc^G(y)$, and, $\zeta_y(x):=0$ if $x \notin \Desc^G(x)$.} $\zeta_y$ on $G$ centred at $y$ and root $r \in \Rt(G)$ we have edge uniformity: \label{enu:intro-flat-hierarchy}
        \begin{align}
            \Sh^G_r(\zeta_y) &=
            \begin{cases}
                \displaystyle\delta_y(r), &\text{ if } y \in \Rt(G),\\
                \displaystyle\frac{|E(r,y)|}{|E(y)|}, &\text{ if }y\notin \Rt(G),
            \end{cases} 
        \end{align}
        where $E(r,y)$ is the set of directed edges in $G$ between nodes $r$ and $y \in V$, and where $E(y)$ is the set of directed edges in $G$ ending at $y \in V$.
    \end{enumerate}
Conversely, if $\Sh$ is any assignment rule that assigns to every DAMG $G=(V,E)$, every $\Q$-vector space $A$, every value function $v:V \to A$ and every $r \in \Rt(G)$ a value $\Sh^G_r(v) \in A$, such that the properties \ref{enu:intro-A-linearity}, \ref{enu:intro-weak-elements}, \ref{enu:intro-flat-hierarchy} from above are satisfied, then $\Sh$ is necessarily given by the explicit formula in $\Cref{eq:shapley-value-def-intro}$.
\end{theorem}
Furthermore, we can show that $\Sh$ also satisfies the corresponding versions of the classical properties \emph{linearity}, \emph{efficiency}, \emph{null player} and \emph{symmetry}.

\begin{corollary}[Classical properties of Shapley values on DAMGs]
\label{thm:shapley-classical-intro}
The Shapley value $\Sh$ assignment, given by \Cref{eq:shapley-value-def-intro}, satisfies the following properties for every DAMG $G=(V,E)$ and $\Q$-vector space $A$:
\begin{enumerate}[resume=intro]
    \item \emph{$\Q$-Linearity}:
    For value functions $v_1,v_2: V \to A$, scalars $c_1,c_2 \in \Q$ and root $r\in \Rt(G)$ we have:
    \begin{align}
        \Sh^G_r(c_1 \cdot v_1 + c_2 \cdot v_2)&= c_1 \cdot \Sh^G_r(v_1) + c_2 \cdot \Sh^G_r(v_2).
    \end{align}
    \item \emph{Efficiency}: For every value function $v: V \to A$ we have\footnote{Note, that, if the DAMG $G$ has only one leaf node $\ell \in V$, the ``grand coalition'', then $\sum_{x \in V} v'_G(x)=v(\ell)$, recovering the classical form of efficiency.}:
    \begin{align}
        \sum_{r \in \Rt(G)} \Sh^G_r(v) = \sum_{x \in V} v'_G(x).
    \end{align}
    \item \emph{Null root}: For every null root\footnote{A \emph{null root} for $(G,v)$ is an element $r \in \Rt(G)$ such that for all $x \in \Desc^G(r)$ we have vanishing synergies:  $v'_G(x)=0$.} $r\in \Rt(G)$ for $(G,v)$ we have:
        \begin{align}
            \Sh^G_r(v)&=0.
        \end{align}
    \item \emph{Symmetry}: For every value function $v: V \to A$, DAMG automorphism\footnote{A \emph{DAMG automorphism} $\alpha=(\alpha_V,\alpha_E):G\bij G$ of the DAMG $G=(V,E)$ consists of two bijective maps $\alpha_V:V\bij V$ and $\alpha_E:E\bij E$ such that for every $e \in E$ we have: $\tl(\alpha_E(e))=\alpha_V(\tl(e))$ and $\hd(\alpha_E(e))=\alpha_V(\hd(e))$.}
    $\alpha: G \bij G$ and root $r\in \Rt(G)$ we have:
    \begin{align}
        \Sh^G_{\alpha(r)}(v)&= \Sh^G_r(v^\alpha),
    \end{align}
    where the value function $v^\alpha: V \to A$ is defined on elements $x\in V$ via: $v^\alpha(x):=v(\alpha(x))$.

    In particular, for fixed $r \in \Rt(G)$, we have the implication:
    \begin{align}
        \lp \forall y \in \Desc^G(r). \;\;  v_G'(\alpha(y)) = v_G'(y) \rp \implies \Sh^G_{\alpha(r)}(v)&= \Sh^G_r(v).
    \end{align}
\end{enumerate}
\end{corollary}

Such properties are the standard axiomatic approach to uniquely define Shapley values. In contrast, we develop a framework to ``project out'' a set of nodes $S$ from a DAMG $G$ that constructs a related DAMG $G^{\sm S}$ (see \Cref{sec:projections_on_damg}) that allows us to state another unique characterization of the same explicit formula:

\begin{theorem}[Shapley values on DAMGs via projection property, simplified]
\label{thm:shapley-main-2-intro}
The Shapley value $\Sh$ assignment, given by \Cref{eq:shapley-value-def-intro}, also satisfies:
\begin{enumerate}[resume=intro]
    \item \emph{Projection}: For every DAMG $G=(V,E)$, $\Q$-vector space $A$, value function $v: V \to A$, subset $S \ins V \sm \Rt(G)$ and root $r\in \Rt(G)$ we have: \label{enu:intro-projection}
    \begin{align}
       \Sh^G_r(v) &= \Sh^{G^{\sm S}}_r(v^{\sm S}),
    \end{align}
    where $v^{\sm S}$ is the projection\footnote{The projection operators are properly defined in \Cref{sec:projections_on_damg}.} of $v$ onto $V\sm S$ in $G$ w.r.t.\ the \emph{path uniform weight function}\footnote{The path uniform weight function of $G$ is formally introduced in \Cref{sec:path_uniform}.} for $G$. 
    \item \emph{Edgeless graph}: For every edgeless DAMG $G=(V,\emptyset)$, every $\Q$-vector space $A$, every value function $v: V \to A$ and every $r \in \Rt(G)=V$ we have: \label{enu:intro-edgeless}
    \begin{align}
       \Sh^G_r(v) &= v(r).
    \end{align} 
\end{enumerate}
Conversely, if $\Sh$ is any assignment rule that assigns to every DAMG $G=(V,E)$, every $\Q$-vector space $A$, every value function $v:V \to A$ and every $r \in \Rt(G)$ a value $\Sh^G_r(v) \in A$, such that the properties \ref{enu:intro-projection} and \ref{enu:intro-edgeless}, from above, are satisfied, then $\Sh$ is necessarily given by the explicit formula in $\Cref{eq:shapley-value-def-intro}$.
\end{theorem}

In \Cref{cor:shapley-main} and \Cref{cor:shapley-further} we generalize \Cref{thm:shapley-main-intro} and the first part of \Cref{thm:shapley-main-2-intro} to value functions $v: V \to A$ with values in a more general abelian group $A$ with a certain module structure. 
c

\paragraph{Conceptual framing}

The line of reasoning above reveals that Shapley values are, at their core, a way to \emph{project higher-order synergy onto atomic parts}. Once seen this way, the lattice assumption is incidental: what matters is (a)~a notion of synergy (Möbius inversion), (b)~an operation that removes nodes with zero synergy (projection), and (c)~a symmetry condition on flat hierarchies (path-proportional splitting). Three axioms, \emph{linearity}, \emph{weak elements}, and \emph{flat hierarchy}, together pin down the Shapley formula uniquely on any DAMG (\Cref{cor:shapley-main}), while automatically implying efficiency, null player, and symmetry. This more abstract setting further allows a formulation of Shapley values purely in terms of elements from an \emph{incidence algebra} (see \Cref{rem:incidence_alg_formulation}). 

\paragraph{Contributions}

We make the following contributions. First, we prove a Möbius inversion theorem for abelian-group-valued functions on DAMGs (\Cref{thm:moebius-trafo}), generalizing the classical result beyond ring-valued functions and beyond partial orders. This includes vector-valued functions as a special case, and lattices, posets, and DAGs as special cases of the underlying structure.

Second, we introduce projection operators on DAMGs (\Cref{sec:projections_on_damg}) and characterize the \emph{path uniform weight function} as the canonical weighting that is compatible with these projections (\Cref{thm:path-uniform-unique}).

Third, we prove that the resulting Shapley values are uniquely determined by three natural axioms (\Cref{cor:shapley-main}), recovering existing lattice-based definitions as special cases and handling non-lattice DAGs, such as the poset game in \Cref{ex:poset_game}, which were previously out of reach. The codomain can be any abelian group, opening application areas in machine learning, natural language processing, and explainable AI \cite{strumbelj2010efficient,lundberg2017unified}.

The weighted generalization to DAMGs with arbitrary edge and root weights is developed in \Cref{thm:shapley-main}; the main text focuses on the canonical (path uniform) case.

\paragraph{Related work}

The history of the Möbius inversion theorem is one of generalization. The Möbius function first came up implicitly in the study of square-free integers by Euler \cite{euler1748introduction} and Gauss \cite{gauss1801article81}, but it was Möbius in 1832 who first used this pattern to invert functions \cite{mobius1832besondere}. Use of the Möbius inversion theorem was mostly confined to analytic number theory, until Gian-Carlo Rota introduced a vast generalization, which extended the domain of the Möbius function beyond the divisibility order to all (locally finite) partial orders \cite{rota1964foundations}, which provided a new way to study combinatorics algebraically. 

More recently, motivated mainly by the link between the Möbius function and the Euler characteristic, Möbius inversions have been defined on monoids \cite{cartier2006problemes}, decomposition spaces \cite{galvez2018decomposition}, and suitably finite categories \cite{leinster2012notions,leroux1975categories,haigh1980mobius}. However, to the best of our knowledge, these generalizations have never been used to actually invert a function. Furthermore, the codomain of the functions that make up the incidence algebra (see Remark \ref{rem:path_algebra}) in all these cases is still a commutative ring (possibly without negatives). The only generalization of the codomain was Rota's original work that generalized Möbius' formula with codomain $\mathbb{C}$ to any commutative ring. An example where one would need value functions with values in an abelian group is given in \cite{lang2025information,lang2024abstract}. There, one can interpret the technique for proving the generalization of Hu's theorem as a M\"obius inversion (up to sign conventions), and the occurring value functions are required to have values in certain general function spaces.

Shapley values similarly have a history of significant generalizations. Shapley defined the attributions for individual players in terms of the full power set of possible coalitions, but Faigle and Kern considered the situation in which some coalitions might be excluded \cite{faigle1992shapley}, and \cite{grabisch1997k} defined Shapley values for intermediate coalitions. Beyond these, formulations for games on distributive lattices \cite{grabisch2007games} and concept lattices \cite{faigle2016games} were developed, the latter being especially general since every finite lattice is isomorphic to a concept lattice. Each of these definitions, however, only concerned games with real-valued value functions. All lattice-based definitions show uniqueness of the attribution with respect to the standard Shapley value axioms (efficiency, linearity, null player, and symmetry), sometimes generalizing the symmetry axiom to the notion of \emph{hierarchical strength}. We replace the hierarchical strength axiom by the \emph{flat hierarchy} axiom (see \Cref{cor:shapley-main}), and show that we reproduce the original Shapley values, but differ from those based on hierarchical strength even in simple examples (see \Cref{ex:treeDAG} and the remark that follows it). 

In \cite{billot2005share}, the authors consider different weightings of the synergies in the calculation of the Shapley value, which is a feature  our formalism also has.

\paragraph{Outline of the paper}

The paper is organized around the example in Figure~\ref{fig:damg-projections}: Sections~\ref{sec:dagm} through~\ref{sec:path_uniform} formalize the ingredients needed to carry out the calculation informally sketched there, DAMGs, Möbius inversion, projection operators, and path uniform weights, and Section~\ref{sec:shap_on_damg} assembles them into the main theorem.

\Cref{sec:dagm} defines directed acyclic multigraphs (DAMGs), which generalize DAGs and partially ordered sets. \Cref{sec:moebius_trafo} presents Möbius inversion for DAMGs, defining the synergy function of a value function and proving its invertibility. \Cref{sec:projections_on_damg} introduces projection operators that remove weak-element nodes while preserving directed-path counts. \Cref{sec:path_uniform} characterizes the \emph{path uniform weight function} as the canonical weighting compatible with these projections. With these ingredients, \Cref{sec:shap_on_damg} establishes the main result (\Cref{cor:shapley-main}): a unique, explicit formula for Shapley values on DAMGs characterized by three natural axioms. \Cref{sec:examples} works through concrete examples, recovering the classical Shapley formula on Boolean algebras and illustrating how our definition handles games on non-lattice DAGs. 

The appendix develops the general theory for edge- and root-weighted DAMGs (including the weighted version of the main theorem, \Cref{thm:shapley-main}), contains full proofs deferred from the main text, and gives a complete catalogue of projection properties.

\section{Directed acyclic multigraphs}
\label{sec:dagm}
We start by defining directed acyclic multigraphs (DAMGs), which generalise (locally finite) posets and DAGs. Our formalism supports Shapley-like values for arbitrary weightings on the graph; the general theory for such weighted DAMGs is developed in \Cref{sec:weighted_DAMGs}. For clarity, the main text focuses on the canonical weighting that leads to Shapley values that are unique with respect to natural axioms.
\begin{definition}[Directed acyclic multigraph]
\label{def:dagm}
\begin{enumerate}
    \item A \emph{directed acyclic multigraph (DAMG)} $G=(V,E,\tlhd)$ consists of:
    \begin{enumerate}
        \item a finite set $V$ of vertices,
        \item a finite set $E$ of (possibly parallel) directed edges,
        \item together with a map:
    \begin{align}
        \tlhd: E &\to V \times V, & e &\mapsto \tlhd(e)=(\tl(e),\hd(e)),
    \end{align}
    assigning to each directed edge $e \in E$ its tail vertex (source) $\tl(e)\in V$ and head vertex (target) $\hd(e) \in V$,
    \end{enumerate}
such that:
    \begin{enumerate}[resume]
        \item there exists a \emph{topological ordering} $\prec$ of $G$, that is a total ordering $\prec$ of $V$, such that:
        \begin{align}
            e \in E & \implies \tl(e) \prec \hd(e).
        \end{align}
    \end{enumerate}
    \item We call $G=(V,E,\tlhd)$ a \emph{directed acyclic graph (DAG)} if it is a DAMG and $\tlhd$ is injective, that is, $E$ can be identified with a subset of $V \times V$. This means that there is at most one directed edge $e \in E$ between each pair of nodes $(x,y) \in V \times V$.
    \item For a DAMG $G=(V,E,\tlhd)$ we call $\ul G =(V, \ul E)$ with $\ul E := \tlhd(E) \ins V \times V$, the \emph{underlying DAG} of $G$.
\end{enumerate}
\end{definition}

\begin{notation}
\label{not:damg}
Let $G=(V,E,\tlhd)$ be a DAMG. We make the following further (abuses of) notations:
\begin{enumerate}
    \item We often just write $G=(V,E)$ instead of $G=(V,E,\tlhd)$ for a DAMG and leave the map $\tlhd$ implicit.
    \item For $x,y \in V$ we abbreviate the set of all directed edges between $x$ and $y$ as:
    \begin{align}
        E(x,y) &:= \lC e\in E \st \tl(e)=x, \hd(e)=y \rC, & E(y) &:=\lC e \in E \st \hd(e)=y \rC.
    \end{align}
    \item For $e \in E(x,y)$ we often just write $x\tuh[e] y \in E$ or just $x\tuh[e] y\in G$.
    \item For $x,y \in V$ we often just write $x\tuh y\notin E$ (or $G$) to mean $E(x,y) = \emptyset$.
    \item For $x \in V$ we define the set of \emph{parents}, \emph{children}, resp., of $x$ in $G$ as:
    \begin{align}
        \Pa^G(x) &:= \lC y \in V \st E(y,x) \neq \emptyset \rC, &
        \Ch^G(x) &:= \lC y \in V \st E(x,y) \neq \emptyset \rC.
    \end{align}
    \item For $x \in V$ we define the set of \emph{ancestors}, \emph{descendants}, resp., of $x$ in $G$ as:
    \begin{align}
        \Anc^G(x) &:= \lC y \in V \st \exists \text{ directed path in }G:\; y \tuh[e_1] \cdots \tuh[e_m] x \rC, \\
        \Desc^G(x) &:= \lC y \in V \st \exists \text{ directed path in }G:\; x \tuh[e_1] \cdots \tuh[e_m] y \rC.
    \end{align}
    Note that by convention we assume that we always have: $x \in \Anc^G(x)$ and $x \in \Desc^G(x)$.
    \item We further define the sets of \emph{roots (sources)} and \emph{leaves (sinks)}, resp., of $G$ as follows:
    \begin{align}
        \Rt(G) &:= \lC x \in V \st \Pa^G(x) = \emptyset  \rC, &
        \Lf(G) &:= \lC x \in V \st \Ch^G(x) = \emptyset  \rC.
    \end{align}
    \item For $x,y \in V$ we define the \emph{set of directed paths} from $x$ to $y$ in $G$,  its cardinality, and the one for directed paths starting from any root $r \in \Rt(G)$ as:
    \begin{align}
        \Pi^G(x,y) &:= \lC \text{directed path } x \tuh[e_1] \cdots \tuh[e_m] y  \in G  \rC, \\
        \pi^G(x,y) &:= \lI\Pi^G(x,y)\rI, \\
        \pi^G(y) &:= \sum_{r \in \Rt(G)} \pi^G(r,y).
    \end{align}
    Note that we consider the trivial directed path, consisting only of one node $x \in V$ (and without any directed edges) still as a directed path, so that we always have: $\pi^G(x,x)=1$.
    \item In the following, we will use the symbol $R$ to mean a ring, which we will always assume to be commutative and with one element, and, the symbol $A$ to mean either an abelian group or (the underlying abelian group of) an $R$-module.
\end{enumerate}  
\end{notation}

\begin{example}[Posets $P$ as DAGs]
\label{eg:poset-as-R-dag}
    Let $(P,\le)$ be a finite partially ordered set (poset).
    Then we can assign to $P$ its \emph{Hasse diagram}, which is the following DAG: $G(P) := (V(P), E(P))$, where: 
    \begin{align}
        V(P) &:= P,\\
        E(P) &:= \lC x\tuh y \st x < y \text{ for which there does not exist a }  z \in P \text{ with } x < z < y  \rC.
    \end{align}
    We can thus turn $(P,\le)$ into a DAG via its Hasse diagram.
\end{example}

\begin{example}[Lattices $L$ as DAGs]
\label{eg:lattice-as-R-dag}
    Let $(L,\land,\lor,\perp)$ be a finite lattice with bottom element $\perp$. 
    Then $P:=L \sm \lC \perp \rC$ is a finite poset. 
    We can thus turn $P$ into a DAG via 
    \Cref{eg:poset-as-R-dag}.
    The reason we remove the bottom element is that we typically get a non-trivial set of roots $\Rt(L):=\Rt(G(P))$, which then corresponds to the usual definition of ``atoms'' and ``players''. Furthermore, with this construction, we will recover the usual concepts of M\"obius inversions and Shapley values for lattices later on.
    Examples will be provided in \Cref{sec:examples}.
\end{example}

\begin{example}[Power sets as DAGs]
    Let $X$ be a finite set and $P:=\two^X\sm \lC \emptyset\rC$ be the power set of $X$ minus the empty set.
    Then $P$ is a finite poset via the inclusion relations $\ins$ .
    Again, we can thus turn $P$ into a DAG via 
    \Cref{eg:poset-as-R-dag}.
    Note that we can identify: $X \cong \Rt(P)$ via $x \mapsto \lC x\rC$.
\end{example}

\begin{definition}[Flat hierarchy]
\label{def:flat-hierarchy}
    A DAMG $G=(V,E)$ is called \emph{hierarchically flat} if: 
    \begin{align}
        V&=\Rt(G) \cup \Lf(G).
    \end{align}
\end{definition}

\section{M\"obius inversion on directed acyclic multigraphs}
\label{sec:moebius_trafo}

A value function $v: V \to A$ on a DAMG assigns an observable quantity to each node. In practice, the value $v(y)$ at a node $y$ typically reflects the \emph{cumulative} contribution of all its ancestors. The \emph{Möbius transform} (or \emph{synergy function}) $w = v'_G$ isolates the \emph{net} contribution at each node, the part of $v(y)$ that cannot be explained by lower-level nodes. The Möbius inversion theorem (\Cref{thm:moebius-trafo}) asserts that this decomposition is invertible: $v$ and $w$ uniquely determine each other. In machine learning terms, $v$ can be thought of as the cumulative feature importance up to a coalition, while $w$ captures the irreducible higher-order synergy introduced by that coalition alone.

\begin{definition}[M\"obius transform on DAMGs]
\label{def:moebius-trafo}
Let $G=(V,E)$ be a DAMG and $A$ be an abelian group.
Furthermore, consider a map $v: V \to A$, which we call \emph{value function}\footnote{Note that we do not impose any positivity, monotonicity or convexity assumptions on $v$, in contrast to other literature.}.
We then define the \emph{M\"obius transform} or \emph{synergy function} $v'_G: V \to A$ of $v$ w.r.t.\ $G$ recursively by following any topological ordering $\prec$ of $G$ as follows:
    \begin{align}
        v_G'(x) &:= \displaystyle v(x) - \sum_{y \in \Anc^G(x) \sm \lC x\rC} v'_G(y).
    \end{align}
Note that this is well-defined and does not depend on the choice of topological ordering $\prec$ of $G$. 
Also note that we have:
\begin{align}
    x \in \Rt(G) & \implies v_G'(x) = v(x).
\end{align}
\end{definition}

\begin{definition}[M\"obius function of a DAMG]
\label{def:moebius-function}
Let $G=(V,E)$ be a DAMG. 
The \emph{M\"obius function $\mu_G$ of $G$} is then recursively defined as follows:
\begin{align}
    \mu_G: V \times V & \to \Z, & \mu_G(x,y) &:= 
    \begin{cases}
        0, & \text{ if } x \notin \Anc^G(y), \\
        1, & \text{ if } x=y, \\
        \displaystyle - \sum_{z \in \Desc^G(x) \cap \Anc^G(y) \sm \lC y\rC} \mu_G(x,z) & \text{ if } x \in \Anc^G(y) \sm \lC y \rC.
    \end{cases}
\end{align}
Note that this is well-defined as we can recursively follow any topological order $\prec$ of $G$.
\end{definition}

\begin{lemma}
\label{lem:moebius-sums}
Let $G=(V,E)$ be a DAMG and $\mu_G$ the M\"obius function of $G$.
For $x,y \in V$  we have the following two equations:
    \begin{align}
        \sum_{z \in \Desc^G(x) \cap \Anc^G(y)} \mu_G(x,z)
        &= \delta_{x,y}, &
       \sum_{z \in \Desc^G(x) \cap \Anc^G(y)} \mu_G(z,y)
        &= \delta_{x,y}.
    \end{align}
\begin{proof}
First note that the index sets in both sums are empty if $x \notin \Anc^G(y)$. So, in this case, the sums are interpreted as zero and the equalities hold. In the following we will now assume that: $x \in \Anc^G(y)$.

For the first equation, first assume the case: $x=y$.
Then the sum degenerates to the one element $\mu_G(x,y)=1$, which solves this case. 
So, now assume the case: $x \neq y$.
Then we get:
\begin{align}
    &\sum_{z \in \Desc^G(x) \cap \Anc^G(y)} \mu_G(x,z)\\
    &= \mu_G(x,y)+\sum_{z \in \Desc^G(x) \cap \Anc^G(y)\sm\lC y\rC} \mu_G(x,z) \\
    &\overset{x\in \Anc^G(y)\sm\lC y\rC}{=} -\sum_{z \in \Desc^G(x) \cap \Anc^G(y)\sm\lC y\rC} \mu_G(x,z) + \sum_{z \in \Desc^G(x) \cap \Anc^G(y)\sm\lC y\rC} \mu_G(x,z) \\
    &=0.
\end{align}
This shows the first equation.

For the second equation, again, first assume the case: $x=y$. Then the sum degenerates to the one element $\mu_G(x,y)=1$, which solves this case.
So, now assume the case: $x \neq y$.
Then we get:
\begin{align}
    &\sum_{z \in \Desc^G(x) \cap \Anc^G(y)} \mu_G(z,y)\\
    &=\mu_G(y,y) + \sum_{z \in \Desc^G(x) \cap \Anc^G(y)\sm\lC y\rC} \mu_G(z,y) \\
    &\overset{z \in \Anc^G(y)\sm\lC y\rC}{=}  1 - \sum_{z \in \Desc^G(x) \cap \Anc^G(y)\sm\lC y\rC} \sum_{u \in \Desc^G(z) \cap \Anc^G(y)\sm\lC y\rC} \mu_G(z,u) \\
    &= 1 - \sum_{u \in \Desc^G(x) \cap \Anc^G(y)\sm\lC y\rC} \sum_{z \in \Desc^G(x) \cap \Anc^G(u)} \mu_G(z,u) \\
    &= 1 - \sum_{u \in \Desc^G(x) \cap \Anc^G(y)\sm\lC y\rC} \delta_{x,u} \\
    &= 1 - 1 \\
    &=0.
\end{align}
This shows the second equation and we are done.
\end{proof}    
\end{lemma}

\begin{theorem}[M\"obius inversion theorem for DAMGs]
\label{thm:moebius-trafo}
Let $G=(V,E)$ be a DAMG and $A$ be an abelian group.
Furthermore, consider two maps $v,w: V \to A$.
Then the following statements are equivalent:
\begin{enumerate}
    \item For all $x \in V$: $\displaystyle v(x) = \sum_{y \in \Anc^G(x)} w(y)$;
    \item For all $x \in V$: $\displaystyle w(x) = \sum_{y \in \Anc^G(x)} \mu_G(y,x) \cdot v(y)$;
\end{enumerate} 
where $\mu_G$ is the M\"obius function of $G$.

In particular, if $v$ is fixed, then the unique $w$ solving either of the above equivalent conditions is exactly the synergy function $v'_G$ of $v$ from \Cref{def:moebius-trafo}.
\begin{proof}
For the M\"obius inversion, we first fix $v$ and abbreviate: 
\begin{align}
     w(y) &:= \sum_{z \in \Anc^G(y)} \mu_G(z,y) \cdot v(z).
\end{align}
Then we get with help of \Cref{lem:moebius-sums}:
\begin{align}
    \sum_{y \in \Anc^G(x)} w(y)
    &= \sum_{y \in \Anc^G(x)} \sum_{z \in \Anc^G(y)} \mu_G(z,y) \cdot v(z). \\
    &= \sum_{z \in \Anc^G(x)} \sum_{y \in \Desc^G(z) \cap \Anc^G(x)} \mu_G(z,y) \cdot v(z) \\
    &= \sum_{z \in \Anc^G(x)} \delta_{z,x} \cdot v(z)\\
    &= v(x).
\end{align}
This shows one direction of the M\"obius inversion.
For the reverse direction, fix $w$ and abbreviate: 
\begin{align}
     v(y) &:= \sum_{z \in \Anc^G(y)} w(z).
\end{align}
Then we get, again, with help of \Cref{lem:moebius-sums}:
\begin{align}
   & \sum_{y \in \Anc^G(x)} \mu_G(y,x) \cdot v(y) \\
   &= \sum_{y \in \Anc^G(x)} \mu_G(y,x) \cdot \sum_{z \in \Anc^G(y)} w(z) \\
   &= \sum_{y \in \Anc^G(x)} \sum_{z \in \Anc^G(y)} \mu_G(y,x) \cdot  w(z) \\
   &= \sum_{z \in \Anc^G(x )} \sum_{y \in \Desc^G(z) \cap \Anc^G(x)} \mu_G(y,x) \cdot  w(z) \\
   &= \sum_{z \in \Anc^G(x )} \delta_{z,x} \cdot w(z) \\
   &= w(x).
\end{align}
This shows the other direction of the M\"obius inversion.
So both directions of the M\"obius inversion are shown.
\end{proof}
\end{theorem}

\begin{remark}
    In the context of \Cref{thm:moebius-trafo}, we call the map $v$ the \emph{value function} and $w$ the \emph{synergy function} of $v$, or more descriptively, $w$ the \emph{M\"obius transform} of $v$ and $v$ the \emph{inverse M\"obius  transform} of $w$. 
    We also often write $v'_G:=w$ for the M\"obius transform of $v$, in the sense of a \emph{discrete derivative of $v$ w.r.t.\ $G$}. Indeed, when $G$ is a single directed chain, \Cref{thm:moebius-trafo} reduces to a discrete version of the fundamental theorem  of calculus.
\end{remark}

\begin{remark}[M\"obius inversion theorem for DAGs vs.\ DAMGs]
Stating the M\"obius inversion theorem for DAMGs $G$ is the same as stating it for the underlying DAGs $\ul G$, because \Cref{thm:moebius-trafo} does not make use of multiple (parallel) directed edges anywhere. Borrowing terminology from \cite{leinster2012notions}, it is therefore a \emph{coarse} Möbius inversion.
\end{remark}

\begin{remark}[Path algebra and path convolution]
    \label{rem:path_algebra}
    The Möbius inversion theorem for posets is usually stated in terms of the incidence algebra of the poset. Our generalization to DAMGs can similarly be phrased in terms of a \emph{path algebra}, which we define for a DAMG $G=(V, E)$ and any fixed ring $R$ as the following $R$-algebra of functions: 
    \begin{align}
        \Acal(G) &:= \lC f: V\times V \to R \st \forall x,y \in V:\; \lp x\notin \Anc^G(y) \implies f(x, y)=0 \rp \rC.
    \end{align}
    Then define the algebra operation $\star$ on $\Acal(G)$ as the following \emph{path convolution}:
    \begin{align}
    \star: \Acal(G) \times \Acal(G) &\to \Acal(G),&
        (f\star g)(x, y) 
        &:= \sum_{z \in V}f(x, z)\cdot g(z, y).
    \end{align}
    We can further introduce the \emph{delta function} $\delta_G$ and \emph{zeta function} $\zeta_G$ of $G$ as follows:
    \begin{align}
        \delta_G: V \times V &\to \Z, & \delta_G(x,y) &:= 
        \begin{cases}
            1, \text{ if } x =y, \\
            0, \text{ if } x \neq y, 
        \end{cases}
        \\
        \zeta_G: V \times V &\to \Z, &
        \zeta_G(x,y) &:= 
        \begin{cases}
            1, \text{ if } x \in \Anc^G(y), \\
            0, \text{ if } x \notin \Anc^G(y).
        \end{cases}
    \end{align}
    The identity function with respect to the operation $\star$ is given by the delta function $\delta_G$, while the constant function is the zeta function $\zeta_G$. Convolving a function $f \in \Acal(G)$ with $\zeta_G$ thus amounts to integrating $f$ over $G$. Distributivity of $R$ ensures associativity of $\star$.  
    Rephrasing \Cref{lem:moebius-sums} in terms of $\zeta_G$, $\mu_G$, $\delta_G$ gives us the following identities:
    \begin{align}
        \zeta_G \star \mu_G &= \delta_G, &
        \mu_G \star \zeta_G &= \delta_G. \label{eq:moebius-inversion-convolution}
    \end{align}
    This shows that $\zeta_G$ and $\mu_G$ are $\star$-inverses to each other.  
    Furthermore, we can introduce the path convolution between functions $u:V \to A$ to an $R$-module $A$ and $g \in \Acal(G)$ as follows:
    \begin{align}
        u \star g : V &\to A, &
        (u \star g)(y)
        &:= \sum_{z \in V} g(z,y)\cdot u(z).
    \end{align}
    With this notation, \Cref{eq:moebius-inversion-convolution} immediately implies the Möbius inversion theorem for DAMGs, see \Cref{thm:moebius-trafo}, in the following form:
    \begin{align}
        v = w \star \zeta_G &\iff w = v \star \mu_G.
    \end{align}
    Fixing a topological ordering $\prec$ on $G$ makes the path algebra $\Acal(G)$ isomorphic to a sub-algebra of the $|V|$-dimensional lower-triangular matrices, which can greatly help with practical computations, see \cite{jansma2025fast}.
\end{remark}

\begin{remark}
\label{rem:unanimity-game}
    Let $G=(V,E)$ be a DAMG and $z \in V$ be a fixed element. Let $A$ be an abelian group. We then consider the following value function $\zeta_y:V \to \Z$, called \emph{unanimity value function} or \emph{unanimity game centred at $y$}, given on elements $x \in V$ as follows:
    \begin{align}
        \zeta_y(x) &:= 
            \begin{cases}
                1, & \text{ if } x \in \Desc^G(y), \\
                0, & \text{ if } x \notin \Desc^G(y).
            \end{cases}
    \end{align}
    Note that $(\zeta_y)_G'=\delta_y$. For any other value function $v: V \to A$ with synergy function $w=v_G'$ we can write by \Cref{thm:moebius-trafo}:
    \begin{align}
        v(x) &= \sum_{y \in V} \underbrace{\zeta_y(x)}_{\in \Z} \cdot \underbrace{w(y)}_{\in A}.
    \end{align}
    By recognising the unanimity game $\zeta_y(x)$ as the zeta function $\zeta_G(y, x)$ on $G$, see Remark \ref{rem:path_algebra}, it becomes clear that the above equation is simply Möbius inversion theorem, cast in the language of game theory. 
\end{remark}

\section{Projecting on directed acyclic multigraphs}
\label{sec:projections_on_damg}

We want Shapley values to be invariant under removing \emph{weak elements}, nodes whose synergy is zero and that carry no information. To make ``removing'' precise, we introduce a \emph{projection operator} that deletes a node from the DAMG while preserving the hierarchical structure above and below it. The key constraint is that directed path counts must be preserved: when an intermediate node $z$ is deleted, every pair of paths $x \tuh z \tuh y$ becomes a single direct path $x \tuh y$, which is why the construction introduces parallel edges and yields a \emph{multi}graph.

To redistribute the synergy of each node to its parents, we assign \emph{projection weights} to the edges and propagate each node's synergy upward to the roots of the graph, the `players'. The weights are normalized so that the total synergy is preserved. One may also start with a graph that has \emph{a priori} weighted edges or roots, which then induce projection weights, as outlined in Appendix~\ref{sec:weighted_DAMGs}.

\begin{definition}[Projectable directed acyclic multigraph ($R$-PDAMG)]
\label{def:weights-pdamg}
Let $R$ be a ring and $G=(V,E)$ be a DAMG and $q:E \to R$ an edge weight function.
By abuse of the same notation, we abbreviate the (corresponding) \emph{pairwise (edge) weight function} as:
\begin{align}
    q: V \times V & \to R, & q(x|y) &:= \sum_{e \in E(x,y)} q(x\tuh[e] y).
\end{align}
Note that $q(x|y)=0$ if $E(x,y)=\emptyset$. 
\begin{enumerate}
\item We call an edge weight function $q: E \to R$ on $G$ \emph{normalized}, if the following normalization condition holds:
    \begin{align}
     y \in V \sm \Rt(G) \iff   \Pa^G(y) \neq \emptyset & \overset{!}{\implies} \sum_{x \in V} q(x|y) = 1.
    \end{align}
\item A tuple $(G,q)$ consisting of a DAMG $G$ together with such a normalized weight function $q$ with values in $R$ will be called a \emph{projectable directed acyclic multigraph (PDAMG)} or \emph{$R$-PDAMG} to indicate the underlying ring $R$.
\end{enumerate}
\end{definition}

A naive but reasonable way to construct a normalized weight function $q$ on DAMGs $G=(V,E)$ and thus turn them into $R$-PDAMGs is the following:

\begin{example}[Edge uniform weights for DAMGs]
\label{eg:edge-uniform-damg}
    Let $G=(V,E)$ be a DAMG. Then we define the  \emph{edge uniform weight function} $q$ for the DAMG $G$ as follows, $x,y \in V$, $x\tuh[e]y \in E$:
    \begin{align}
        q(x\tuh[e] y) &:= \frac{1}{|E(y)|}, &
        q(x|y) &= \sum_{e \in E(x,y)} q(x\tuh[e] y) = \frac{|E(x,y)|}{|E(y)|}.
    \end{align}
    The values of $q$ lie in the smallest ring $R$ inside the rational numbers $\Q$ that contains all the fractions $|E(y)|^{-1}$ for $y \in V\sm\Rt(G)$:
    \begin{align}
        R &:= \Z\lB\lC|E(y)|^{-1} \st y \in V\sm\Rt(G)\rC\rB \ins \Q.
    \end{align}
    Note that $q$ is a normalized weight function of $G$ and has non-negative values.
    So, $(G,q)$ is an $R$-PDAMG.
\end{example}

On the DAMG corresponding to the Hasse diagram of the powerset, edge uniform weights are obviously of the form $q(T|U) = q(T\tuh[e] U) := \frac{1}{|U|}$. However, recovering the classic Shapley values from Equation \eqref{eq:classic_shapley} requires setting the \emph{total path weight} to $\frac{1}{|U|}$. We first define this total path weight here, before revealing the \emph{path uniform weight function} that achieves this goal, which will be covered in \Cref{def:path-uniform-damg}.

\begin{definition}[Total path weights of a PDAMG]
\label{not:total-weights-q}
Let $G=(V,E)$ be a DAMG and $q:E \to R$ be an edge weight function on $G$.
We then denote the \emph{total path weight function} for $q$ as $s=s^{(G,q)}$ of $(G,q)$:
\begin{align}
    s=s^{(G,q)}: V \times V & \to R, &
    s(x|y) &= \begin{cases}
        0, &\text{ if } y \notin \Desc^G(x), \\
        1, &\text{ if } y=x, \\
        \displaystyle\sum_{z \in \Pa^G(y)} s(x|z) \cdot q(z|y), &\text{ if } y \in \Desc^G(x) \sm \lC x\rC.
    \end{cases}
\end{align}
(For total path weights on arbitrarily weighted DAMGs, see \Cref{def:total-weights})
\end{definition}

\begin{definition}[Projection operator on $R$-PDAMGs w.r.t.\ one element]
\label{def:proj-damg-one}
    Let $G=(V,E)$ be a DAMG and $z \in V$ be a fixed element of $G$.
\begin{enumerate}
\item We then define the \emph{projection of $G$ onto the complement of $z$} as the following DAMG $G^{\sm z} =(V^{\sm z},E^{\sm z})$ with:
    \begin{align}
        V^{\sm z} &:= V \sm \{z\}, \\
        E^{\sm z} &:= \lC x \tuh[e] y \in G \st x,y \in V^{\sm z} \rC \dcup \lC x \tuh[e_1e_2] y \st x,y \in V^{\sm z}, x \tuh[e_1] z \tuh[e_2] y \in G \rC, \label{eq:proj-edges}
    \end{align}
    where for every pair of edges $x \tuh[e_1] z$ and $z \tuh[e_2] y$ in $G$ we introduce a new edge $x\tuh[e_1e_2] y$ in $G^{\sm z}$. 
    For $x,y \in V^{\sm z}$ we thus have:
    \begin{align}
        E^{\sm z}(x,y) &\cong E(x,y) \dcup \lp E(x,z) \times E(z,y) \rp.
    \end{align}
    Note that if $z \in \Rt(G) \cup \Lf(G)$ then we have for all $x,y \in V^{\sm z}$:
    \begin{align}
        E^{\sm z}(x,y) &\cong E(x,y).
    \end{align}
\item If, furthermore, $R$ is a ring and $q:E\to R$ a weight function for $G$ then we also define its projection as follows:
    \begin{align}
        q^{\sm z}: E^{\sm z} &\to R, & 
        q^{\sm z}(x\tuh[e]y) &:=\begin{cases}
            q(x\tuh[e]y) &\text{ if }x\tuh[e]y \in G, \\
            q(x\tuh[e_1]z)\cdot q(z\tuh[e_2]y) &\text{ if } e=e_1e_2,\; x\tuh[e_1]z\tuh[e_2]y \in G.
        \end{cases}
    \end{align}
    Note that for the underlying pairwise weights $q^{\sm z}(x|y)$ we get:
    \begin{align}
        q^{\sm z}: V^{\sm z} \times V^{\sm z} & \to R, & q^{\sm z}(x|y) 
        &= \sum_{e \in E(x,y)} q(x\tuh[e]y) + \sum_{\substack{e_1 \in E(x,z)\\e_2 \in E(z,y)}} q(x\tuh[e_1]z) \cdot q(z\tuh[e_2]y) \\
        &&&= q(x|y) + q(x|z) \cdot q(z|y). 
    \end{align}
    Note that if $z \in \Rt(G) \cup \Lf(G)$ then we have for all $x,y \in V^{\sm z}$:
    \begin{align}
        q^{\sm z}(x|y) &= q(x|y). 
    \end{align}
\item Now let $A$ be an $R$-module and $v: V \to A$ be a value function on $G$ and $w=v'_G: V \to A$ its synergy function w.r.t.\ $G$, then we can also define the corresponding projections with help of $q$:
    \begin{align}
        w^{\sm z}: V^{\sm z} &\to A, & w^{\sm z}(x)&:= w(x) + q(x|z) \cdot w(z), \label{eq:proj-synergy}\\
        v^{\sm z}: V^{\sm z} &\to A, & v^{\sm z}(y)&:= \sum_{x \in \Anc^{G^{\sm z}}(y)} w^{\sm z}(x).
    \end{align}
    Note that if $z \in \Rt(G)$ then we have for all $x,y \in V^{\sm z}$:
    \begin{align}
        w^{\sm z}(x) &= w(x), 
        & v^{\sm z}(y) &=\begin{cases}
            v(y)&\text{ if }z \notin \Anc^G(y),\\
            v(y) - v(z) &\text{ if }z \in \Anc^G(y).
        \end{cases}
    \end{align}
\end{enumerate}
\end{definition}

\begin{definition}[Projection operators w.r.t.\ subsets]
\label{def:projection-subset}
Let $G=(V,E)$ be a DAMG and $q:E \to R$ a weight function on $G$, $A$ be an $R$-module and $v:V \to A$ be a value function with its synergy function $w=v_G'$. 
Let $S \ins V$ be a subset with (unique) elements $S=\lC z_1,\dots,z_m  \rC$. Then the following \emph{projections onto the complement of the subset $S$} is well-defined by \Cref{lem:projections-commute} and does not depend on the order of single element projections:
    \begin{align}
        G^{\sm S} &:= ((G^{\sm z_m})\dots)^{\sm z_1}, &
        q^{\sm S} &:= ((q^{\sm z_m})\dots)^{\sm z_1}, &
        w^{\sm S} &:= ((w^{\sm z_m})\dots)^{\sm z_1}.
    \end{align}
    We, again, define the value function $v^{\sm S}$ via the M\"obius transform of $w^{\sm S}$:
    \begin{align}
        v^{\sm S}: V^{\sm S} &\to A, & v^{\sm S}(y)&:= \sum_{x \in \Anc^{G^{\sm S}}(y)} w^{\sm S}(x).
    \end{align}
Complementary, for a subset $T \ins V$ and complement $T^\cmpl := V \sm T$ we define:
\begin{align}
    G^{T} &:= G^{\sm T^\cmpl}, & q^{T} &:= q^{\sm T^\cmpl}, & w^{T} &:= w^{\sm T^\cmpl}, & v^{T} &:= v^{\sm T^\cmpl}.
\end{align}
It is important to note that the projections depend on the choice of the weight function $q$, which is suppressed in the notation.
\end{definition}

\begin{lemma}[Key properties of the projection operators]
\label{lem:proj-key-props}
    Let $G=(V,E)$ be a DAMG, $q:E \to R$ a weight function, $A$ an $R$-module, $v: V \to A$ a value function, and $w=v'_G$ its synergy function.
    For any subset $S \ins V$ the following properties hold:
    \begin{enumerate}
        \item \emph{DAMG}: $G^{\sm S}$ is a DAMG (\Cref{lem:proj-damg}).
        \item \emph{Directed paths}: For $x,y \in V^{\sm S}$ we have: $\pi^{G^{\sm S}}(x,y) = \pi^G(x,y)$ (\Cref{lem:proj-anc}).
        \item \emph{Synergy}: $w^{\sm S}$ is the synergy function of $v^{\sm S}$ w.r.t.\ $G^{\sm S}$ (\Cref{rem:proj-moebius-trafo}).
        \item \emph{Commutativity}: If $U \ins V$ is another subset disjoint from $S$, then
            $(G^{\sm S})^{\sm U} = (G^{\sm U})^{\sm S}$,
            $(q^{\sm S})^{\sm U} = (q^{\sm U})^{\sm S}$,
            $(w^{\sm S})^{\sm U} = (w^{\sm U})^{\sm S}$ (\Cref{lem:projections-commute}).
    \end{enumerate}
    A complete list of projection properties, including normalized weights, efficiency, null elements, and element weights for admissible projection sets (\Cref{def:admissible-proj-set}), is given in \Cref{thm:proj-stable-prop}.
\end{lemma}

\section{The path uniform weight function of directed acyclic multigraphs}
\label{sec:path_uniform}

The projection weights $q$ determine how synergy is distributed from each node to its parents. While any normalized weight function yields a valid projection (\Cref{def:weights-pdamg}), we seek the \emph{canonical} choice that is intrinsically determined by the graph structure alone. The key insight is that on a hierarchically flat graph, where there is no intermediate structure to inform any preference, the only natural, uninformative choice is to weight edges proportionally to the number of directed paths they carry. The path uniform weight function defined below is the unique extension of this flat-graph prescription to arbitrary DAMGs, as shown in \Cref{thm:path-uniform-unique}.

\begin{definition}[Path uniform weights for DAMGs]
\label{def:path-uniform-damg}
    Let $G=(V,E)$ be a DAMG. 
We define the \emph{path uniform (edge/projection) weight function} $q$ for the DAMG $G$ for directed edges $x\tuh[e]y\in E$ and $x,y \in V$ as follows:
    \begin{align}
        q_G(x\tuh[e] y) &:= \frac{\pi^G(x)}{\pi^G(y)}, &
        q_G(x|y) = \sum_{e \in E(x,y)} q_G(x\tuh[e] y) =
            \frac{\pi^G(x)}{\pi^G(y)} \cdot |E(x,y)|.
    \end{align}
The values of $q_G$ lie in the smallest ring $\Z_G$ inside the rational numbers $\Q$ that contain all the fractions $\pi^G(x)\cdot\pi^G(y)^{-1}$ for $x\tuh[e]y \in E$:
    \begin{align}
        \Z_G &:= \Z\lB\lC\pi^G(x)\cdot \pi^G(y)^{-1} \st x\tuh[e]y \in E\rC\rB \ins \Q. \label{eq:zg-ring}
    \end{align}
More generally, we may define for any ring $R$:
    \begin{align}
        R_G &:= R\lB\lC\pi^G(x)\cdot \pi^G(y)^{-1} \st x\tuh[e]y \in E\rC\rB. \label{eq:Rg-ring}
    \end{align}
We have thus constructed a weight function $q_G:E \to \Z_G$ on the DAMG $G$.
In \Cref{lem:path-uniform-recursion} we see that $q_G$ is also normalized, and thus $(G,q_G)$ a $\Z_G$-PDAMG.
\end{definition}

\begin{lemma}[Path uniform total path weights]
    \label{lem:path-uniform-total-weights}
Let $G=(V,E)$ be a DAMG and $q_G$ the path uniform weight function from \Cref{def:path-uniform-damg}.  
 We now claim that the total path weight function $s_G=s^{(G,q_G)}$ for $(G,q_G)$ is given  for all $x,y \in V$ by the simple formula:
\begin{align}
    s_G(x|y) &= \frac{\pi^G(x)}{\pi^G(y)}\cdot \pi^G(x,y)\label{eq:total_path_uniform_weights}.
\end{align}
Note that for $r \in \Rt(G)$ and $y \in V$ this simplifies to:
 \begin{align}
     s_G(r|y)&= \frac{\pi^G(r,y)}{\pi^G(y)}.
 \end{align}
\begin{proof}
See \Cref{lem:path-uniform-total-weights-proof}.
\end{proof}
\end{lemma}

In fact, the path uniform weights are the \emph{unique} weights that are compatible with the projection operators and treat elements on hierarchically flat graphs uniformly:

\begin{theorem}[Uniqueness of path uniform weights for a fixed DAMG]\label{thm:path-uniform-unique}
    Let $G=(V,E)$ be a DAMG and $q:E \to R_G$ be an edge weight function.
    Assume that for every subset $S \ins V$ for which $G^{\sm S}$ is hierarchically flat, we have for all $x\in V\sm S$ and $y \in \Desc_G(x)\sm \{x\}$:
    \begin{align}
        q^{\sm S}(x|y) &= \frac{|E^{\sm S}(x,y)|}{|E^{\sm S}(y)|} \label{eq:path_uniform_unique_hypothesis}
    \end{align}
then $q$ must necessarily coincide with the path uniform weight function from \Cref{def:path-uniform-damg}: $q=q_G$.
\begin{proof}
See \Cref{thm:path-uniform-unique-proof}.
\end{proof}
\end{theorem}

\begin{remark}
By \Cref{lem:proj-path-unif-comm}, the path uniform weight function $q_G$ satisfies $(q_G)^{\sm S}=q_{G^{\sm S}}$ and therefore \Cref{eq:path_uniform_unique_hypothesis}, so \Cref{thm:path-uniform-unique} characterizes it as the unique such weight function. Equivalently, one may thus characterize $q_G$ as the unique weight assignment satisfying projection compatibility and edge uniformity on hierarchically flat graphs across all DAMGs.
\end{remark}

\begin{remark}
 Note that in \Cref{lem:path-uniform-total-weights} we can compute all $\pi^G(r,y)$ for $r \in \Rt(G)$ and $y \in V$ with the recursion formula from \Cref{eq:path-recursion} with time complexity $O(K\cdot L \cdot D)$ and memory complexity $O(K \cdot D)$, where $K:=|\Rt(G)|$ and $L:=\max_{z \in V} |\Pa^G(z)|$ and $D:=|V|$.
\end{remark}

\begin{remark}
    Let $\Xcal \ins V$ be a horizontal subset (\Cref{def:horizontal-subset}) for $y \in V$, then we have the recursion:
    \begin{align}
        \pi^G(y)&=\sum_{x \in \Xcal} \pi^G(x) \cdot \pi^G(x,y).
    \end{align}
\end{remark}

\section{Shapley values on directed acyclic multigraphs}
\label{sec:shap_on_damg}

We now have all ingredients to state and prove the main result. The Shapley value for a root $r$ is defined as the projection of the synergy function $w = v'_G$ down to $r$, using the path uniform weights from \Cref{sec:path_uniform}. The central question is whether this definition is the \emph{unique} reasonable one. \Cref{cor:shapley-main} answers yes: three simple axioms already force the formula. The axioms are:
\begin{itemize}
    \item \emph{Linearity}: playing multiple games simultaneously yields the sum of individual payoffs.
    \item \emph{Weak elements}: coalitions with zero synergy can be projected out without changing any attribution.
    \item \emph{Flat hierarchy}: on graphs with no intermediate structure, attributions are distributed proportionally to the number of directed edges.
\end{itemize}
These axioms uniquely force the path uniform weighting, and with it the explicit formula (\ref{eq:shapley-value-final}), for all DAMGs $G$ and all value functions $v: V \to A$ with $A$ any $\Z_G$-module (e.g.\ any $\Q$- or $\R$-vector space). The weighted generalization to edge- and root-weighted $R$-DAMGs is stated in \Cref{thm:shapley-main}.

\begin{definition}[Weak elements in a DAMG]
\label{def:weak-element}
Let $G=(V,E)$ be a DAMG and $v:V\to A$ be a value function with its synergy function $v_G'$.
An element $z \in V$ is called a \emph{weak element} for $(G,v)$ if $v_G'(z)=0$.
\end{definition}

\begin{definition}[Null elements in DAMGs w.r.t.\ a value function]
\label{def:null-element}
    Let $G=(V,E)$ be a DAMG and $v:V \to A$ a value function. An element $y \in V$ is called \emph{null element} for $(G,v)$ if for every $x \in \Desc^G(y)$ we have vanishing synergies:  $v'_G(x)=0$. 
    If $r \in \Rt(G)$ is a null element for $(G,v)$ then we also call it a \emph{null root} for $(G,v)$.
\end{definition}

\begin{theorem}[Main result: Shapley values on DAMGs]
\label{cor:shapley-main}
    Let $\Sh$ be an assignment rule that assigns to every DAMG $G=(V,E)$, every $\Z_G$-module\footnote{The ring $\Z_G$ was introduced in \Cref{eq:zg-ring} in \Cref{def:path-uniform-damg}.} $A$, every value function $v:V \to A$ and every $r \in \Rt(G)$ a value $\Sh^G_r(v) \in A$.
    Further, assume that we have the following properties:
    \begin{enumerate}
        \item \emph{$A$-Linearity}: For every DAMG $G=(V,E)$, $\Z_G$-algebra $R$, $R$-module $A$, finite index set $I$, elements $a_i \in A$ and value functions $v_i: V \to R$, for $i \in I$, and root $r \in \Rt(G)$ we have:
        \begin{align}
            \Sh^G_r\lp  \sum_{i \in I} v_i \cdot a_i\rp&=  \sum_{i \in I} \Sh^G_r(v_i) \cdot a_i \in A.
        \end{align}
        \item \emph{Weak elements}: For every DAMG $G=(V,E)$, $\Z_G$-module $A$, value function $v: V \to A$ and subset $W \ins V \sm \Rt(G)$ of weak elements for $(G,v)$ and every root $r \in \Rt(G)$ we have:
        \begin{align}
            \Sh^G_r(v) &=\Sh^{G^{\sm W}}_r(v|_{V\sm W}),
        \end{align}
        where $v|_{V\sm W}$ is the restriction of $v$ to $V \sm W$.
        \item \emph{Flat hierarchy}: For every hierarchically flat DAMG $G=(V,E)$, every $y \in V$, unanimity value function $\zeta_y$ on $G$ centred at $y$ and root $r \in \Rt(G)$ we have edge uniformity:
        \begin{align}
            \Sh^G_r(\zeta_y) &=
            \begin{cases}
                \displaystyle\delta_y(r), &\text{ if } y \in \Rt(G),\\
                \displaystyle\frac{|E(r,y)|}{|E(y)|}, &\text{ if }y\notin \Rt(G).
            \end{cases} 
        \end{align}
    \end{enumerate}
    Then $\Sh$ necessarily satisfies the following:
    \begin{enumerate}[resume]
        \item \emph{Path uniform projection weights}: For every DAMG $G=(V,E)$, $\Z_G$-module $A$, value function $v:V\to A$ and root $r \in \Rt(G)$ we have:
    \begin{align}
        \Sh_r^G(v) &= \sum_{y \in V} s_G(r|y) \cdot v'_G(y) \label{eq:shapley-DAMG-shapley-PDAMG},
    \end{align}
    where $s_G$ is the path-uniform total path weight from \Cref{lem:path-uniform-total-weights}.
    \end{enumerate}
    This then leads to the explicit formula:
    \begin{align}
        \boxed{\Sh_r^G(v) = \sum_{y \in V} \frac{\pi^G(r,y)}{\pi^G(y)} \cdot v'_G(y).} \label{eq:shapley-value-final}
    \end{align}
    The Shapley value for root $r$ is thus a weighted sum of the synergies $v'_G(y)$ over all nodes $y$, where the weight $\frac{\pi^G(r,y)}{\pi^G(y)}$ is the fraction of all directed paths to $y$ that pass through $r$.

Conversely, if we define $\Sh$ through the explicit formula in \Cref{eq:shapley-value-final} then $\Sh$ satisfies all the properties in this theorem.
\begin{proof}
By \Cref{rem:unanimity-game} we can always write any value function $v:V \to A$ as follows:
    \begin{align}
        v(x) &= \sum_{y \in V} \underbrace{\zeta_y(x)}_{\in \Z} \cdot \underbrace{v_G'(y)}_{\in A},
    \end{align}
    with the unanimity value function $\zeta_y$ on $G$ centred at $y \in V$. 
    By $A$-linearity we then can write for $r \in \Rt(G)$:
 \begin{align}
     \Sh^G_r(v) & = \sum_{y \in V} \Sh^G_r(\zeta_y) \cdot v'_G(y).
 \end{align}
For fixed $y \in V$ let $W:=V\sm \lp\Rt(G) \cup \lC y\rC\rp$. Then $W$ consists only of weak elements for $(G,\zeta_y)$. Note that $V^{\sm W} = \Rt(G) \cup \lC y\rC$, which shows that $G^{\sm W}$ is a hierarchically flat DAMG. Further note, that $\zeta_y|_{V \sm W}$ is the unanimity game of $G^{\sm W}$ centred at $y \in V^{\sm W}$. We then get:
\begin{align}
\Sh^G_r(\zeta_y)
    &\overset{\text{weak elements}}{=} \Sh^{G^{\sm W}}_r(\zeta_y|_{V \sm W})
    &\overset{\text{flat hierarchy}}{=} \frac{|E^{\sm W}(r,y)|}{|E^{\sm W}(y)|}
    &= \frac{\pi^{G^{\sm W}}(r,y)}{\pi^{G^{\sm W}}(y)}
    &\overset{\text{\Cref{lem:proj-key-props}}}{=} \frac{\pi^G(r,y)}{\pi^G(y)}.
\end{align}
By plugging this back into the above equation we get the explicit formula and the identification:
 \begin{align}
     \Sh^G_r(v) & = \sum_{y \in V} \Sh^G_r(\zeta_y) \cdot v'_G(y)
     = \sum_{y \in V} \frac{\pi^G(r,y)}{\pi^G(y)} \cdot v'_G(y) \overset{\text{\ref{lem:path-uniform-total-weights}}}{=}  \sum_{y \in V} s_G(r|y) \cdot v'_G(y).
 \end{align}
 This shows the main claim. 
Conversely, if we define $\Sh$ by \Cref{eq:shapley-value-final} one can check that $\Sh$ satisfies all the properties in this theorem. Details can be found in \Cref{thm:properties-shapley-values}.
\end{proof}
\end{theorem}

\begin{corollary}[Further properties of Shapley values on DAMGs]
\label{cor:shapley-further}
Under the assumptions of \Cref{cor:shapley-main}, $\Sh$ also satisfies:
\begin{enumerate}
    \item \emph{$R$-Linearity}:
    For every DAMG $G=(V,E)$, $\Z_G$-algebra $R$, $R$-module $A$, value functions $v_1,v_2: V \to A$, scalars $c_1,c_2 \in R$ and root $r\in \Rt(G)$ we have:
    \begin{align}
        \Sh^G_r(c_1 \cdot v_1 + c_2 \cdot v_2)&= c_1 \cdot \Sh^G_r(v_1) + c_2 \cdot \Sh^G_r(v_2).
    \end{align}
    \item \emph{Efficiency}: For every DAMG $G=(V,E)$, $\Z_G$-module $A$ and value function $v: V \to A$ we have:
    \begin{align}
        \sum_{r \in \Rt(G)} \Sh^G_r(v) = \sum_{x \in V} v'_G(x).
    \end{align}
    \item \emph{Null root}: For every DAMG $G=(V,E)$, $\Z_G$-module $A$, value function $v: V \to A$ and every null root $r\in \Rt(G)$ for $(G,v)$ we have:
        \begin{align}
            \Sh^G_r(v)&=0.
        \end{align}
    \item \emph{Symmetry}: For every DAMG $G=(V,E)$, $\Z_G$-module $A$, value function $v: V \to A$, DAMG automorphism\footnote{A \emph{DAMG automorphism} $\alpha=(\alpha_V,\alpha_E):G\bij G$ of the DAMG $G=(V,E)$ consists of two bijective maps $\alpha_V:V\bij V$ and $\alpha_E:E\bij E$ such that for every $e \in E$ we have: $\tl(\alpha_E(e))=\alpha_V(\tl(e))$ and $\hd(\alpha_E(e))=\alpha_V(\hd(e))$.}
    $\alpha: G \bij G$ and root $r\in \Rt(G)$ we have:
    \begin{align}
        \Sh^G_{\alpha(r)}(v)&= \Sh^G_r(v^\alpha),
    \end{align}
    where the value function $v^\alpha: V \to A$ is defined on elements $x\in V$ via: $v^\alpha(x):=v(\alpha(x))$.

    In particular, for fixed $r \in \Rt(G)$, we have the implication:
    \begin{align}
        \lp \forall y \in \Desc^G(r). \;\;  v_G'(\alpha(y)) = v_G'(y) \rp \implies \Sh^G_{\alpha(r)}(v)&= \Sh^G_r(v).
    \end{align}
    \item \emph{Projection}: For every DAMG $G=(V,E)$, $\Z_G$-module $A$, value function $v: V \to A$, subset $S \ins V \sm \Rt(G)$ and root $r\in \Rt(G)$ we have:
    \begin{align}
       \Sh^G_r(v) &= \Sh^{G^{\sm S}}_r(v^{\sm S}),
    \end{align}
    where $v^{\sm S}$ is the projection of $v$ onto $V\sm S$ w.r.t.\ the $\Z_G$-PDAMG $(G,q_G)$, where $q_G$ is the path uniform weight function for $G$.
\end{enumerate}
\begin{proof}
Follows from \Cref{cor:shapley-main} and \Cref{thm:properties-shapley-values} together with the fact that $q_{G^{\sm S}}=(q_G)^{\sm S}$ by \Cref{lem:proj-path-unif-comm} and \Cref{rem:shapley-path-uniform}.
\end{proof}
\end{corollary}

\begin{remark}
    The weighted generalization of \Cref{cor:shapley-main} to $R$-DAMGs with arbitrary edge weights $\varsigma$ and root weights $\tau$ is given in \Cref{thm:shapley-main}. The three axioms take the same form, except that the flat hierarchy axiom becomes a $\varsigma$- and $\tau$-weighted uniformity condition, and the explicit formula replaces $\pi^G(r,y)/\pi^G(y)$ by $(\tau(r)/\tau(y)) \cdot \sigma(r,y)$, where $\sigma$ is the total path weight function of $\varsigma$ and $\tau(y) = \sum_{u \in \Rt(G)} \tau(u) \cdot \sigma(u,y)$. All other properties of \Cref{cor:shapley-further} carry over unchanged.
\end{remark}

\begin{remark}
    The three properties from \Cref{cor:shapley-main} have natural interpretations in game theoretic terms: $A$-linearity means that if you play multiple games at the same time, your reward is simply the sum of the rewards of the individual games. The weak elements property says that players should not get rewarded for forming coalitions that do not add anything. The flat hierarchy property states that player-coalition bonds should \emph{a priori} be treated equally in simple (hierarchically flat) graphs.
\end{remark}

\begin{remark}\label{rem:incidence_alg_formulation}
    Note that the total path weight function $s$ from \Cref{not:total-weights-q} is an element of the \emph{path algebra} $\Acal(G)$ of $G$ (see \Cref{rem:path_algebra}). This means that Shapley values (written in the form of Equations \ref{eq:shapley-DAMG-shapley-PDAMG} and \ref{eq:shapley-value-total-weight-formula}) can be efficiently written in terms of the path algebra as $\Sh_r^{(G, q)}(v) = (s \star (v \star \mu)) (r)$. 
    
    Furthermore, $s$ under uniform path weights itself is also a natural element from the path algebra. Let $\alpha_G \in \Acal(G)$ be the covering relation, which for DAMGs is written $\alpha_G(x, y) = |E(x, y)|$. We can then define the path-counting function $\pi_G(x, y) = \sum_{k\geq 0} \alpha_G^{\star k}(x, y)$, which converges by finiteness and acyclicity of $G$ (or local finiteness of posets), so that $\pi_G = (\delta_G - \alpha_G)^{\star(-1)}$. Let $\rho$ be the root indicator such that $\rho(x)=1\iff x\in \Rcal(G)$, and $0$ otherwise. Note that we can embed a function $f:V\to R$ diagonally as $\Delta_f(x, y) = \delta_{xy}f(x)$. We then have that $s = \Delta_{\rho\star \pi_G}\star \pi_G \star \Delta_{\rho\star \pi_G}^{-1}$, which is the path algebra equivalent of \Cref{eq:total_path_uniform_weights}. This yields a general expression for Shapley values in path/incidence algebras:
    \begin{align}
        \Sh_r(v) = \left(\Delta_{\rho\star \pi_G}\star \pi_G \star \Delta_{\rho\star \pi_G}^{-1} \star (v\star \mu)\right)(r)
    \end{align}
    
    For a general $R$-DAMGs $(G,\varsigma,\tau)$, $\alpha_G$ is replaced by $\varsigma$ and $\pi^G(r,y)$ by $\sigma(r,y)$ and $\rho$ by $\tau$ (extended by $0$ outside of $\Rt(G)$).

    Since $s$ and $\mu$ can be precomputed for a given DAMG, calculating Shapley values for different games can be implemented very efficiently as lower-diagonal matrix multiplication. 
\end{remark}

\begin{remark}
    The main text derives the projection weights from the graph structure, but one can derive similar uniqueness theorems for arbitrary normalized weight functions, which are presented in \Cref{sec:shap_on_rdamg}. In that setting, \Cref{thm:pdamg-shapley-unique-flat} gives an analogous three-axiom characterization, and \Cref{thm:pdamg-shapley-unique-proj} provides a two-axiom characterization via the projection property alone, of which \Cref{thm:shapley-main-2-intro} is the path-uniform special case. 
\end{remark}

\section{Examples, applications and recovering corner cases}
\label{sec:examples}

\begin{example}[Recovering Shapley's definition on Boolean algebras]
\label{eg:recover-shapley-original}
Let $L = \two^I$ be the power set of an index set $I$ (of ``players''), ordered by inclusion, and $v:L \to \R$ be any function (``game''). $L$ forms the inclusion lattice in Shapley's original definition, where the following formula for the Shapley values was derived for $i \in I$:
\begin{align}
   \Sh_i(v) &= \sum_{i\in y \ins I} \frac{1}{|y|} \cdot w(y). \label{eq:Shapley-original}
\end{align}
Now consider the partially ordered subset 
$V:=\two^I\sm\lC\emptyset\rC \ins L$, and let $G:=(V,E)$ be the DAG of the corresponding Hasse diagram.
Note that we can identify the set of players $I$ with the set of roots $\Rt(G)$ of $G$:
\begin{align}
    I & \cong \Rt(G), & i \mapsto \lC i \rC.
\end{align}
Since the Hasse diagram of a Boolean algebra, like the inclusion lattice $L$, forms a hypercube, the DAG $G$ corresponds to a hypercube where the node corresponding to the empty set has been removed. 
For our formulation of Shapley values, first note that $v(\emptyset)=w(\emptyset)=0$, so removing the bottom element of $L$ does not affect the value function $v$ or its synergy function $w$.
That said, and with the path uniform weighting, our Shapley value is for $r \in \Rt(G) \cong I$ written as:
\begin{align}
    \Sh_r^G(v) &= \sum_{y \in V} \frac{\pi^G(r, y)}{\pi^G(y)} \cdot w(y) = \sum_{y \in \Desc^G(r)} \frac{\pi^G(r, y)}{\pi^G(y)} \cdot w(y).   
\end{align}
Let $y \in V$ and $r \in \Rt(G)\cong I$ a root with $r \ins y$. Since $y$ is a subset of $I$, and $r$ is a singleton set, there are exactly $(|y|-1)!$ directed paths from $r$ to $y$ in $G$, each edge corresponding to adding another element of $y$. 
By the symmetry of the hypercube, this number is equal for every $r \in \Anc^G(y)$. We thus get:
\begin{align}
    \pi^G(r,y) &= (|y|-1)!, &
    \pi^G(y) &=\sum_{\substack{r \in \Rt(G)\\ r \ins y}} \pi^G(r,y) = |y| \cdot (|y|-1)!.
\end{align}
With this we then get the formula for the Shapley values for $r \in \Rt(G) \cong I$:
\begin{align}
    \Sh_r^G(v) &= \sum_{y \in \Desc^G(r)} \frac{\pi^G(r, y)}{\pi^G(y)} \cdot w(y) = \sum_{\substack{y \ins I \\ r\ins y}} \frac{(|y|-1)!}{|y| \cdot (|y|-1)!} \cdot w(y) 
    = \sum_{r\ins y \ins I} \frac{1}{|y|} \cdot w(y).
\end{align}
This shows that our Shapley values $\Sh^G_{\lC i\rC}(v)$ with the path uniform weighting on $G$ exactly matches Shapley's original definition $\Sh_i(v)$ on Boolean algebras $L$. 
\end{example}

\begin{example}[Shapley values for coalitions]
\label{eg:recover-shapley-coalitions}
Grabisch \cite{grabisch1997k} generalized Shapley values to coalitions, instead of only individual players, to introduce an interaction index in the context of fuzzy measure theory. For a finite set $I$ of players and any function $v:L \to \R$ with $v(\emptyset)=0$ and Möbius transform $w$ (calculated on $L$) he derived the following Shapley value formula for a coalition $r \ins I$:
\begin{align}
    \Sh_r(v) = \sum_{r \ins y \ins I}\frac{1}{|y|-|r|+1} \cdot w(y). \label{eq:Shapley-Grabisch-coalition}
\end{align}
In terms of game theory, this corresponds to the situations in which the players in $r$ decide to act like a single player, only ever joining or leaving coalitions as a group, even though $w$ is calculated on the full inclusion lattice. 
To compare this to the approach in this paper, we more generally decompose the whole set $I$ of players into $K\ge 1$ non-empty disjoint subsets (coalitions), $r_k \ins I$, $k\in [K]$:
\begin{align}
   I&= r_1 \dcup r_2 \dcup \dots \dcup r_K, & \tilde \Rt&:=\lC r_k \st k \in [K] \rC.
\end{align}
We then introduce the following set of subsets of $I$:
\begin{align}
    \tilde V &:= \lC y \ins I \st y\neq \emptyset, \forall r \in \tilde \Rt. r \ins y \lor r \cap y=\emptyset  \rC,
\end{align}
where every $y \in \tilde V$ either fully contains a coalition $r \in \tilde \Rt$ or is disjoint from it (for each $r \in \tilde \Rt$ individually).
Note that $\tilde V$ is ordered by inclusion and is isomorphic to a Boolean algebra, where the bottom element was removed. The DAG of its Hasse diagram is now denoted by $\tilde G=(\tilde V, \tilde E)$. 
We are thus in a similar situation as in \Cref{eg:recover-shapley-original}, where $\tilde \Rt$ is replacing $I$. So, if we abbreviate for $y \in \tilde V$:
\begin{align}
    \|y\| &:= \lI\lC r \in \tilde\Rt \st r \ins y \rC\rI,
\end{align}
then we can use the formula from \Cref{eg:recover-shapley-original} and get for $r \in \tilde \Rt$:
\begin{align}
    \Sh^{\tilde G}_r(v) &= \sum_{\substack{y \in \tilde V\\r \ins y}} \frac{1}{\|y\|}\cdot v_{\tilde G}'(y).
\end{align}
If we further assume that $|r_k|=1$ for $k \ge 2$, 
%and $\tau(r)=1$ for all $r \in \tilde \Rt$, 
then for $y \in \tilde V$ we get:
\begin{align}
    \|y\|&= 
    \begin{cases}
        |y|, &\text{ if } r_1 \cap y = \emptyset, \\
        |y|-|r_1|+1, &\text{ if } r_1 \ins y.
    \end{cases}
\end{align}
In this case, we then get the following explicit formulas:
\begin{align}
    \Sh^{\tilde G}_{r_1}(v) &= \sum_{\substack{y \in \tilde V\\r_1 \ins y}} \frac{1}{|y|-|r_1|+1}\cdot v_{\tilde G}'(y), \\
    \Sh^{\tilde G}_{r_k}(v) &= \sum_{\substack{y \in \tilde V\\r_k \ins y\\r_1 \ins y}} \frac{1}{|y|-|r_1|+1}\cdot v_{\tilde G}'(y) + \sum_{\substack{y \in \tilde V\\ r_k \ins y\\r_1 \cap y = \emptyset}} \frac{1}{|y|}\cdot v_{\tilde G}'(y), & k& \ge 2.
\end{align}
We see that this Shapley value $\Sh^{\tilde G}_r(v)$ for coalition $r \ins I$ with the path uniform weighting, again, recovers the definition $\Sh_r(v)$ from \cite{grabisch1997k}. It is important to note that the synergy function $v_{\tilde G}'$ of $v$ needs to be computed on the newly defined Hasse diagram $\tilde G$ and not on the Hasse diagram $G$ from \Cref{eg:recover-shapley-original}, where all players are treated individually. In this final but crucial detail, our proposal differs from that of \cite{grabisch1997k}. 
\end{example}

\begin{example}[Shapley values for coalitions with coalition strength]
\label{eg:recover-shapley-coalitions-str}
    Consider again the situation from \Cref{eg:recover-shapley-coalitions}.
    Now we assume that, a priori, we have different coalition strengths, e.g.\ $\tau(r)=|r|$ for $r \in \tilde \Rt$. Then the Shapley value formula changes according to \Cref{thm:shapley-main}, to:
\begin{align}
    \Sh^{(\tilde G,\tau)}_r(v) 
    &= \sum_{\substack{y \in \tilde V\\r \ins y}} \frac{\displaystyle\tau(r) \cdot \pi^{\tilde G}(r,y)}{\displaystyle\sum_{\substack{u \in \tilde \Rt\\u \ins y}}\tau(u)\cdot \pi^{\tilde G}(u,y)}\cdot v_{\tilde G}'(y) 
    = \sum_{\substack{y \in \tilde V\\r \ins y}} \frac{\displaystyle\tau(r) \cdot (\|y\|-1)!}{\displaystyle\sum_{\substack{u \in \tilde \Rt\\u \ins y}}\tau(u)\cdot (\|y\|-1)!}\cdot v_{\tilde G}'(y) \\
    &= \sum_{\substack{y \in \tilde V\\r \ins y}} \frac{\displaystyle\tau(r)}{\displaystyle\sum_{\substack{u \in \tilde \Rt\\u \ins y}}\tau(u)}\cdot v_{\tilde G}'(y).
\end{align}
If we now plug in the natural choice of $\tau(r)=|r|$ we get the following Shapley value formula for coalitions (with player count coalition strength):
\begin{align}
    \Sh^{(\tilde G,\tau)}_r(v) = \sum_{\substack{y \in \tilde V\\r \ins y}} \frac{|r|}{|y|}\cdot v_{\tilde G}'(y).
\end{align}
Note the similarity to the original single player version of Shapley values in \Cref{eq:Shapley-original} 
and again the differences to the formula for coalitions in \Cref{eq:Shapley-Grabisch-coalition} from \cite{grabisch1997k}.
\end{example}

\begin{example}[Shapley values for a reverse tree]\label{ex:treeDAG} Consider the following DAG $G$ with value function $v$ and synergy function $w$. 
We are going to compute the corresponding Shapley values for $(G,v)$.
\tikzset{> = stealth}
\begin{center}
    $G = $  
\begin{tikzpicture}[baseline=(current bounding box.center)]
    \node (a) at (0,0) {$a$};
    \node (b) at (1,0) {$b$};
    \node (c) at (2,0) {$c$};
    \node (d) at (0.5,1) {$d$};
    \node (e) at (1,2) {$e$};
    \path[->] (a) edge (d);
    \path[->] (b) edge (d);
    \path[->] (d) edge (e);
    \path[->] (c) edge (e);
\end{tikzpicture}
\quad \quad $v = $ 
\begin{tikzpicture}[baseline=(current bounding box.center)]
    \node (a) at (0,0) {$1$};
    \node (b) at (1,0) {$1$};
    \node (c) at (2,0) {$1$};
    \node (d) at (0.5,1) {$2$};
    \node (e) at (1,2) {$7$};
    \path[->] (a) edge (d);
    \path[->] (b) edge (d);
    \path[->] (d) edge (e);
    \path[->] (c) edge (e);
\end{tikzpicture}
\quad \quad $w = $ 
\begin{tikzpicture}[baseline=(current bounding box.center)]
    \node (a) at (0,0) {$1$};
    \node (b) at (1,0) {$1$};
    \node (c) at (2,0) {$1$};
    \node (d) at (0.5,1) {$0$};
    \node (e) at (1,2) {$4$};
    \path[->] (a) edge (d);
    \path[->] (b) edge (d);
    \path[->] (d) edge (e);
    \path[->] (c) edge (e);
\end{tikzpicture}
\end{center}
For the path uniform weighting we need to compute the number of directed paths $\pi_r$ from each root $r \in \Rt(G)$ to every $y \in V$ and also sum them up via $\pi=\sum_{r \in \Rt(G)} \pi_r$. We then get:
\begin{center}
    $\pi_a = $  
\begin{tikzpicture}[baseline=(current bounding box.center)]
    \node (a) at (0,0) {$1$};
    \node (b) at (1,0) {$0$};
    \node (c) at (2,0) {$0$};
    \node (d) at (0.5,1) {$1$};
    \node (e) at (1,2) {$1$};
    \path[->] (a) edge (d);
    \path[->] (b) edge (d);
    \path[->] (d) edge (e);
    \path[->] (c) edge (e);
\end{tikzpicture}
\qquad $\pi_b = $ 
\begin{tikzpicture}[baseline=(current bounding box.center)]
    \node (a) at (0,0) {$0$};
    \node (b) at (1,0) {$1$};
    \node (c) at (2,0) {$0$};
    \node (d) at (0.5,1) {$1$};
    \node (e) at (1,2) {$1$};
    \path[->] (a) edge (d);
    \path[->] (b) edge (d);
    \path[->] (d) edge (e);
    \path[->] (c) edge (e);
\end{tikzpicture}
\qquad $\pi_c = $ 
\begin{tikzpicture}[baseline=(current bounding box.center)]
    \node (a) at (0,0) {$0$};
    \node (b) at (1,0) {$0$};
    \node (c) at (2,0) {$1$};
    \node (d) at (0.5,1) {$0$};
    \node (e) at (1,2) {$1$};
    \path[->] (a) edge (d);
    \path[->] (b) edge (d);
    \path[->] (d) edge (e);
    \path[->] (c) edge (e);
\end{tikzpicture}
\qquad
$\pi =$ 
\begin{tikzpicture}[baseline=(current bounding box.center)]
    \node (a) at (0,0) {$1$};
    \node (b) at (1,0) {$1$};
    \node (c) at (2,0) {$1$};
    \node (d) at (0.5,1) {$2$};
    \node (e) at (1,2) {$3$};
    \path[->] (a) edge (d);
    \path[->] (b) edge (d);
    \path[->] (d) edge (e);
    \path[->] (c) edge (e);
\end{tikzpicture}
\end{center}
To compute the total path weights $s_r$ for each root $r \in \Rt(G)$ we need to divide $s_r=\frac{\pi_r}{\pi}$. As a sanity check, we also sum them up via $s=\sum_{r \in \Rt(G)}s_r$, which should equal $1$ everywhere. 
We get:
\begin{center}
    $s_a =$  
\begin{tikzpicture}[baseline=(current bounding box.center)]
    \node (a) at (0,0) {$1$};
    \node (b) at (1,0) {$0$};
    \node (c) at (2,0) {$0$};
    \node (d) at (0.5,1) {$\frac{1}{2}$};
    \node (e) at (1,2) {$\frac{1}{3}$};
    \path[->] (a) edge (d);
    \path[->] (b) edge (d);
    \path[->] (d) edge (e);
    \path[->] (c) edge (e);
\end{tikzpicture}
\qquad $s_b =$ 
\begin{tikzpicture}[baseline=(current bounding box.center)]
    \node (a) at (0,0) {$0$};
    \node (b) at (1,0) {$1$};
    \node (c) at (2,0) {$0$};
    \node (d) at (0.5,1) {$\frac{1}{2}$};
    \node (e) at (1,2) {$\frac{1}{3}$};
    \path[->] (a) edge (d);
    \path[->] (b) edge (d);
    \path[->] (d) edge (e);
    \path[->] (c) edge (e);
\end{tikzpicture}
\qquad $s_c =$ 
\begin{tikzpicture}[baseline=(current bounding box.center)]
    \node (a) at (0,0) {$0$};
    \node (b) at (1,0) {$0$};
    \node (c) at (2,0) {$1$};
    \node (d) at (0.5,1) {$0$};
    \node (e) at (1,2) {$\frac{1}{3}$};
    \path[->] (a) edge (d);
    \path[->] (b) edge (d);
    \path[->] (d) edge (e);
    \path[->] (c) edge (e);
\end{tikzpicture}
\qquad $s =$ 
\begin{tikzpicture}[baseline=(current bounding box.center)]
    \node (a) at (0,0) {$1$};
    \node (b) at (1,0) {$1$};
    \node (c) at (2,0) {$1$};
    \node (d) at (0.5,1) {$1$};
    \node (e) at (1,2) {$1$};
    \path[->] (a) edge (d);
    \path[->] (b) edge (d);
    \path[->] (d) edge (e);
    \path[->] (c) edge (e);
\end{tikzpicture}
\end{center}
We then get the Shapley values with path uniform weights as the scalar product of the above diagrams of $s_r$ and $w$ for each root $r \in \Rt(G)$ as follows:
\begin{align}
    \Sh^G_r(v) &= \langle s_r, w \rangle.
\end{align} 
 We then compute those values to:
\begin{align}
    \Sh^G_a(v)&= \frac{7}{3}, & 
    \Sh^G_b(v)&= \frac{7}{3}, & 
    \Sh^G_c(v)&= \frac{7}{3}.
\end{align}
\end{example}
\begin{remark}
    The previous example illustrates how our approach differs from some of the previous generalizations. In particular, consider the definition for Shapley values on \textit{concept lattices} from \cite{faigle2016games}. Since every lattice is isomorphic to a concept lattice, their definition on concept lattices can be applied to the DAG from Example \ref{ex:treeDAG} when we add an arbitrary minimal element $\bot$. By setting $v(\bot) = w(\bot)=0$, this does not affect any of the calculations. The authors of \cite{faigle2016games} define the Shapley value as the average marginal contribution of root $i$ across all maximal chains $\text{ch(G)}$:
    \begin{align}
        \tilde{\Sh}^G_i(v) = \frac{1}{|\text{ch}(G)|}\sum_{C \in \text{ch}(G)} |S_C^i\setminus T_C^i|^{-1} (v(S_C^i) - v(T_C^i)),\label{eq:lattice_shapley_faigle}
    \end{align}
    where $S_C^i$ denotes the first element $a \in C$ for which $a\geq i$, and $T_C^i$ the direct ancestor of $S_C^i$ in $C$. We turn the graph $G$ from Example \ref{ex:treeDAG} into a lattice by adding a $\bot$, so that there are three maximal chains: $\bot \to a\to ab \to abc$, $\bot \to b \to ab \to abc$, and $\bot \to c \to abc$, where we identified $d=ab$ and $e=abc$. This results in 
    \begin{align}
        \tilde{\Sh}^G_a(v)= \tilde{\Sh}^G_b(v) &= \frac{1}{3}(1+1+\frac{6}{2})=\frac{5}{3},\\
        \tilde{\Sh}^G_c(v) &= \frac{1}{3}(5+5+1) = \frac{11}{3}.
    \end{align}
    Note that even though each of the three players have equal bargaining power (since there is zero synergy for any subcoalition), the players get a different payout under $\tilde{\Sh}$. We would argue that any coalitions that offer no marginal contribution might as well be excluded, and that the resulting symmetry should be reflected by the Shapley values. As illustrated by Example \ref{ex:treeDAG}, only the path uniform weighting achieves this. Note, however, that our approach is flexible enough to still reproduce the calculation of $\tilde{\Sh}$ by introducing an own weight function $q$. Setting $q(ab|abc):=\frac{1}{3}$, $q(c|abc):=\frac{2}{3}$, and $q(a|ab):=q(b|ab):=\frac{1}{2}$ recovers the $\tilde\Sh^G(v)=(\frac{5}{3}, \frac{5}{3}, \frac{11}{3})$ attributions. 

    While we have based our generalization of Shapley values on the synergy-based formulation, the authors of \cite{faigle2016games} have chosen to extend the ordering-based formulation:
    \begin{align}
        \Sh_i(v) = \frac{1}{n!} \sum_{R}\lp v(P_R^i)\cup\{i\} - v(P_R^i)\rp,
    \end{align}
    where the sum is over all orderings $R$ of all players, and $P_R^i$ is the set of players that precede $i$ in $R$. Our definition has the benefit that it does not depend on maximal chains at all, which means it is insensitive to adding zero-synergy elements that can modify the number of maximal chains. 
\end{remark}

\begin{example}
    \label{ex:poset_game}
    \tikzset{> = stealth}
    Consider the following DAG. Note that it can be interpreted as the Hasse diagram of a partial order, but not of a lattice. We in addition specify a value function $v: G \to \mathbb{R}$. 
\begin{center}
    $G$ = \begin{tikzpicture}[baseline=(current bounding box.center)]
    \node (a) at (0,0) {$a$};
    \node (b) at (1,0) {$b$};
    \node (c) at (0,1) {$c$};
    \node (d) at (1,1) {$d$};
    \path[->] (a) edge (c);
    \path[->] (a) edge (d);
    \path[->] (b) edge (c);
    \path[->] (b) edge (d);
\end{tikzpicture}
\quad \quad $v$ = \begin{tikzpicture}[baseline=(current bounding box.center)]
    \node (a) at (0,0) {$2$};
    \node (b) at (1,0) {$1$};
    \node (c) at (0,1) {$4$};
    \node (d) at (1,1) {$5$};
    \path[->] (a) edge (c);
    \path[->] (a) edge (d);
    \path[->] (b) edge (c);
    \path[->] (b) edge (d);
\end{tikzpicture}
\end{center}

Note that $\mu_G(a, c) = \mu_G(a, d) = \mu_G(b, c) = \mu_G(b, d) = -1$. This allows us to also calculate the Möbius inversion $w$ of $v$ over $G$. The weights $s_a$ are also straighforwardly calculated to be:

\begin{center}
$w =$ \begin{tikzpicture}[baseline=(current bounding box.center)]
    \node (a) at (0,0) {$2$};
    \node (b) at (1,0) {$1$};
    \node (c) at (0,1) {$1$};
    \node (d) at (1,1) {$2$};
    \path[->] (a) edge (c);
    \path[->] (a) edge (d);
    \path[->] (b) edge (c);
    \path[->] (b) edge (d);
\end{tikzpicture}
\quad$s_a = $ \begin{tikzpicture}[baseline=(current bounding box.center)]
    \node (a) at (0,0) {$1$};
    \node (b) at (1,0) {$0$};
    \node (c) at (0,1) {$\frac{1}{2}$};
    \node (d) at (1,1) {$\frac{1}{2}$};
    \path[->] (a) edge (c);
    \path[->] (a) edge (d);
    \path[->] (b) edge (c);
    \path[->] (b) edge (d);
\end{tikzpicture}
\quad $s_b = $ \begin{tikzpicture}[baseline=(current bounding box.center)]
    \node (a) at (0,0) {$0$};
    \node (b) at (1,0) {$1$};
    \node (c) at (0,1) {$\frac{1}{2}$};
    \node (d) at (1,1) {$\frac{1}{2}$};
    \path[->] (a) edge (c);
    \path[->] (a) edge (d);
    \path[->] (b) edge (c);
    \path[->] (b) edge (d);
\end{tikzpicture}
\end{center}
The Shapley values can then be calculated through the scalar product $\langle s_r, w \rangle$:
\begin{align}
    \Sh_a^G(v) = 3.5\quad \quad \quad
    \Sh_b^G(v) = 2.5
\end{align}
Note that this indeed seems like the fairest attribution in this case: `players' $a$ and $b$ are fully symmetric, except for the fact that $a$ contributes 1 unit more by herself. That unit of difference is precisely reflected by the difference in Shapley values. 

Since $G$ is not a lattice, it was previously not possible to calculate these Shapley values. Still, $G$ can be naturally interpreted as the structure underlying a cooperative game. Let the goal of the game be to solve two puzzles $P_1$ and $P_2$, and allow each player to only solve one of them. Let $c\coloneq (a,b)$ denote the situation in which $a$ solves $P_1$ and $b$ solves $P_2$, and $d\coloneq (b, a)$ denote the alternative. Then we can interpret the value function above as the situation in which player $a$ is the better puzzle solver, and $P_2$ is the harder of the two puzzles, so that $v((b,a))>v((a, b))$. Our definition ensures that the better puzzler gets the most credit. 
\end{example}

\begin{example}[Shapley value for a spin in the Ising model]
\label{ex:ising}
The (generalized) Ising model for a set of variables $S$ is given by the Boltzmann distribution
\begin{align}
    p(S=s) &= \mathcal{Z}^{-1} \exp{(-\beta E(S=s))}
    \intertext{where $\mathcal{Z}$ is a normalisation constant, $\beta$ the inverse temperature, and the energy $E$ in its most general form is written in terms of interactions among subsets of spins:}
    E(S=s) &= \sum_{T\subseteq S} J_T \prod_{i=1}^{|T|} T_i
\end{align}
When the spin variables take values $\{0, 1\}$, the interactions can be expressed in terms of the energies using the Möbius inversion theorem \cite{jansma2023higher}:
\begin{align}
    I_S(E) = \beta \sum_{T\subseteq S} (-1)^{|S\setminus T|} E(T=1, S\setminus T=0)
\end{align}
The implicit mereology is thus that of the power set of $S$, which means Shapley values are straightforwardly calculated by summing (higher-order) Ising interactions that $i$ takes part in:
\begin{align}
    \Sh_i(E) = \sum_{i\subseteq T\subseteq S} \frac{I_T}{|T|}
\end{align}
The Shapley value for spin $i$ now represents that part of the total energy that can be attributed to spin $i$. This might be useful is you have a limited budget to flip spins. For example, let $S=\{a, b, c, d\}$, with $I_{ab}=I_{ac}=I_{ad}=1$, $I_{bcd}=x$, and all other interactions set to zero. This Ising model can be represented by the following hypergraph:

\begin{center}
\begin{tikzpicture}[
        spin/.style={circle, draw, thick, minimum size=8mm, font=\small, inner sep=0pt, fill=white},
        interaction/.style={thick},
        scale=2
      ]
    
      % Spins (square arrangement) - only coordinates first
      \coordinate (a) at (0,0);
      \coordinate (b) at (1,0);
      \coordinate (c) at (1,1);
      \coordinate (d) at (0,1);
    
      % Three-point interaction: shaded triangle (drawn first = behind)
      \fill[blue!30] (b) -- (c) -- (d) -- cycle;
    
      % Pairwise interactions from a
      \draw[interaction] (a) -- (b) node[midway, below] {$I_{ab}$};
      \draw[interaction] (a) -- (c) node[near start, right] {$I_{ac}$};
      \draw[interaction] (a) -- (d) node[midway, left] {$I_{ad}$};
    
      % Nodes on top
      \node[spin] at (a) {$a$};
      \node[spin] at (b) {$b$};
      \node[spin] at (c) {$c$};
      \node[spin] at (d) {$d$};
    
      % Region label
      \node[blue!30] at (1.25,0.65) {$I_{bcd}$};
    
    \end{tikzpicture}
\end{center}
    
First, note that the Shapley values all go to zero as $\beta\to 0$. Indeed, at infinite temperature, all energy is thermal and cannot be attributed to individual spins. However, let us fix $\beta=1$ and consider the ratio $\frac{\Sh_a(E)}{\Sh_d(E)}$ as a function of the 3-point interaction $I_{bcd}$ in Figure \ref{fig:ising_shap}. It can be seen that $\Sh_d\geq \Sh_a \iff I_{bcd}\geq 3$. The Shapley value of an Ising spin is thus a measure of `importance' of a spin to the energy of the fully magnetized state $s=\vec{1}$. While this example did not require generalizing beyond the power set mereology, it is included to illustrate that Shapley values are an appropriate technique to project down higher-order structure in contexts beyond game theory.

\begin{figure}
    \centering
    \begin{tikzpicture}
      \begin{axis}[
        width=8cm, height=6cm,
        xlabel={$J_{bcd}$},
        ylabel={$\frac{\Sh_{a}(E)}{\Sh_{d}(E)}$},
        xlabel style={font=\Large},
        ylabel style={font=\Large, rotate=270, xshift=-0.5cm, yshift=0cm},
        xtick={0,1,2,3,4},
        ytick={0,1,2,3},
        ticklabel style={font=\Large},
        domain=-1:6,
        samples=100,
        ymin=-0.3,
        ymax=3.3,
        xmin=-0.3,
        xmax=4.3
      ]
        \addplot[blue, thick] {(3/2)/(x/3+1/2)};
        % horizontal line at y=1
        \addplot[gray, dashed] {1};
        % vertical line at x=2
      \end{axis}
    \end{tikzpicture}
    \caption{The Shapley value of spin $d$ in the higher-order Ising model from Example \ref{ex:ising} becomes larger than the Shapley value for spin $a$ whenever the 3-point interaction $J_{bcd}$ exceeds $3$.}
    \label{fig:ising_shap}
\end{figure}
\end{example}

\section{Conclusion}

We have presented a double generalization of two different, but related quantities: Möbius inversions and Shapley values. Möbius inversions are a principled way to define and calculate higher-order quantities of ring-valued functions, where the `order' in higher-order has historically been a partial order. We generalized this to Abelian group-valued functions on weighted directed acyclic multigraphs. A similar generalization was introduced for Shapley values, which hitherto were only studied for real-valued functions on lattices. We believe that Shapley values are so closely associated with game theory for mostly historical reasons, and aim to understand them much more generally as the `right' way to project down higher order structure, be it in coalitional games, Ising models, information theory, machine learning, or any other context where higher-order quantities are commonly encountered. We have shown that there are natural settings for Shapley values on non-lattices (see e.g. \Cref{ex:poset_game}), and believe that also the more general codomain will lead to new insights. Note that Abelian group-valued functions in particular include vector-valued functions. Shapley values were already a popular tool in the machine learning community for feature attribution of real-valued predictions, but with modern transformer-based machine learning models embedding all kinds of data in vector spaces, we expect our generalization to extend this to vector-based decomposition and attribution methods. 

Throughout the literature, Shapley values are commonly defined through an \emph{axiomatic} approach: One demands linearity, null player, efficiency, and some notion of symmetry, which uniquely constrains the Shapley values. We instead first defined Shapley values as a systematic way to project down higher-order structure (as the projection of the synergy function onto the roots, Definition~C.1 in the appendix), and then showed that this projection is in fact the unique one that satisfies the generally accepted axioms of linearity, null player, and efficiency for a given set of normalized weights. Usually, the symmetry axiom is then invoked to uniquely constrain the weights as well, but we showed that an arguably simpler set of axioms already suffices. This definition already obeys what we called the \emph{weak elements} property: zero-synergy coalitions can be ignored (see \Cref{thm:properties-shapley-values} in the appendix). After imposing linearity and the \emph{flat hierarchy} axioms (which we imagine are hard to argue against), the introduction of the weak elements axiom already uniquely fixes the Shapley values (\ref{cor:shapley-main}). The resulting definition agrees with Shapley's on the power set lattice, but disagrees with other generalizations that were based on generalized symmetry axioms. Given that the weak elements property is already satisfied by all projections that satisfy linearity, null player, and efficiency, we do not think it is a very strong demand. Furthermore, in the context of game theory it makes intuitive sense: The introduction of a new coalition that does not achieve anything should not change the payout to players.

Crucial to the widespread adoption of Shapley values in the machine learning community have been efficient algorithms for their estimation. While an exact calculation typically scales only linearly in the number $|V|$ of elements in the graph, $|V|$ itself commonly scales at least exponentially in the number of roots. This is the case, for example, for the original inclusion lattice, as well as for partition lattices and the free distributive lattice, both of which are commonly encountered mereologies \cite{jansma2025mereological,jansma2025decomposing}. We expect efficient methods to approximate or bound the Shapley values on similarly complex graphs to be developed in the future. However, there are graphs which allow for more efficient estimation than hitherto possible. On full binary tree graphs (which arise naturally in the context of syntax trees for example) the time complexity of Möbius inversion and Shapley value calculation scales linearly with $|\Rt(G)|$. Some graphs might appear often, in which case one could pre-compute the Möbius function and the matrix $\pi^G(r, y)$, which would reduce the calculation of Shapley values to a matrix multiplication. 

We hope that this generalized framework opens up new possibilities to study complex systems \emph{the Rota way.}

\bibliographystyle{alpha}
\bibliography{references}

\appendix

\section{Projection operator properties}
\label{sec:projection-operators}

\begin{definition}[Admissible projection set]
\label{def:admissible-proj-set}
    Let $G=(V,E)$ be a DAMG and $S \ins V$.
    \begin{enumerate}
    \item We call $S$ an \emph{admissible projection set for $G$} if, with $T := S \sm \Rt(G)$ and $U:=S \cap \Rt(G)$, for $y \in V^{\sm T}$ we have:
    \begin{align}
        \Pa^{G^{\sm T}}(y) \cap U\neq \emptyset &\implies \Pa^{G^{\sm T}}(y) \ins U.
    \end{align}
    \item We call $S$ a \emph{restricted admissible projection set for $G$} if it is an admissible projection set for $G$ and does not (fully) contain any directed path $r \tuh[e_1] \cdots \tuh[e_m] l$ from any root $r \in \Rt(G)$ to any leaf $l \in \Lf(G)$ in $G$.
    \end{enumerate}
\end{definition}

\begin{theorem}[Properties of the projection operators]
\label{thm:proj-stable-prop}
    Let $R$ be a ring and $A$ be an $R$-module.
    Let $G=(V,E)$ be a DAMG and $q:E \to R$ an edge weight function, $v: V \to A$ be a value function and $w=v'_G: V \to A$ its synergy function w.r.t.\ $G$.
    Furthermore, let $S \ins V$ be any subset. Then the following properties hold:
    \begin{enumerate}
        \item \emph{DAMG}: $G^{\sm S}$ is a DAMG.
        \item \emph{Roots}: We have the inclusion: $\Rt(G) \sm S \ins \Rt(G^{\sm S})$, with equality if $S \cap \Rt(G)=\emptyset$.
        \item \emph{Ancestors}: For every $y\in V^{\sm S}$ we have: $\Anc^{G^{\sm S}}(y) = \Anc^G(y)\sm S$.
        \item \emph{Directed paths}: For $x,y \in V^{\sm S}$ we have the bijection between the sets of directed paths: $\Pi^{G^{\sm S}}(x,y) \cong \Pi^G(x,y)$ and thus equality of their cardinalities: $\pi^{G^{\sm S}}(x,y) = \pi^G(x,y)$.
        \item \emph{Normalized weights}: If $q$ is normalized and $S$ an admissible projection set for $G$ then $q^{\sm S}$ is a normalized weight function on $G^{\sm S}$ and $(G^{\sm S},q^{\sm S})$ forms an $R$-PDAMG.
        \item \emph{Total path weights}: For every $x,y \in V^{\sm S}$ we have for the total path weights: $s^{\sm S}(x|y)=s(x|y)$.
        \item \emph{Synergy}: $w^{\sm S}$ is the synergy function of $v^{\sm S}$ w.r.t.\ $G^{\sm S}$.
        \item \emph{Linearity}: $w^{\sm S}$ is $R$-linear and $A$-linear in $v$ (and also in $v^{\sm S}$).
        \item \emph{Null element}: If $y \in V^{\sm S}$ is a null element for $(G,v)$ then it is also a null element for $(G^{\sm S},v^{\sm S})$.
        \item \emph{Efficiency}: If $q$ is normalized and $S \cap \Rt(G)$ is empty (or consists only of null elements for $(G,v)$) then: \begin{align}
            \sum_{x \in V^{\sm S}} w^{\sm S}(x) = \sum_{x \in V} w(x).
        \end{align}
        \item \emph{Commutativity}: If $U \ins V$ is another subset, disjoint from $S$, then we have:
            \begin{align}
        (G^{\sm S})^{\sm U} &= (G^{\sm U})^{\sm S}, &
        (q^{\sm S})^{\sm U} &= (q^{\sm U})^{\sm S}, &
        (w^{\sm S})^{\sm U} &= (w^{\sm U})^{\sm S}, &
        (v^{\sm S})^{\sm U} &= (v^{\sm U})^{\sm S}.
    \end{align}
        \item \emph{Weak elements}: Let $S \dcup U \ins V$ be disjoint subsets of weak elements for $(G,v)$ then $U$ is also a subset of weak elements for $(G^{\sm S},v^{\sm S})$.
    \item \emph{Element weights}: Let $\tau: \Rt(G) \to R$ be a root weight function (see \Cref{def:weighted-damg}). If $S$ is an admissible projection set for $G$, then: $\tau^{\sm S}=\tau|_{V \sm S}$.
    \end{enumerate}
    \begin{proof}
        Recursively applying \Cref{lem:proj-damg}, \Cref{lem:proj-roots}, \Cref{lem:proj-anc}, \Cref{lem:proj-norm-weights}, \Cref{lem:proj-total-weights},  \Cref{lem:proj-efficiency}, \Cref{lem:proj-null-player},  \Cref{rem:proj-linearity}, \Cref{lem:projections-commute}, \Cref{lem:proj-str} the corresponding statements follow immediately.
    \end{proof}
\end{theorem}

\begin{proof}[Proof of \Cref{lem:path-uniform-total-weights}]
\label{lem:path-uniform-total-weights-proof}
First note that for $y \notin \Desc^G(x)$ both sides are zero.
So, we can assume $y \in \Desc^G(x)$.

To show the claim, we use the recursion formulas from \Cref{not:total-weights-q} and \Cref{lem:path-uniform-recursion}.

As initialization, we start with $y=x$ and get:
\begin{align}
    \frac{\pi^G(x)}{\pi^G(y)}\cdot \pi^G(x,y) &= 1 = s(x|y).
\end{align}
For $y \in \Desc^G(x)\sm\lC x\rC$ we recursively get:
\begin{align}
 s_G(x|y) &=    \sum_{z \in \Pa^G(y)} s_G(x|z) \cdot q_G(z|y) \\
 &= \sum_{z \in \Pa^G(y)} \frac{\pi^G(x)}{\pi^G(z)}\cdot \pi^G(x,z) \cdot \frac{\pi^G(z)\cdot |E(z,y)|} {\pi^G(y)} \\
 &= \frac{\pi^G(x)} {\pi^G(y)} \cdot \sum_{z \in \Pa^G(y)} \pi^G(x,z) \cdot |E(z,y)| \\
 &= \frac{\pi^G(x)} {\pi^G(y)} \cdot \pi^G(x,y).
\end{align}
This shows the claim.
\end{proof}

\begin{proof}[Proof of \Cref{thm:path-uniform-unique}]
\label{thm:path-uniform-unique-proof}
    
Let $x,y \in V$ and $S:=V \sm  \lC x,y \rC$. By \Cref{lem:proj-total-weights}:
\begin{align}
    s^G(x, y) &= (s^G)^{\setminus S}(x, y)
    \intertext{When there is no path $x\to y$ in $G$ this evaluates to zero, and when $x=y$ this evaluates to one. In all other cases, $G^{\setminus S}$ is hierarchically flat, $x$ being a root and $y$ a leaf. Since total path weights on hierarchically flat graphs are just the weights, this reduces to}
    &= (q^G)^{\setminus S}(x, y)\\
    &= \frac{|E^{\setminus S}(x, y)|}{|E^{\setminus S}(y)|}\\
    &= \frac{\pi^G(x, y)}{\pi^G(y)}\\
    &= s_G(x, y)
\end{align}
That is, for all $x, y \in V$, $s^G(x, y) = s_G(x, y)$, and thus, by \Cref{lem:q_s_equiv}
\begin{align}
    q^G(x|y) &= q_G(x|y).
\end{align}
\end{proof}

\begin{lemma}[Directed acyclic multigraph stable under projection]
\label{lem:proj-damg}
Let $G=(V,E)$ be a DAMG and $z \in V$ be a fixed element of $G$.
Then $G^{\sm z}$ is again a DAMG.
\begin{proof}
To check that $G^{\sm z}$ is a DAMG we are only left to check acyclicity. For this, let $\prec$ be a topological ordering of $G$. Then $\prec$ induces a total ordering of $V^{\sm z}$. 
If now $x\tuh[e] y \in G^{\sm z}$, then, by definition, we either have $x \tuh[e] y \in G$, which implies $x\prec y$, or, $e=e_1e_2$ and  $x\tuh[e_1] z \in G$ and $z \tuh[e_2] y \in G$, which implies $x \prec z \prec y$, which implies $x \prec y$. This shows that $\prec$ is also a topological ordering for $G^{\sm z}$ and shows that $G^{\sm z}$ is a acyclic and thus DAMG.
\end{proof}
\end{lemma}

\begin{lemma}[Roots under projection]
\label{lem:proj-roots}
Let $G=(V,E)$ be a DAMG and $z \in V$ be a fixed element of $G$.
Then we have the following inclusions and equalities for the set of roots: 
\begin{align}
  \Rt(G) \sm \lC z \rC \ins \Rt(G^{\sm z}) & = \begin{cases}
       \Rt(G) , &\text{ if } z \notin \Rt(G),\\
       \lp \Rt(G) \sm \lC z \rC \rp \dcup \lC u \in V^{\sm z} \st \Pa^G(u)=\lC z\rC \rC,&\text{ if } z \in \Rt(G).
   \end{cases} 
\end{align}
A dual statement holds for the set of leaves.
\begin{proof}
Let $y \in V^{\sm z}$.
Assume that either $y \in \Rt(G) \sm \lC z \rC$, or, $z \in \Rt(G)$ and $\Pa^G(y)=\lC z\rC$.
By way of contradiction, assume that
 $y \notin \Rt(G^{\sm z})$. Then $\Pa^{G^{\sm z}}(y) \neq \emptyset$ and there exists an edge $x \tuh[e]y \in E^{\sm z}_{\hd}(y)$. This implies, by the definition of $E^{\sm z}$, that $x\tuh[e] y \in G$ or $e=e_1e_2$ with $x \tuh[e_1] z \tuh[e_2] y \in G$, which shows that $\Pa^G(y) \neq \emptyset$ or $\Pa^G(z) \neq 0$, in contradiction to $y \in \Rt(G)$ or $z \in \Rt(G)$.
    
For the reverse inclusion, assume that  $y \in \Rt(G^{\sm z})$.
By way of contradiction, let either $y \in V^{\sm z} \sm \Rt(G)$ and $z \notin \Rt(G)$, or, $\Pa^G(y) \neq \lC z \rC$ and $z \in \Rt(G)$. 
So, there either exists an edge $z \tuh[e] y \in G$ (with tail vertex $z$) and $z \notin \Rt(G)$ or there is an edge $x \tuh[e'] y \in G$ (with any other tail vertex $x$). If $z \notin \Rt(G)$ ten there exists an $u \in V^{\sm z}$ and an edge $u \tuh[\tilde{e}] z \in G$. This implies either $u \tuh[\tilde{e}] z \tuh[e] y \in G$ or $x \tuh[e'] y \in G$. By definition of $G^{\sm z}$ we get that either $u \in \Pa^{G^{\sm z}}(y)$ or $x \in \Pa^{G^{\sm z}}(y)$. So, $\Pa^{G^{\sm z}}(y)\neq \emptyset$, in contradiction to $y \in \Rt(G^{\sm z})$.  
\end{proof}
\end{lemma}

\begin{lemma}[Directed paths and ancestral relationships stable under projection]
\label{lem:proj-anc}
Let $G=(V,E)$ be a DAMG and $z \in V$ be a fixed element of $G$. 
For $x,y \in V^{\sm z}$ we claim that we have a bijection between the set of directed paths:
    \begin{align}
        \Pi^G(x,y) & \cong \Pi^{G^{\sm z}}(x,y), & \rho = (x \cdots u\tuh[e_1] z \tuh[e_2] t \cdots y) &\mapsto (x \cdots u \tuh[e_1e_2] t \cdots y)=\rho^{\sm z}, 
    \end{align} 
which maps $\rho \in G$ to $\rho \in G^{\sm z}$ if $\rho$ does not contain $z$, and, otherwise, replaces the pair of edges $u\tuh[e_1] z \tuh[e_2] t \in G$ with the one edge $u \tuh[e_1e_2] t \in G^{\sm z}$, preserving the pair of edge labels $(e_1, e_2)$.

In particular, the following properties hold for $x,y \in V^{\sm z}$:
\begin{align}
        \Anc^{G^{\sm z}}(y) &= \Anc^G(y) \sm \lC z\rC,  &
        \Desc^{G^{\sm z}}(x) &= \Desc^G(x) \sm \lC z\rC, &
        \pi^{G^{\sm z}}(x,y) &=\pi^G(x,y).
    \end{align}
\begin{proof}
First not that every directed path $\rho \in G$ can have at most one occurrence of the node $z$, since $G$ is acyclic. Similarly, every directed path $\rho \in G^{\sm z}$ can have at most one edge of the type $\tuh[e_1e_2]$, since $G^{\sm z}$ is acyclic by \Cref{lem:proj-damg}. 
So the inverse to the map above is given just by mapping $u\tuh[e_1e_2] t \in G^{\sm z}$ to $u\tuh[e_1]z\tuh[e_2] t \in G$.
This already shows all the claims.  
\end{proof}
\end{lemma}

\begin{lemma}[Normalized weight function stable under projection]
\label{lem:proj-norm-weights}
Let $G=(V,E)$ be a DAMG and $q:E \to R$ a weight function. Let $z \in V$ be a fixed element of $G$. Then $q^{\sm z}$ is a weight function on $G^{\sm z}$.
If $q$ is normalized and $z \notin \Rt(G) \sm \Lf(G)$ then $q^{\sm z}$ is also a normalized weight function on $G^{\sm z}$. 
So, $(G^{\sm z},q^{\sm z})$ forms an $R$-PDAMG.
\begin{proof}
$q^{\sm z}$ is a well-defined map, so it is a weight function on $G^{\sm z}$.
To show the normalization of $q^{\sm z}$, let $y \in V^{\sm z}$ with $y \notin \Rt(G^{\sm z})$. 
The by \Cref{lem:proj-roots} we have that: $y \notin \Rt(G)$.
We now look at the sum:
    \begin{align}
        \sum_{x \in V^{\sm z}} q^{\sm z}(x|y) 
        &= \sum_{x \in V^{\sm z}} \lp  q(x|y) +  q(x|z) \cdot q(z|y) \rp \\
        &= \sum_{x \in V^{\sm z}} q(x|y) + \sum_{x \in V^{\sm z}} q(x|z) \cdot q(z|y) + \overbrace{q(z|z)}^{=0} \cdot q(z|y) \\
        &= \sum_{x \in V^{\sm z}} q(x|y) + \underbrace{\lp \sum_{x \in V} q(x|z)  \rp}_{\substack{=1,\\\text{ if } z \notin \Rt(G)}} \cdot \underbrace{q(z|y)}_{\substack{=0,\\\text{ if } z \in \Rt(G) \cap \Lf(G)}}\\
        &= \sum_{x \in V^{\sm z}} q(x|y) + q(z|y)\\
        &= \underbrace{\sum_{x \in V} q(x|y)}_{=1, \text{ as } y \notin \Rt(G)} \\
        &=1.
    \end{align}
    This shows that $q^{\sm z}$ is also normalized.
\end{proof}
\end{lemma}

\begin{lemma}[Total path weight function stable under projection]
\label{lem:proj-total-weights}
Let $G=(V,E)$ be a DAMG and $q:E \to R$ a weight function and $z \in V$ be a fixed element of $G$.
Let $s=s^{(G,q)}$ be the total path weight function of $(G,q)$ and $s^{\sm z}=s^{(G^{\sm z},q^{\sm z})}$ the total path weight function of the projection $(G^{\sm z},q^{\sm z})$. Then we have the equality:
\begin{align}
    s^{\sm z} = s|_{V^{\sm z}}: V^{\sm z} \times V^{\sm z } & \to R, & s^{\sm z}(x|y)&=s(x|y).
\end{align}
\begin{proof}
Let $x \in V^{\sm z}$ be fixed.
We then recursively choose $y \in \Desc^G(x)$ according to some topological ordering $\prec$ of $G$.
For $y=x$ we get:
\begin{align}
    s^{\sm z}(x|y) &= 1 = s(x|y).
\end{align}
Now consider $y \in \Desc^G(x) \sm \lC x\rC$.
We then recursively get the following: 
    \begin{align}
        s^{\sm z}(x|y) 
        & =\sum_{ u \in \Pa^{G^{\sm z}}(y)} s^{\sm z}(x|u) \cdot q^{\sm z}(u|y) \\
        & =\sum_{ u \in V^{\sm z}} s^{\sm z}(x|u) \cdot q^{\sm z}(u|y) \\
        &\overset{\text{rec.}}{=} \sum_{ u \in V^{\sm z}} s(x|u) \cdot \lp q(u|y) + q(u|z) \cdot q(z|y) \rp \\
        &= \sum_{ u \in V^{\sm z}} s(x|u) \cdot  q(u|y) +  \sum_{ u \in V^{\sm z}} s(x|u)\cdot q(u|z)  \cdot q(z|y) + s(x|z) \cdot \overbrace{q(z|z)}^{=0}  \cdot q(z|y) \\
        &= \sum_{ u \in V^{\sm z}} s(x|u) \cdot  q(u|y) +  \underbrace{\sum_{ u \in V} s(x|u) \cdot q(u|z)}_{=s(x|z)}  \cdot q(z|y)  \\
        &= \sum_{ u \in V^{\sm z}} s(x|u) \cdot  q(u|y) + s(x|z) \cdot q(z|y) \\
        &=\sum_{ u \in V} s(x|u) \cdot  q(u|y) \\
        &= s(x|y).
    \end{align}
This shows the claim.
\end{proof}
\end{lemma}

\begin{remark}[M\"obius transform under projection]
\label{rem:proj-moebius-trafo}
Let $G=(V,E)$ be a DAMG and $q:E \to R$ a weight function on $G$, $A$ be an $R$-module and $v:V \to A$ be a value function with its synergy function $w=v_G'$. Let $z \in V$ be a fixed element of $G$.
Then by \Cref{thm:moebius-trafo} the projected synergy function $w^{\sm z}$ is the M\"obius transform of $v^{\sm z}$ w.r.t.\ $G^{\sm z}$, because $v^{\sm z}$ is defined through the first property in that theorem and does not depend on any properties of $w^{\sm z}$ at all.
\end{remark}

\begin{lemma}[Efficiency stable under projection]
\label{lem:proj-efficiency}
Let $G=(V,E)$ be a DAMG and $q:E \to R$ a normalized weight function on $G$, $A$ be an $R$-module and $v:V \to A$ be a value function with its synergy function $w=v_G'$. 
Let $z \in V$ be a fixed element of $G$ with $z \notin \Rt(G)$ or $w(z)=0$ (weak element for $v$). 
Then we have the equality:
\begin{align}
    \sum_{x \in V^{\sm z}} w^{\sm z}(x) &= \sum_{x \in V} w(x).
\end{align}
\begin{proof}
We now have:
\begin{align}
    \sum_{x \in V^{\sm z}} w^{\sm z}(x) 
    &= \sum_{x \in V^{\sm z}} w(x) + \sum_{x \in V^{\sm z}} q(x|z) \cdot w(z) + \overbrace{q(z|z)}^{=0} \cdot w(z) \\
    &= \sum_{x \in V^{\sm z}} w(x) + \underbrace{\lp \sum_{x \in V} q(x|z)  \rp}_{=1, \text{ if }z \notin \Rt(G) } \cdot \underbrace{w(z)}_{\substack{=0,\\\text{ if }w(z)=0}}  \\
    &= \sum_{x \in V^{\sm z}} w(x) + w(z)  \\
    &= \sum_{x \in V} w(x).
\end{align}
This shows the claim.
\end{proof}
\end{lemma}

\begin{lemma}[Null elements stable under projection]
\label{lem:proj-null-player}
Let $G=(V,E)$ be a DAMG and $q:E \to R$ a weight function on $G$, $A$ be an $R$-module and $v:V \to A$ be a value function with its synergy function $w=v_G'$.  
Let $z \in V$ be fixed and let $y \in V^{\sm z}$ be a \emph{null element} for $(G,v)$.  Then $y$ is also a null element for $(G^{\sm z},v^{\sm z})$.
\begin{proof}
Let $x \in \Desc^{G^{\sm z}}(y)$ then by \Cref{lem:proj-anc} we have that: $x \in \Desc^G(y)$. By \Cref{eq:proj-synergy} we then have for $w=v_G'$:
\begin{align}
    w^{\sm z}(x) &= \underbrace{w(x)}_{=0} + q(x|z) \cdot w(z)
    = q(x|z) \cdot w(z). 
\end{align}
If $z \in \Desc^G(y)$ then also $w(z)=0$ and we are done.
If $z \notin \Desc^G(y)$ then $z \notin \Desc^G(x)$ and thus $E(x,z)=\emptyset$, which implies $q(x|z)=0$ and we are also done.
This shows the claim.
\end{proof}
\end{lemma}

\begin{lemma}[Projections commute]
\label{lem:projections-commute}
Let $G=(V,E)$ be a DAMG and $q:E \to R$ a weight function on $G$, $A$ be an $R$-module and $v:V \to A$ be a value function with its synergy function $w=v_G'$.
Let $z,u \in V$ with $z \neq u$. Then we have:
    \begin{align}
        (G^{\sm z})^{\sm u} &= (G^{\sm u})^{\sm z}, &
        (q^{\sm z})^{\sm u} &= (q^{\sm u})^{\sm z}, &
        (w^{\sm z})^{\sm u} &= (w^{\sm u})^{\sm z}, &
        (v^{\sm z})^{\sm u} &= (v^{\sm u})^{\sm z}.
    \end{align}
Note that if $q$ is normalized and $z,u \in V \sm \Rt(G)$ then $(q^{\sm z})^{\sm u}$ is also normalized. 
\begin{proof}
It is clear that:
    \begin{align}
        (V^{\sm z})^{\sm u} &= V\sm \lC z, u \rC = (V^{\sm u})^{\sm z}.
    \end{align}
Regarding the set of directed edges, for $x,y \in V\sm \lC z, u \rC$ we have:

\begin{align}
    E^{\sm z}(x,y) &\cong E(x,y) \dcup \lp E(x,z) \times E(z,y) \rp, \\
    (E^{\sm z})^{\sm u}(x,y) & \cong E^{\sm z}(x,y) \dcup \lp E^{\sm z}(x,u) \times E^{\sm z}(u,y) \rp \\
    &\cong \lB E(x,y)\dcup\lp E(x,z)\times E(z,y)\rp\rB  \notag\\
    & \qquad \dcup \lB\lp E(x,u) \dcup \lp E(x,z) \times E(z,u) \rp\rp \times \lp E(u,y) \dcup \lp E(u,z) \times E(z,y) \rp\rp\rB \\
    &\cong E(x,y)\dcup\lp E(x,z)\times E(z,y)\rp \dcup \lp E(x,u)\times E(u,y) \rp \notag\\
    &\quad\dcup \lp E(x,u)\times E(u,z)\times E(z,y) \rp
      \dcup \lp E(x,z)\times E(z,u)\times E(u,y) \rp \notag\\
    &\quad \dcup \bigg( E(x,z)\times \underbrace{E(z,u)\times E(u,z)}_{=\emptyset, \text{ as $G$ acyclic}}\times E(z,y) \bigg) \\
    &\cong E(x,y)\dcup\lp E(x,z)\times E(z,y)\rp \dcup \lp E(x,u)\times E(u,y) \rp \notag\\
    &\quad\dcup \lp E(x,u)\times E(u,z)\times E(z,y)\rp
      \dcup\lp E(x,z)\times E(z,u)\times E(u,y)\rp  \label{eq:proj-edge-decomp}\\
    &\overset{\text{symmetry}}{\cong} (E^{\sm u})^{\sm z}(x,y).
\end{align}
This shows that: $(G^{\sm z})^{\sm u} \cong (G^{\sm u})^{\sm z}$.
    
For the weight function: 
\begin{align}
    (q^{\sm z})^{\sm u}: (E^{\sm z})^{\sm u} \to R,
\end{align}
we use the disjoint decomposition of $(E^{\sm z})^{\sm u}$ from \Cref{eq:proj-edge-decomp} into multiplicative parts and $q$'s multiplicativity from \Cref{def:proj-damg-one}. For $x,y \in V\sm \lC z, u \rC$ we then get:
\begin{align}
   (q^{\sm z})^{\sm u}(x\tuh[e]y) &= \begin{cases}
        q(x\tuh[e]y) &\text{ if } x\tuh[e]y \in G, \\
        q(x\tuh[e_1]z)\cdot q(z\tuh[e_2]y) &\text{ if } e=e_1e_2, x\tuh[e_1] z\tuh[e_2]y \in G,\\
        q(x\tuh[e_1]u)\cdot q(u\tuh[e_2]y) &\text{ if } e=e_1e_2, x\tuh[e_1] u\tuh[e_2]y \in G,\\
        q(x\tuh[e_1]u)\cdot q(u\tuh[e_2]z)\cdot q(z\tuh[e_3]y) &\text{ if } e=e_1e_2e_3, x\tuh[e_1] u\tuh[e_2]z\tuh[e_3]y \in G,\\
        q(x\tuh[e_1]z)\cdot q(z\tuh[e_2]u)\cdot q(u\tuh[e_3]y) &\text{ if } e=e_1e_2e_3, x\tuh[e_1] z\tuh[e_2]u\tuh[e_3]y \in G,
    \end{cases} \notag\\
    &=(q^{\sm u})^{\sm z}(x\tuh[e]y).
\end{align}
If we wanted to compute this more explicitly for the pairwise weights $(q^{\sm z})^{\sm u}(x|y)$ we would get :
\begin{align}
    (q^{\sm z})^{\sm u}(x|y) 
    &= q^{\sm z}(x|y) + q^{\sm z}(x|u)\cdot q^{\sm z}(u|y) \\
    &= \lB q(x|y) + q(x|z)\cdot q(z|y) \rB + \lB \lB q(x|u) + q(x|z)\cdot q(z|u) \rB \cdot \lB q(u|y) + q(u|z)\cdot q(z|y)\rB \rB \\
    &= q(x|y) + q(x|z)\cdot q(z|y) + q(x|u)\cdot q(u|y) + q(x|u)\cdot q(u|z) \cdot q(z|y)\notag \\
    & \qquad + q(x|z)\cdot q(z|u) \cdot q(u|y) +  q(x|z)\cdot \underbrace{q(z|u) \cdot q(u|z)}_{=0, \text{ as $G$ acyclic}}\cdot q(z|y)  \\ 
    &= q(x|y) + q(x|z)\cdot q(z|y) + q(x|u)\cdot q(u|y) \notag\\
    & \qquad + q(x|u)\cdot q(u|z) \cdot q(z|y) + q(x|z)\cdot q(z|u) \cdot q(u|y)\\
    &\overset{\text{symmetry}}{=} (q^{\sm u})^{\sm z}(x|y).
\end{align}
For the synergy functions we have for $x\in V\sm \lC z, u \rC$:
    \begin{align}
        (w^{\sm z})^{\sm u}(x) 
        &= w^{\sm z}(x) + q^{\sm z}(x|u) \cdot w^{\sm z}(u) \\
        &= \lB w(x) + q(x|z) \cdot w(z) \rB + \lB q(x|u) + q(x|z)\cdot q(z|u)  \rB \cdot \lB  w(u) + q(u|z) \cdot w(z)\rB \\
        &= w(x) + q(x|z) \cdot w(z) + q(x|u) \cdot w(u) + q(x|u) \cdot q(u|z)\cdot w(z) \notag\\
        &\qquad + q(x|z) \cdot q(z|u)\cdot w(u) +  q(x|z) \cdot \underbrace{q(z|u)\cdot q(u|z)}_{=0,\text{ as $G$ acyclic}} \cdot w(z) \\
        &= w(x) + q(x|z) \cdot w(z)  + q(x|u) \cdot w(u)  
        + q(x|u) \cdot q(u|z)\cdot w(z) + q(x|z) \cdot q(z|u)\cdot w(u) \\
        &\overset{\text{symmetry}}{=} (w^{\sm u})^{\sm z}(x). 
    \end{align}
    The equality between the value functions are clear, as they only add up the synergy functions.
    So all claims are shown.
\end{proof}
\end{lemma}

\begin{definition}[Horizontal subset]
    \label{def:horizontal-subset}
Let $G=(V,E)$ be a DAMG. A subset $\Xcal \ins V$ is called \emph{horizontal subset of $G$} if  every directed path $r \tuh[e_1] \cdots \tuh[e_m] l$ from any root $r \in \Rt(G)$ to any leaf $l \in \Lf(G)$ in $G$ contains exactly one $x \in \Xcal$.
\end{definition}

\begin{remark}
    For every DAMG $G=(V,E)$ the sets $\Xcal=\Rt(G)$ and $\Xcal=\Lf(G)$ are always horizontal subsets.
\end{remark}

\begin{lemma}
    Let $G=(V,E)$ be a DAMG and $ \Xcal \ins V$ be subset.
    Then the following are equivalent:
    \begin{enumerate}
        \item $\Xcal$ is a horizontal subset of $G$, see \Cref{def:horizontal-subset}.
        \item There exists a restricted admissible projection set $S$ of $G$ such that: $\Xcal =\Rt(G^{\sm S})$.
    \end{enumerate}
\begin{proof}
    Let $\Xcal$ be a horizontal subset. Note that for $x_1,x_2 \in \Xcal$ with $x_1 \neq x_2$ we have: $x_1 \notin \Anc^G(x_2)$. Otherwise, there would exists a directed paths with $x_1$ and $x_2$ on it.
    Now put $S:=\Anc^G(\Xcal)\sm \Xcal$, $T:=S \sm \Rt(G)$, $U:=S \cap \Rt(G)$. We now claim that $S$ is a restricted admissible projection set of $G$. 
    For this first note that $\Xcal \cap S=\emptyset$ and $S$ thus cannot fully contain any directed path from a root to a leaf in $G$.
    Now consider $y \in V^{\sm T}$ and $z_0 \in \Pa^{G^{\sm T}}(y)\cap U$ then there exists a directed path $z_0 \tuh[e_1] z_1 \cdots z_{m-1} \tuh[e_m] y$ in $G$ with $z_1,\dots, z_{m-1} \in T$. 
    We can now extend it to a  directed path to a leaf $l \in \Lf(G)$:
    \begin{align}
        \underbrace{z_0}_{\in S \cap \Rt(G)} \tuh[e_1] \underbrace{z_1 \tuh[e_2] \cdots \tuh[e_{m-1}] z_{m-1} }_{ \in S\sm\Rt(G)} \tuh[e_m] \underbrace{y}_{\notin S} \tuh[e_{m+1}] \cdots \tuh[e_n] l.
    \end{align} 
    By assumption, the joint directed path contains exactly one $x \in \Xcal$. Since $x \notin S$, $\Anc^G(x)\sm\lC x\rC \ins S$ and $y \notin S$ and $z_0,\dots,z_{m-1} \in S$ we have: $y=x \in \Xcal$.
    Now consider any $z \in V^{\sm T}$ with $z \in \Pa^{G^{\sm T}}(y)$.
    Then there exists a directed path 
    $z \tuh[f_1] u_1 \cdots u_{k-1} \tuh[f_k] y$ in $G$ with $u_1,\dots, z_{k-1} \in S \sm \Rt(G)$.
    Since $y \in \Xcal$ and $\Anc^G(y) \sm \lC y\rC \ins S$ we have $z \in S \sm T = S \cap \Rt(G)=U$. 
    So $z \in U$. This shows that $S$ is an admissible projection set. 
    We now claim that $\Rt(G^{\sm S})=\Xcal$. Clearly, $\Xcal \ins \Rt(G^{\sm S})$.
    For the reverse inclusion, let $y \in \Rt(G^{\sm S})$. Then $y \notin S$.
    Then there exists a directed path $\rho=(r \tuh[e] \dots y \dots \tuh[f] l)$ in $G$ from some root $r \in \Rt(G)$ to some leaf $l \in \Lf(G)$. By assumption this path contains exactly one $x \in \Xcal$. Since $x \notin S$, $x$ cannot occur before $y$ on $\rho$. Otherwise, $\Pa^{G^{\sm S}}(y)\neq \emptyset$, in contradiction to $y \notin \Rt(G^{\sm S})$. $x$ can
    also not occur after $y$ on $\rho$, since otherwise $y \in \Anc^G(x) \sm \lC x\rC \ins S$, in contradiction to $y \notin S$. It follows $y=x \in \Xcal$. This completes the claim: $\Rt(G^{\sm S}) = \Xcal$.

For the reverse implication, let $S$ be a restricted admissible projection set of $G$, $T:=S \sm \Rt(G)$, $U:=S \cap \Rt(G)$ and $\Xcal:=\Rt(G^{\sm S})$. We now claim that $\Xcal$ is a horizontal subset of $G$. For this, consider a directed path $\rho=(z_0 \tuh[e_1] z_1 \dots z_{m-1} \tuh[e_m] z_m) \in G$
from a root $z_0 \in \Rt(G)$ to a leaf $z_m \in \Lf(G)$. First note that it is clear that $\rho$ can contain at most one $x \in \Xcal$. 
Then consider the projected directed path $\rho^{\sm T}$ in $G^{\sm T}$. 
We then have: $z_0 \in \Rt(G)=\Rt(G^{\sm T})$. 
We now either have $z_0 \in S$ or $z \notin S$. 
If $z_0 \notin S$ then $z_0 \in \Rt(G) \sm S \ins \Rt(G^{\sm S})=\Xcal$ and we are done. 
So now assume $z_0 \in S$ then $z_0 \in S \cap \Rt(G)=U$. Then let $k$ be the smallest index such that $z_k \notin 
S$, which exists since $S$ is a restricted admissible projection set of $G$. Then $z_1,\dots,z_{k-1} \in S \sm \Rt(G) =T$. Furthermore, we have: $U \ni z_0 \in \Pa^{G^{\sm T}}(z_k)$, which implies by assumption on $S$ that $\Pa^{G^{\sm T}}(z_k) \ins U$.
From this follows that $\Pa^{G^{\sm S}}(z_k) = \Pa^{(G^{\sm T})^{\sm U}}(z_k)=\emptyset$ and thus $z_k \in \Rt(G^{\sm })=\Xcal$, which shows the claim.
\end{proof}
\end{lemma}

\begin{lemma}[Normalized weights under projection w.r.t.\ subsets]
\label{lem:proj-norm-weights-sets}
    Let $G=(V,E)$ be a DAMG and $q:E \to R$ a normalized weight function on $G$. 
Let $S \ins V$ be an admissible projection set for $G$.
    Then $q^{\sm S}$ is a normalized weight function for $G^{\sm S}$ and $(G^{\sm S},q^{\sm S})$ forms an $R$-PDAMG.
\begin{proof}
    Let $T := S \sm \Rt(G)$ and $U:=S \cap \Rt(G)$. 
    By \Cref{lem:proj-norm-weights} we already know that $q^{\sm T}$ is normalized. By the made assumption on $S$, we then have for $y \in V^{\sm T}$ only the two options:
    \begin{align}
        \Pa^{G^{\sm T}}(y) \cap U = \emptyset \qquad \lor \qquad \Pa^{G^{\sm T}}(y) \ins U.
    \end{align}
    For the latter case we then have:
    \begin{align}
        \Pa^{G^{\sm S}}(y) &=\Pa^{(G^{\sm T})^{\sm U}}(y) = \emptyset,
    \end{align}
    and no normalization condition has to be checked for such $y$.
    For the former case we have:
    \begin{align}
        \Pa^{G^{\sm S}}(y) &=\Pa^{(G^{\sm T})^{\sm U}}(y) = \Pa^{G^{\sm T}}(y),
    \end{align}
    and we get:
    \begin{align}
        \sum_{x \in V^{\sm S}} q^{\sm S}(x|y) &=
        \sum_{x \in \Pa^{(G^{\sm T})^{\sm U}}(y)} (q^{\sm T})^{\sm U}(x|y) \\
        &\overset{U \ins \Rt(G^{\sm T})}{=} \sum_{x \in \Pa^{(G^{\sm T})^{\sm U}}(y)} q^{\sm T}(x|y) \\
        &= \sum_{x \in \Pa^{G^{\sm T}}(y)} q^{\sm T}(x|y) \\
        &=1.
    \end{align}
    This shows the claim.
\end{proof}
\end{lemma}

We now show how the total path weights $s$ of an $R$-PDAMG can be used to evaluate the projections $w^{\sm S}$ of synergy functions $w$ at elements $x$ whose proper descendants have been projected out.

\begin{lemma}[Projection of synergy via total path weights]
\label{lem:proj-synergy-total-weights}
    Let $G=(V,E)$ be a DAMG and $q:E \to R$ a weight function on $G$, $v : V \to A$ a value function and $w: V \to A$ its synergy function w.r.t.\ $G$. Let $x \in V$ be fixed and $S \ins V$ any subset such that:\footnote{Note that $S=V \sm \lC x \rC$ would be a maximal subset and $S=\Desc^G(x)\sm \lC x\rC$ be a minimal subset with this property.}
    \begin{align}
        \Desc^G(x)\cap S &= \Desc^G(x) \sm \lC x \rC.
    \end{align}
    Then we have the formula (for this one fixed $x$):
    \begin{align}
        w^{\sm S}(x)&= \sum_{y \in V} s(x|y) \cdot w(y) = \sum_{y \in \Desc^G(x)} s(x|y) \cdot w(y).
    \end{align}
\begin{proof}
    First consider the case $\Desc^G(x)=\lC x \rC$. Then $w^{\sm S}(x) = w(x)$ and we are done.
    So, now assume that there exists a leaf $z \in \Desc^G(x)$ of $G$ with $z \neq x$. 
    Then we can write $S=T \dcup \{z \}$.
    Note that:
    \begin{align}
        \Desc^{G^{\sm z}}(x) \cap T & = \Desc^{G^{\sm z}}(x) \sm \lC x \rC,
    \end{align}
    which implies that $T$ satisfies the same condition for $G^{\sm z}$ as $S$ for $G$. 
    By \Cref{lem:proj-total-weights} we have for all $y \in V^{\sm z}$ the equality:
    \begin{align}
        s^{\sm z}(x|y) &:= s^{(G^{\sm z},q^{\sm z})}(x|y) = s(x|y).
    \end{align}
    By induction we then get:
    \begin{align}
        w^{\sm S}(x) & = (w^{\sm z})^{\sm T}(x) \\
        &= \sum_{y \in \Desc^{G^{\sm z}}(x)} s^{\sm z}(x|y) \cdot w^{\sm z}(y) \\
        &= \sum_{y \in \Desc^{G^{\sm z}}(x)} s(x|y) \cdot \lp w(y) + q(y|z) \cdot w(z) \rp \\
        &= \sum_{y \in \Desc^{G^{\sm z}}(x)} s(x|y) \cdot  w(y) + \sum_{y \in \Desc^{G^{\sm z}}(x)} s(x|y) \cdot \underbrace{q(y|z)}_{\substack{=0\\\text{ if }y \notin \Pa^G(z)}} \cdot  w(z) \\
        &= \sum_{y \in \Desc^{G^{\sm z}}(x)} s(x|y) \cdot  w(y) + \sum_{y \in \Desc^{G^{\sm z}}(x)\cap \Pa^G(z)} s(x|y) \cdot q(y|z) \cdot  w(z) \\
        &= \sum_{y \in \Desc^{G^{\sm z}}(x)} s(x|y) \cdot  w(y) + s(x|z) \cdot  w(z) \\
        &=\sum_{y \in \Desc^G(x)} s(x|y) \cdot  w(y).
    \end{align}
This shows the claim.
\end{proof}
\end{lemma}

We now want to highlight that the path uniform weight function, see \Cref{def:path-uniform-damg}, \Cref{lem:path-uniform-recursion} and \Cref{lem:path-uniform-total-weights} have further convenient properties. One of them is that projecting and re-computing lead to the same weights.

\begin{lemma}[Path uniform weights commute with projections]
\label{lem:proj-path-unif-comm}
Let $G=(V,E)$ be a DAMG and $q_G$ the path uniform weight function on $G$ from \Cref{def:path-uniform-damg}.
Let $S \ins V \sm \Rt(G)$ be a subset.
Then we have the equality:
\begin{align}
    q_{G^{\sm S}} &= (q_G)^{\sm S} : V^{\sm S} \to \Z_{G^{\sm S}} \ins \Z_G.
\end{align}
\begin{proof}
By \Cref{thm:proj-stable-prop} we can reduce to the case: $S=\lC z \rC$.
 Let $x\tuh[e]y \in E^{\sm z}$ then we have:
 \begin{align}
     (q_G)^{\sm z}(x\tuh[e]y) &= 
     \begin{cases}
         q_G(x\tuh[e]y), & \text{ if } x\tuh[e]y \in G,\\
         q_G(x\tuh[e_1]z)\cdot q_G(z\tuh[e_2]y), & \text{ if }e=e_1e_2,\;x\tuh[e_1]z\tuh[e_2]y \in G,
     \end{cases} \\
     &=
      \begin{cases}
         \frac{\pi^G(x)}{\pi^G(y)}, & \text{ if } x\tuh[e]y \in G,\\
         \frac{\pi^G(x)}{\pi^G(z)}\cdot \frac{\pi^G(z)}{\pi^G(y)}, & \text{ if }e=e_1e_2,\; x\tuh[e_1]z\tuh[e_2]y \in G,
     \end{cases} \\
     &= \frac{\pi^G(x)}{\pi^G(y)}.\\
     &\overset{\ref{lem:proj-anc}}{=} \frac{\pi^{G^{\sm z}}(x)}{\pi^{G^{\sm z}}(y)}, \\
      &=q_{G^{\sm z}}(x\tuh[e]y).
 \end{align}
This shows the claim.
\end{proof}
\end{lemma}

We now want to investigate the interaction between projection and restriction of value functions. For this we introduce further notions.

\begin{definition}[Ancestrally closed subsets]
    Let $G=(V,E)$ be a DAMG and $\tilde V \ins V$ a subset. We call $\tilde V$ an \emph{ancestrally closed subset} of $G$ if:
    \begin{align}
        \bigcup_{x \in \tilde V} \Anc^G(x) \ins \tilde V,
    \end{align}
    for which then equality holds.
    Let $\tilde G=(\tilde V, \tilde E)$ be a subgraph of $G$. We call $\tilde G$ an \emph{ancestrally closed subgraph} (or sub-DAMG, resp.) of $G$ if $\tilde V \ins V$ is an ancestrally closed subset of $G$ and:
    \begin{align}
        \tilde E &= \bigcup_{x,y \in \tilde V} E(x,y).
    \end{align}
\end{definition}

\begin{lemma}[M\"obius transform on ancestrally closed subsets]
\label{lem:moebius-ancestrally-closed}
    Let $G=(V,E)$ be a DAMG and $\tilde G =(\tilde V, \tilde E)$ be an ancestrally closed subgraph of $G$.
    Let $v: V \to A$ be a value function and $w=v_G'$ be the synergy function of $v$ w.r.t.\ $G$.
    Furthermore, consider the restriction of $v$ to $\tilde V$:
    \begin{align}
        \tilde v:=v|_{\tilde V}: \tilde V &\to A, & \tilde v(x) &=v(x),
    \end{align}
    and let $\tilde w:=\tilde v_{\tilde G}'$ be the synergy function of $\tilde v$ w.r.t.\ $\tilde G$.
    Then we have the equality:
    \begin{align}
        \tilde w = w|_{\tilde V}: \tilde V & \to A, & \forall x \in \tilde V. \qquad \tilde w(x)=w(x).
    \end{align}
\begin{proof}
    This follows from the recursive formula in \Cref{def:moebius-trafo} and the observation that the formula only invokes the sets $\Anc^G(x)\sm\lC x \rC$, which then fully lie in $\tilde V$ at each step, since $\tilde V$ is ancestrally closed. We also need to use that: $\Anc^G(x)\sm\lC x \rC = \Anc^{\tilde G}(x)\sm\lC x \rC$, which also holds since $\tilde G$ is an ancestrally closed subgraph of $G$. We then get by induction for $x \in \tilde V$:
    \begin{align}
        w(x) &= v(x) - \sum_{y \in \Anc^G(x) \sm \lC x\rC} w(y) \\
        &= \tilde v(x) - \sum_{y \in \Anc^{\tilde G}(x) \sm \lC x\rC} \tilde w(y) \\
        &=\tilde w(x).
    \end{align}
    This shows the claim.
\end{proof}
\end{lemma}

\begin{lemma}[M\"obius transforms and projections ignore weak elements]
\label{lem:moebius-weak-elements}
    Let $G=(V,E)$ be a DAMG and $v:V\to A$ be a value function with values in any $R$-module $A$ and with synergy function $w=v_G'$. Let $W \ins V$ be a set of weak elements w.r.t.\ $(G,v)$. Let $G^{\sm W}=(V^{\sm W},E^{\sm W})$ be the projection of $G$ onto the complement of $W$.
    Define the restriction $v|_{\sm W}$ of $v$ to $V^{\sm W}$:
    \begin{align}
        v|_{\sm W}: V^{\sm W} &\to A, & v|_{\sm W}(x) &:= v(x).
    \end{align}
    Further, let $\tilde w:=(v|_{\sm W})_{G^{\sm W}}'$ be the synergy function of $v|_{\sm W}$ w.r.t.\ $G^{\sm W}$. 
    Then we have the equality:
    \begin{align}
         \tilde w =w|_{\sm W}:  V^{\sm W} &\to A, &\forall x \in V^{\sm W}. \qquad \tilde w(x) &= w(x).
    \end{align}
    If, furthermore, $q$ is any normalized weight function for $G$ with values in $R$, and, $w^{\sm W}$ the projection of $w$ onto the complement of $W$ w.r.t.\ $(G,q)$, then we also have the equality:
    \begin{align}
         w^{\sm W} =w|_{\sm W}: V^{\sm W} &\to A, &\forall x \in V^{\sm W}. \qquad w^{\sm W}(x) &= w(x).
    \end{align}
    In particular, we have the equality of projection and restriction of the following value functions\footnote{Note that the restriction of $v$ does not depend on the weight function $q$, while the projection a priori does.}:
    \begin{align}
        v^{\sm W}=v|_{\sm W}: V^{\sm W} & \to A, &\forall x \in V^{\sm W}. \qquad v^{\sm W}(x) &= v(x).
    \end{align}
\begin{proof}
W.l.o.g.\ we can assume: $W=\lC z \rC$.
Since $G$ is a DAMG there exists a topological ordering $\prec$ of $G$, which then also induces a topological ordering on $G^{\sm z}$. Let $x \in V^{\sm z}$.
By induction we can now assume that $\tilde w(y)=w(y)$ for all $y \prec x$.
If $x \prec z$ then we see that $\Anc^{G^{\sm z}}(x)=\Anc^G(x)$ and thus, by the formula for the M\"obius transforms from \Cref{def:moebius-trafo}:
\begin{align}
    w(x) &= v(x) - \sum_{y \in \Anc^G(x) \sm \lC x\rC} w(y) \\
    & = v|_{\sm z}(x) - \sum_{y \in \Anc^{G^{\sm z}}(x) \sm \lC x\rC} \tilde w(y)\\
    &= \tilde w(x).
\end{align}
If $z\prec x$ then: $\Anc^{G\sm z}(x) = \Anc^G(x)\sm\lC z\rC$ by \Cref{lem:proj-anc} and we get: 
\begin{align}
    w(x) 
    &= v(x) - \sum_{y \in \Anc^G(x) \sm \lC x\rC} w(y) \\
    &= v(x) - \sum_{y \in \Anc^G(x) \sm \lC z,x\rC}  w(y) - \underbrace{w(z)}_{=0} \\
    &= v|_{\sm z}(x) - \sum_{y \in \Anc^{G^{\sm z}}(x) \sm \lC x\rC} \tilde w(y) \\
    & =\tilde w(x).
\end{align}
This shows the first claim. For the last claim we invoke the projection formula from \Cref{def:proj-damg-one} for $x \in V^{\sm z}$:
\begin{align}
    w^{\sm z}(x) &= w(x) + q(x|z) \cdot \underbrace{w(z)}_{=0} =w(x).
\end{align}
This already shows the claim. 
For the final claim about the value functions, we just use the equality $w^{\sm z}=\tilde w$ and sum up:
\begin{align}
    v^{\sm z}(x) 
    &= \sum_{y \in \Anc^{G^{\sm z}}(x)} w^{\sm z}(y) 
    =\sum_{y \in \Anc^{G^{\sm z}}(x)} \tilde w(y) 
    = v|_{\sm z}(x).
\end{align}
This shows the claim.
\end{proof}
\end{lemma}

\begin{remark}[Projections are $R$-linear and $A$-linear]
\label{rem:proj-linearity}
    By \Cref{thm:moebius-trafo} we see that the synergy function $w$ is $R$-linear in $v$.
    \Cref{eq:proj-synergy} shows that the projection $w^{\sm z}$ is $R$-linear in $w$ and thus in $v$.

    It is also crucial to note that we have another form of linearity w.r.t.\ the $R$-module $A$: If $v: V \to R$ is a value function and $a \in A$ a fixed element. Then we can construct a value function $\tilde v: V \to A$ via $\tilde v(x):= v(x) \cdot a$. Again, by \Cref{thm:moebius-trafo} we see that we have for its synergy function: 
    \begin{align} 
      \tilde v = v \cdot a &\iff  \tilde w = w \cdot a.
    \end{align}
    In addition note that the projection w.r.t.\ the $R$-PDAMG $(G,q)$ in \Cref{eq:proj-synergy} also gives:
    \begin{align} 
      \tilde w = w \cdot a &\implies  \tilde w^{\sm z} = w^{\sm z} \cdot a.
    \end{align}
    We thus get the following $A$-linearity for a finite index set $I$, $v_i: V \to R$ and $a_i \in A$, for $i\in I$:
    \begin{align} 
      \tilde v = \sum_{i \in I} v_i \cdot a_i & \iff  \tilde w = \sum_{i \in I} w_i \cdot a_i \implies \tilde w^{\sm z} = \sum_{i \in I} w_i^{\sm z} \cdot a_i 
      \iff \tilde v^{\sm z} = \sum_{i \in I} v_i^{\sm z} \cdot a_i.
    \end{align}
\end{remark}

\begin{remark}
    \Cref{def:admissible-proj-set} ensures that projections preserve normalization (\Cref{lem:proj-norm-weights-sets}). 
    A typical example, where normalization can fail to hold after projecting out $S \ins V$, is when $\Pa^G(y)=\lC u,x \rC \ins \Rt(G)$ with $u \in S$ and $x \notin S$ and $q(x|y) \neq 1$. Then we have: $\Pa^{G^{\sm S}}(y)=\lC x\rC$ and possibly: $\sum_{z \in V^{\sm S}}q^{\sm S}(z|y)=q(x|y)\neq 1$.
\end{remark}

\begin{remark}
Let $G=(V,E)$ be a DAMG. Then every subset $S \ins V \sm \Rt(G)$ is always a restricted admissible projection set for $G$. 
\end{remark}

\begin{lemma}[Injection from weights to total path weights]
\label{lem:q_s_equiv}
    Let $q^G: V \times V \to R$ and $\tilde{q}^G: V \times V \to R$ be two projection weight functions on a DAMG $G=(V,E)$ with total path weights $s^G$ and $\tilde{s}^G$, respectively. If 
    \begin{align}
        s^G(x|y) &= \tilde{s}^G(x|y)
        \intertext{for all $x, y \in V$, then}
        q^G(x|y) &= \tilde{q}^G(x|y)
    \end{align}
    \begin{proof}
        Note that (by \Cref{not:total-weights-q})
        \begin{align}
            s^G &= \delta^G + \sum_{k\geq 1} (q^G)^{*k}\\
            &= (\delta^G - q^G)^{*(-1)}
            \intertext{such that}
            q^G &= \delta^G - (s^G)^{*(-1)}
        \end{align}
    \end{proof}
\end{lemma}

\begin{lemma}[Path count recursion formulas]
\label{lem:path-uniform-recursion}
Let the notation be like in \Cref{def:path-uniform-damg}.
Then we have the following recursion formulas for all $x,y \in V$:
\begin{align}
    \pi^G(x,y) &= 
    \begin{cases}
        0, &\text{ if } y \notin \Desc^G(x), \\
        1, &\text{ if }y =x, \\
        \displaystyle \sum_{z \in \Pa^G(y)}  \pi^G(x,z) \cdot |E(z,y)|, &\text{ if } y\in \Desc^G(x)\sm\lC x\rC,
    \end{cases}    \label{eq:path-recursion}
    \\
    \pi^G(y) &= 
    \begin{cases}
        1, &\text{ if }y \in \Rt(G), \\
        \displaystyle \sum_{z \in \Pa^G(y)} \pi^G(z)\cdot |E(z,y)|, &\text{ if } y\notin \Rt(G). \label{eq:path-recursion-aggr}
    \end{cases}
\end{align}
This shows that $q_G$ is a normalized weight function on $G$ and has non-negative values.
In particular, $(G,q_G)$ forms a $\Z_G$-PDAMG.
\begin{proof}
Since every directed path from an element $x \in V$ to $z \in \Pa^G(y)$ can be extended to a unique directed path to $y$ by concatenating any edge $z \tuh[e] y \in E(z,y)$ the recursion formula follows immediately.
\end{proof}
\end{lemma}

\section{Shapley values on projectable directed acyclic multigraphs}
\label{sec:shap_on_rdamg}
We here study Shapley values on directed acyclic multigraphs (DAMGs) $G$ where the normalized weight function $q$ is pre-determined and fixed, that is, for the case of $R$-PDAMGs. The case of DAMGs, where the weight functions are not pre-determined, but inferred from the DAMG $G$ itself, is treated in \Cref{sec:shap_on_damg}.

\subsection{Definition and properties of Shapley values on PDAMGs}

We start with an ad hoc definition of Shapley values on projectable directed acyclic graphs ($R$-PDAMGs), based on the previous sections of M\"obius transforms and projection operators. Then we will study properties of their Shapley values that hold for general normalized weights functions $q$.

\begin{definition}[Shapley values on $R$-PDAMGs]
\label{def:shapley-values}
Let $R$ be a ring and $A$ be an $R$-module. Let $G=(V,E)$ be a DAMG and $q$ a normalized weight function on $G$ with values in $R$, $v: V \to A$ be a value function and $w=v'_G: V \to A$ its synergy function w.r.t.\ $G$.

We define the \emph{Shapley values} for $v$ w.r.t.\ $(G,q)$ by projecting its synergy function $w=v'_G$ onto the set of roots $\Rt(G)$ of $G$ via \Cref{def:projection-subset} as follows:
\begin{align}
 \forall r \in \Rt(G).\qquad   \Sh^{(G,q)}_r(v) &:= (v'_G)^{\Rt(G)}(r) \in A. \label{eq:shapley-def}
\end{align}
Note again that the projection onto the set of roots $\Rt(G)$ on the right hand side implicitly depends on the weight function $q$ (and $G$), which the notation does not indicate.
\end{definition}

\begin{lemma}[Normalization of total path weights]
\label{lem:norm-total-weights}
Let $G=(V,E)$ be a DAMG and $q:E \to R$ be a normalized weight function on $G$.
Let $\Xcal \ins V$ be a subset and $y \in V$ be any element such that every directed path $r \tuh[e_1] \cdots \tuh[e_k] y$ from any root $r \in \Rt(G)$ to $y$ in $G$ contains exactly one $x \in \Xcal$.
Then we have the following normalization of total path weights for $y$:
\begin{align}
    \sum_{x \in \Xcal} s(x|y) &=1.
\end{align}
Note that the condition holds for every horizontal subset $\Xcal$ of $G$ and every $y \in \Desc^G(\Xcal)$.
In particular, this holds for the set of roots $\Xcal=\Rt(G)$ of $G$ and every $y \in V$.
\begin{proof}
Note that by the made assumption, we necessarily have that $y \in \Desc^G(\Xcal)$.

First, assume that $y \in \Xcal$.
Then, by the made assumption, we necessarily have for $x \in \Xcal$:
\begin{align}
    s(x|y) &=\delta_y(x).
\end{align}
Summing over $x \in \Xcal$ gives the normalization claim. 

Now, let $y \in \Desc^G(\Xcal)\sm \Xcal$. Then for every $z \in \Pa^G(y)$ we see that the same condition for $y$ w.r.t.\ $\Xcal$ also holds for $z$.
Then by recursion we have:
\begin{align}
    \sum_{x \in \Xcal} s(x|y) 
    &= \sum_{x \in \Xcal} \sum_{z \in \Pa^G(y)} s(x|z) \cdot q(z|y) 
    = \sum_{z \in \Pa^G(y)} \underbrace{\sum_{x \in \Xcal} s(x|z)}_{=1} \cdot q(z|y) 
    = \sum_{z \in \Pa^G(y)} q(z|y) 
    = 1.
\end{align}
This shows the claim.
\end{proof}
\end{lemma}

\begin{definition}[$R$-PDAMG automorphism]
\label{def:r-pdamg-automorphism}
Let $(G=(V,E),q)$ be an $R$-PDAMG.
An \emph{$R$-PDAMG automorphism of $(G,q)$}:
\begin{align}
    \alpha=(\alpha_V,\alpha_E): (G,q) &\bij (G,q),
\end{align}
consists of two bijective maps:
\begin{align}
    \alpha_V: V&\bij V, & \alpha_E: E &\bij E,
\end{align}
such that for every $e \in E$ we have:
\begin{align}
    \tl(\alpha_E(e))&=\alpha_V(\tl(e)), &\hd(\alpha_E(e))&=\alpha_V(\hd(e)), &
    q(\alpha_E(e)) &= q(e).
\end{align}
\end{definition}

\begin{theorem}[Properties of Shapley values]
\label{thm:properties-shapley-values}
Let $R$ be a ring and $A$ be an $R$-module. Let $(G=(V,E),q)$ be an $R$-PDAMG, $v: V \to A$ be a value function and $w=v'_G: V \to A$ its synergy function w.r.t.\ $G$. The Shapley values from \Cref{def:shapley-values} then satisfy the following properties:
\begin{enumerate}   
    \item \emph{Efficiency}: $\displaystyle\sum_{r \in \Rt(G)} \Sh^{(G,q)}_r(v) = \sum_{x \in V} v'_G(x)$.
    \item \emph{$R$-Linearity}: For all $c_1,c_2 \in R$ and value functions $v_1,v_2: V \to A$ and $r \in \Rt(G)$ we have:
    \begin{align}
        \Sh^{(G,q)}_r(c_1 \cdot v_1 + c_2 \cdot v_2)&= c_1 \cdot \Sh^{(G,q)}_r(v_1) + c_2 \cdot \Sh^{(G,q)}_r(v_2).
    \end{align}
    \item \emph{$A$-Linearity}: For any finite index set $I$, $a_i \in A$ and value functions $v_i: V \to R$, $i \in I$, and $r \in \Rt(G)$ we have:
    \begin{align}
        \Sh^{(G,q)}_r\lp  \sum_{i \in I} v_i \cdot a_i\rp&=  \sum_{i \in I} \Sh^{(G,q)}_r(v_i) \cdot a_i \in A.
    \end{align}          
    \item \emph{Null root}\footnote{Classically this would be called \emph{null player}.}: If $r \in \Rt(G)$ is a null root for $(G,v)$ then: $\Sh^{(G,q)}_r(v)=0$.
    \item \emph{Weak elements}: If $W \ins V\sm \Rt(G)$ is a subset of weak elements for $(G,v)$ then we have for every $r \in \Rt(G)$:
    \begin{align}
       \Sh^{(G,q)}_r(v) &= \Sh^{(G^{\sm W},q^{\sm W})}_r(v|_{V \sm W}),
    \end{align}
    where $v|_{V \sm W}$ is the restriction of $v$ onto the complement of $W$.
    \item \emph{Projection}: For every subset $S \ins V \sm \Rt(G)$ and every $r \in \Rt(G)$ we have:
    \begin{align}
       \Sh^{(G,q)}_r(v) &= \Sh^{(G^{\sm S},q^{\sm S})}_r(v^{\sm S}).
    \end{align}
    \item \emph{Total path weights}: We have the explicit formula for $r \in \Rt(G)$ in terms of total path weights and synergy function\footnote{A similar formula was introduced by \cite{billot2005share}, who referred to $s$ as a \emph{sharing system} for \emph{Möbius values}.}:
    \begin{align}
        \Sh^{(G,q)}_r(v) &= \sum_{y \in V} s(r|y) \cdot v_G'(y),
        \label{eq:shapley-value-total-weight-formula}
    \end{align}
    where $s$ denotes the total path weight function of $(G,q)$ from \Cref{not:total-weights-q}. 
    Recall that for $r \in \Rt(G)$ and $y \in V$ the total path weights satisfy:
    \begin{align}
        \sum_{r \in \Rt(G)} s(r|y) &=1, & y \notin \Desc^G(r) &\implies s(r|y) =0.
    \end{align}
    \item \emph{Symmetry}: For every $R$-PDAMG automorphism\footnote{See \Cref{def:r-pdamg-automorphism}.} $\alpha: (G,q) \bij (G,q)$ and root $r\in \Rt(G)$ we have: 
    \begin{align}
        \Sh^{(G,q)}_{\alpha(r)}(v)&= \Sh^{(G,q)}_r(v^\alpha),
    \end{align}
    where the value function $v^\alpha: V \to A$ is defined on elements $x\in V$ via: $v^\alpha(x):=v(\alpha(x))$.
    
    In particular, for fixed $r \in \Rt(G)$, we have the implication:
    \begin{align}
        \lp \forall y \in \Desc^G(r). \;\;  v_G'(\alpha(y)) = v_G'(y) \rp \implies \Sh^{(G,q)}_{\alpha(r)}(v)&= \Sh^{(G,q)}_r(v).
    \end{align}
    \item \emph{Edgeless graph}: If $G=(V,E)$ is edgeless, i.e.\ $E=\emptyset$, then we have:  
    \begin{align}
        \Sh^{(G,q)}_r(v) &= v(r).
    \end{align}
    \item \emph{Flat hierarchy}: If $G=(V,E)$ is  hierarchically flat, i.e.\ $V=\Rt(G) \cup \Lf(G)$,  then for the unanimity value function $\zeta_y$ on $G$ centred at $y \in V$ and root $r \in \Rt(G)$ we have the following:
    \begin{align}
        \Sh^{(G,q)}_r(\zeta_y) &=
        \begin{cases}
            \displaystyle\delta_y(r), &\text{ if } y \in \Rt(G),\\
            q(r|y), &\text{ if }y\notin \Rt(G).
        \end{cases} 
    \end{align}
\end{enumerate}
\begin{proof}
Let $w:=v'_G$ be the synergy function of $v$ w.r.t.\ $G$. All properties directly follow from \Cref{thm:proj-stable-prop} and \Cref{def:shapley-values}. In more details:

Efficiency: We just apply \Cref{lem:proj-efficiency} recursively. 
For this, note that: $V^{\Rt(G)}=\Rt(G)$. We then get:
\begin{align}
     \sum_{r \in \Rt(G)} \Sh^{(G,q)}_r(v) 
    &= \sum_{x \in V^{\Rt(G)}} w^{\Rt(G)}(x) 
    \overset{\ref{lem:proj-efficiency}}{=} \sum_{x \in V} w(x).
\end{align}

$R$-Linearity and $A$-linearity follow from recursively following \Cref{rem:proj-linearity} and the defining equation of Shapley values \Cref{eq:shapley-def}.

Null root: Let $r \in \Rt(G)$ be a null root for $(G,v)$. Then recursively applying \Cref{lem:proj-null-player} we see that $r$ is also a null root for $(G^{\Rt(G)},v^{\Rt(G)})$. 
In particular this implies, as $r \in \Desc^{G^{\Rt(G)}}(r)$:
\begin{align}
    \Sh^{(G,q)}_r(v) &= w^{\Rt(G)}(r) = 0.
\end{align}

Projection: Let $S \ins V \sm \Rt(G)$ and $T:= (V \sm \Rt(G)) \sm S$. Then $S \dcup T = V \sm \Rt(G)$.
We then have by \Cref{lem:projections-commute}:
\begin{align}
    \Sh^{(G,q)}_r(v) &= w^{\Rt(G)}(r) = w^{\sm (S \dcup T)}(r) = (w^{\sm S})^{\sm T}(r) = (w^{\sm S})^{\Rt(G^{\sm S})} = 
    \Sh^{(G^{\sm S},q^{\sm S})}_r(v^{\sm S}).
\end{align}

Weak elements: This follows from the projection property together with \Cref{lem:moebius-weak-elements}, which shows that the projection and restriction of $v$ agree:
\begin{align}
    v^{\sm W} &= v|_{V \sm W}.
\end{align}

The explicit formula via the total path weights immediately follows from \Cref{lem:proj-synergy-total-weights}. Also note their  normalization from \Cref{lem:norm-total-weights}.

Symmetry: First note that for every $R$-PDAMG automorphisms $\alpha: (G,q) \bij (G,q)$ and $r \in \Rt(G)$ and $y \in V$ we have:
\begin{align}
  \alpha: \Anc^G(y) &\bij \Anc^G(\alpha(y)), &  s(\alpha(r)|\alpha(y)) &= s(r|y), & 
  (v_G')^\alpha &=(v^\alpha)_G'.
\end{align}
With this we get:
\begin{align}
    \Sh^{(G,q)}_{\alpha(r)}(v) 
    &=\sum_{z\in V} s(\alpha(r)|z) \cdot v_G'(z) \\
    &\overset{z=\alpha(y)}{=} \sum_{y\in V} s(\alpha(r)|\alpha(y)) \cdot v_G'(\alpha(y)) \\
    &=\sum_{y\in V} s(r|y) \cdot (v^\alpha)_G'(y) \\
    &=\Sh^{(G,q)}_r(v^\alpha).
\end{align}
This shows the general formula for symmetry. Now assume that for fixed $r \in \Rt(G)$ and all $y \in \Desc^G(r)$ we have that: $v_G'(\alpha(y))=v_G'(y)$, then we get:
\begin{align}
   \Sh^{(G,q)}_r(v^\alpha) &= \sum_{y\in \Desc^G(r)} s(r|y) \cdot v_G'(\alpha(y)) = \sum_{y\in \Desc^G(r)} s(r|y) \cdot v_G'(y) = \Sh^{(G,q)}_r(v).
\end{align}
All other properties immediately follow from the definition.
\end{proof}
\end{theorem}

\begin{remark}[Shapley values with path uniform weights]
\label{rem:shapley-path-uniform}
Regarding \Cref{thm:properties-shapley-values} we encounter Shapley values w.r.t.\ the projected $R$-PDAMG $(G^{\sm S},q^{\sm S})$.
Note if we choose the weight function $q$ to be the path uniform edge weight function $q_G$ from \Cref{def:path-uniform-damg}, then by \Cref{lem:proj-path-unif-comm} we have the equality:
\begin{align}
    (q_G)^{\sm S} &= q_{G^{\sm S}}.
\end{align}
This shows that $(q_G)^{\sm S}$ is directly determined by the structure of $G^{\sm S}$.
We could thus define:
\begin{align}
    \Sh^G_r(v) &:= \Sh^{(G,q_G)}_r(v).
\end{align}
Then $\Sh^G$ still satisfies all the properties from \Cref{thm:properties-shapley-values}, without explicitly mentioning $q_G$, $(q_G)^{\sm S}$ or $(q_G)^{\sm W}$, as it will implicitly choose the correct weight function.

In particular, we can then state the projection and weak elements property, resp., for path uniform weights also as:
    \begin{align}
    \Sh^G_r(v)&= \Sh^{G^{\sm S}}_r(v^{\sm S}), &
    \Sh^G_r(v)&= \Sh^{G^{\sm W}}_r(v|_{V \sm W}),
    \end{align}
for any subsets $S, W \ins V \sm \Rt(G)$, where $W$ is only allowed to consist of weak elements for $(G,v)$.

\Cref{cor:shapley-main} shows the unique characterization of $\Sh^G$ axiomatically. 
\end{remark}

\begin{remark}[Computing Shapley values via recursive projection]
\label{rem:comp-shapley}
Let $(G,q)$ be an $R$-PDAMG and $A$ be an $R$-module and $v: V \to A$ any value function with synergy function $w: V \to A$. 

We have seen that we can (pre-)compute the total path weights $s(r|y)$ with the recursion formula in \Cref{not:total-weights-q} and then plug them into the formula for the Shapley values in \Cref{eq:shapley-value-total-weight-formula}.

Alternatively, we can use recursive projections to compute the Shapley values for a given fixed value function $v$. For this, note that, since $G$ is acyclic, we have a topological ordering $\prec$ of $G$, where the roots $r \in \Rt(G)$ are ordered first. 
We can thus write:
    \begin{align}
        V &= \Rt(G) \dcup \lC z_1,\dots,z_M \rC,
    \end{align}
    with $z_{m-1} \prec z_m$ for all $m$.
    By \Cref{def:shapley-values} and \Cref{lem:projections-commute} we can compute the Shapley values of $v$ by recursive projection operations:
    \begin{align}
     \forall r \in \Rt(G) \qquad   \Sh^{(G,q)}_r(v) &= w^{\Rt(G)}(r) = ( (w^{\sm z_M})\dots )^{\sm z_1}(r).
    \end{align}
    For this, abbreviate for $m=0,\dots,M$, the set:
    \begin{align}
        V^{(m)} &:= \Rt(G) \dcup \lC z_1,\dots,z_m\rC.
    \end{align}
    We then define functions recursively for $m=M,\dots,1$ as follows:
    \begin{align}
        w^{(M)} : V^{(M)}=V & \to A, & w^{(M)}(x) &:= w(x), \\
        w^{(m-1)} : V^{(m-1)} &\to A, & w^{(m-1)}(x) &:= \begin{cases}
            w^{(m)}(x), &\text{ if }x \notin \Pa^G(z_m),\\
            w^{(m)}(x) + q(x|z_m) \cdot w^{(m)}(z_m), &\text{ if }x \in \Pa^G(z_m).
        \end{cases} \label{eq:shapley-rec-comp}
    \end{align}
    Finally, we then have:
    \begin{align}
    w^{(0)}: V^{(0)}=\Rt(G) &\to A, &
        w^{(0)}(r) \overset{!}{=} w^{\Rt(G)}(r) = \Sh_r^{(G,q)}(v) \in A.
    \end{align}
    Note that we do not need to update the pairwise weights $q(x|u)$, since at each step we have that $\Ch^{G^{V^{(m)}}}(z_m)=\emptyset$ and thus:
    \begin{align}
        q^{(m-1)}(x|u) &= q^{(m)}(x|u) + q^{(m)}(x|z_m)\cdot \underbrace{q^{(m)}(z_m|u)}_{=0} 
                =q^{(m)}(x|u) 
                = \dots 
                =  q(x|u).
    \end{align}
    If $D:=|V|$ then computing the M\"obius transform $w$ of $v$ via \Cref{def:moebius-trafo} has $O(D^2)$ time complexity and $O(D)$ memory complexity. Furthermore, to compute all the values $w^{(m)}(x)$ for $x \in V^{(m)}$ and $m=M,\dots,0$ with the above methods has $O(M \cdot L)$ time complexity and $O(D)$ memory complexity, where $L:=\max_{z \in V} |\Pa^G(z)|$, with $M,L \le D$. 

    In case of path uniform weights $q=q_G$, see \Cref{def:path-uniform-damg}, we can use the recursive formula from \Cref{eq:path-recursion-aggr} to compute all the values $\pi^G(y)$ for $y \in V$ in $O(M \cdot L + D)$ time complexity and $O(D)$ memory complexity, which then determine the occurring values $q_G(x|y)=\frac{\pi^G(x) \cdot |E(x,y)|}{\pi^G(y)}$ in \Cref{eq:shapley-rec-comp}.
    
    So, even if neither the synergy function $w$, nor the path uniform weights function $q_G$ are pre-computed, then computing all Shapley values of $v$ overall has still $O(D^2)$ time complexity and $O(D)$ memory complexity.
\end{remark}

\subsection{Uniqueness of Shapley values on PDAMGs}

We now want to investigate how far certain properties/axioms can uniquely restrict the form of the Shapley values and their total path weights.

We start here with a result that based on the analogues of the classical axioms \emph{linearity}, \emph{efficiency} and \emph{null player}, allows us to express the Shapley values as a linear combination of the synergy values, where the scalar coefficients in front are the corresponding total path weights in this setting.  

\begin{proposition}
\label{prp:prop-total-weights}
Let $R$ be a ring and $(G,q)$ an $R$-PDAMG.
Let $\tilde{\Sh}^{(G,q)}$ be an assignment rule that assigns to every value function $v:V \to A$ for any $R$-module $A$ and every $r \in \Rt(G)$ a value $\tilde{\Sh}_r^{(G,q)}(v) \in A$.
Then the following two points are equivalent:
\begin{enumerate}
    \item $\tilde{\Sh}^{(G,q)}$ satisfies the following three properties:
    \begin{enumerate}
        \item \emph{$A$-Linearity}: For every $R$-module $A$, finite index set $I$, $a_i \in A$ and value functions $v_i: V \to R$, for $i \in I$, and $r \in \Rt(G)$ we have:
    \begin{align}
        \tilde{\Sh}^{(G,q)}_r\lp  \sum_{i \in I} v_i \cdot a_i\rp&=  \sum_{i \in I} \tilde{\Sh}^{(G,q)}_r(v_i) \cdot a_i \in A.
    \end{align}
        \item \emph{Efficiency}: For every $R$-module $A$ and value function $v: V \to A$ we have: 
    \begin{align}
        \sum_{r \in \Rt(G)} \tilde{\Sh}^{(G,q)}_r(v) = \sum_{x \in V} v'_G(x).
    \end{align}
        \item \emph{Null root}: For every $R$-module $A$, value function $v: V \to A$ and every null root $r \in \Rt(G)$ for $(G,v)$ we have: 
    \begin{align}
        \tilde{\Sh}^{(G,q)}_r(v)=0.
    \end{align}
    \end{enumerate}
    \item There exists a map:
    \begin{align}
        \tilde s=\tilde s^{(G,q)}: \Rt(G) \times V & \to R, & (r,y) & \mapsto \tilde s(r|y), 
    \end{align}
    such that the following three properties hold:
    \begin{enumerate}
        \item  For every $R$-module $A$, value function $v:V \to A$ and $r \in \Rt(G)$ we have: 
        \begin{align}
            \tilde{\Sh}^{(G,q)}_r(v) = \sum_{y \in V} \tilde s(r|y) \cdot v_G'(y).
        \label{eq:shapley_ito_total_path_weights}
        \end{align}
        \item For all $y \in V$: $\displaystyle \sum_{r \in \Rt(G)} \tilde s(r|y)=1$,
        \item For every $r \in \Rt(G)$ and $y \in V$ we have: $y \notin \Desc^G(r)  \implies \tilde s(r|y)=0$.
    \end{enumerate}
\end{enumerate}
\begin{proof}
``$\Longleftarrow$'': Assume the existence of such function $\tilde s$. The first equation directly implies $A$-linearity by \Cref{thm:moebius-trafo}. We further get:
\begin{align}
    \sum_{r \in \Rt(G)} \tilde{\Sh}^{(G,q)}_r(v) 
    & = \sum_{r \in \Rt(G)} \sum_{y \in V} \tilde s(r|y) \cdot v_G'(y) \\
    & = \sum_{y \in V} \underbrace{\sum_{r \in \Rt(G)}  \tilde s(r|y)}_{=1} \cdot v_G'(y) \\
    &= \sum_{y \in V} v_G'(y).
\end{align}
This shows efficiency. 
Now, let $r \in \Rt(G)$ be a null root for $(G,v)$. Then for $y\in \Desc^G(r)$ we have $v_G'(y)=0$. If on the other hand, $y \notin \Desc^G(r)$, then the last property gives us: $\tilde s(r|y)=0$. We thus get for such null root $r \in \Rt(G)$:
\begin{align}
    \tilde{\Sh}^{(G,q)}_r(v) &= \sum_{y \in V} \tilde s(r|y) \cdot v_G'(y) \\
    & = \sum_{y \in \Desc^G(r)} \tilde s(r|y) \cdot \underbrace{v_G'(y)}_{=0} +  \sum_{y \notin \Desc^G(r)} \underbrace{\tilde s(r|y)}_{=0} \cdot v_G'(y) \\
    &=0.
\end{align}
This shows the null root property.

``$\implies$'':
    By \Cref{rem:unanimity-game} we can always write any value function $v:V \to A$ as follows:
    \begin{align}
        v(x) &= \sum_{y \in V} \underbrace{\zeta_y(x)}_{\in R} \cdot \underbrace{w(y)}_{\in A},
    \end{align}
    with the unanimity value function $\zeta_y$ centred at $y \in V$. 
    By $A$-linearity we then can write for $r \in \Rt(G)$:
    \begin{align}
        \tilde{\Sh}_r^{(G,q)}(v) &= \sum_{y \in V} \underbrace{\tilde{\Sh}_r^{(G,q)}(\zeta_y)}_{\in R} \cdot \underbrace{w(y)}_{\in A}.
    \end{align}
    If we now define the function $\tilde s$ as follows:
    \begin{align}
        \tilde s: \Rt(G) \times V & \to R, &
        \tilde s(r|y) &:= \tilde{\Sh}_r^{(G,q)}(\zeta_y),
    \end{align}
    then the first property is already satisfied.
    We further get by efficiency:
    \begin{align}
        \sum_{r \in \Rt(G)} \tilde s(r|y) &= \sum_{r \in \Rt(G)} \tilde{\Sh}_r^{(G,q)}(\zeta_y) 
        = \sum_{x \in V} (\zeta_y)_G'(x) 
        = \sum_{x \in V} \delta_y(x) =1.
    \end{align}
    This shows the second property.
    For the third property, let $y \notin \Desc^G(r)$. Then $r$ is a null root for $(G,\zeta_y)$ and we get by the null root property:
    \begin{align}
        \tilde s(r|y) &= \tilde{\Sh}_r^{(G,q)}(\zeta_y) =0.
    \end{align}
    This shows the last property.
\end{proof}
\end{proposition}

\begin{remark}[Fixing the total path weights of an $R$-PDAMG]
\label{rem:total-weights}
In the setting of \Cref{prp:prop-total-weights} we have seen that the \emph{linearity}, \emph{efficiency} and \emph{null root} axiom reduce the problem of uniquely determining the Shapley values $\tilde{\Sh}^{(G,q)}_r(v)$ for arbitrary $R$-PDAMGs $(G,q)$ to the problem of finding further constraints on $\tilde{\Sh}$ that then fix the corresponding total path weight functions:
\begin{align}
    \tilde s=\tilde s^{(G,q)}: \Rt(G) \times V &\to R, &
    \tilde s(r|y) &= \tilde{\Sh}^{(G,q)}_r(\zeta_y),
\end{align}
which appear in the formula:
\begin{align}
    \tilde{\Sh}^{(G,q)}_r(v) &= \sum_{y \in V} \tilde{s}(r|y) \cdot v_G'(y).
\end{align}
The classical \emph{symmetry} axiom will in this general setting not be strong enough to constrain the total path weights sufficiently. 
In \Cref{thm:pdamg-shapley-unique-proj} we will instead achieve this by imposing the \emph{projection} axiom on the Shapley values, which says that Shapley values should not change under projection operations. This axiom is so strong that the other axioms can even be relaxed. 
Note that all these properties are satisfied for the proposed definition of Shapley values in \Cref{def:shapley-values} by \Cref{thm:properties-shapley-values}.

In \Cref{thm:shapley-main}, where we consider the case of DAMGs $G$ (without prior given edge weight function), we instead build on the \emph{weak elements} axiom. Also here, this axiom is so strong such that other axioms can be relaxed.
\end{remark}

We now show that the projection property is already strong enough in order to fully determine the Shapley values for $R$-PDAMGs (up to a basic standardization property on edgeless DAMGs).

\begin{theorem}[Uniqueness of Shapley  values on PDAMGs via the projection property]
\label{thm:pdamg-shapley-unique-proj}
Let $R$ be a ring. 
Let $\tilde{\Sh}$ be an assignment rule that assigns to every $R$-PDAMG $(G=(V,E),q)$ and value function $v:V \to A$ for any $R$-module $A$ and every $r \in \Rt(G)$ a value $\tilde{\Sh}_r^{(G,q)}(v) \in A$.
We further assume the following properties for $\tilde{\Sh}$:
\begin{enumerate}   
    \item \emph{Leaf projection}: For every $R$-PDAMG $(G,q)$, every $R$-module $A$, every value function $v: V \to A$ and every (non-root) leaf $z \in \Lf(G) \sm \Rt(G)$ and every root $r \in \Rt(G)$ we have:
    \begin{align}
      \tilde{\Sh}^{(G,q)}_r(v) &= \tilde{\Sh}^{(G^{\sm z},q^{\sm z})}_r(v^{\sm z}).
    \end{align}
    \item \emph{Edgeless graph}: For every edgeless DAMG $G=(V,\emptyset)$\footnote{On an edgeless DAMG, the normalization condition is vacuous. We by convention write $q=0$ for the trivial weight function.}, every $R$-module $A$, every value function $v: V \to A$ and every $r \in \Rt(G)=V$ we have: 
    \begin{align}
       \tilde{\Sh}^{(G,0)}_r(v) &= v(r).
    \end{align}
\end{enumerate} 
Then we necessarily have that $\tilde{\Sh}$ agrees with the Shapley value assignment $\Sh$ from \Cref{def:shapley-values}. It then will also have all the properties mentioned in \Cref{thm:properties-shapley-values}. For the converse, note that by \Cref{thm:properties-shapley-values} the Shapley values $\Sh$ from \Cref{def:shapley-values} satisfies all the above properties.
\begin{proof}
Let $\tilde \Sh$ be an assignment rule satisfying all these properties. We want to show that necessarily: $\tilde \Sh=\Sh$. For this, consider the $R$-PDAMG $(G=(V,E),q)$ and $R$-module $A$ and a general value function $v: V \to A$. 
By the leaf projection property we then recursively get:
\begin{align}
    \tilde \Sh^{(G,q)}_r(v) &= \tilde \Sh^{(G^{\Rt(G)},q^{\Rt(G)})}_r(v^{\Rt(G)}).
\end{align}
Note that $G^{\Rt(G)}=(\Rt(G),\emptyset)$ is an edgeless DAMG and $q^{\Rt(t)}=0$.
We then further get by the edgeless graph property:
\begin{align}
    \tilde \Sh^{(G^{\Rt(G)},q^{\Rt(G)})}_r(v^{\Rt(G)}) &= v^{\Rt(G)}(r) = w^{\Rt(G)}(r) = \Sh^{(G,q)}_r(v),
\end{align}
which shows the claim.
\end{proof}
\end{theorem}

\begin{remark}
    Consider the situation from \Cref{thm:pdamg-shapley-unique-proj}.
    Note that the \emph{efficiency} and \emph{null root} and \emph{$A$-linearity} properties from 
    \Cref{prp:prop-total-weights} (even when restricted to edgeless DAMGs)
    together will imply the \emph{edgeless graph} property from 
    \Cref{thm:pdamg-shapley-unique-proj}.
    So together with \emph{leaf projection} they will also fix the Shapley values to be the ones from \Cref{def:shapley-values}.
    Their total path weights $s(r|y)$, see \Cref{not:total-weights-q}, can then be computed via the recursion formula in  \Cref{eq:total-weights-recursion}; also see \Cref{lem:proj-synergy-total-weights}.
\end{remark}

We now show in the following \Cref{thm:pdamg-shapley-unique-flat} that if we fix the Shapley values for \emph{hierarchically flat} PDAMGs via the projection weights $q(r|y)$ and allow for ignoring \emph{weak elements} then together with \emph{$A$-linearity} we get a unique characterization, which agrees with the Shapley value assignment from \Cref{def:shapley-values} for all $R$-PDAMGs $(G,q)$.

\Cref{thm:shapley-main} and \Cref{cor:shapley-main} show that these properties generalize to the case of $R$-DAMGs and DAMGs where the projection weights $q$ have to be inferred from the graphical structure itself.

\begin{theorem}[Main theorem for Shapley values on PDAMGs]
\label{thm:pdamg-shapley-unique-flat}
Let $R$ be a ring. 
Let $\tilde{\Sh}$ be an assignment rule that assigns to every $R$-PDAMG $(G=(V,E),q)$ and value function $v:V \to A$ for any $R$-module $A$ and every $r \in \Rt(G)$ a value $\tilde{\Sh}_r^{(G,q)}(v) \in A$.
We further assume the following properties for $\tilde{\Sh}$:
\begin{enumerate}   
    \item \emph{$A$-Linearity}: For every $R$-PDAMG $(G=(V,E),q)$, every $R$-module $A$, for any finite index set $I$, $a_i \in A$ and value functions $v_i: V \to R$, $i \in I$, and $r \in \Rt(G)$ we have:
    \begin{align}
        \tilde{\Sh}^{(G,q)}_r\lp  \sum_{i \in I} v_i \cdot a_i\rp&=  \sum_{i \in I} \tilde{\Sh}^{(G,q)}_r(v_i) \cdot a_i \in A.
    \end{align}
    \item \emph{Weak elements}: For every $R$-PDAMG $(G=(V,E),q)$, every $R$-module $A$, every value function $v: V \to A$ and every  subset $W \ins V \sm \Rt(G)$ of weak elements for $(G,v)$ and every root $r \in \Rt(G)$ we have:
    \begin{align}
        \tilde{\Sh}^{(G,q)}_r(v) &=\tilde{\Sh}^{(G^{\sm W},q^{\sm W})}_r(v|_{V\sm W}),
    \end{align}
    where $v|_{V\sm W}$ is the restriction of $v$ to $V \sm W$.
    \item \emph{Flat hierarchy}: For every hierarchically flat $R$-PDAMG $(G=(V,E),q)$, every $y \in V$, unanimity value function $\zeta_y$ on $G$ centred at $y$ and root $r \in \Rt(G)$ we have the following:
    \begin{align}
        \tilde{\Sh}^{(G,q)}_r(\zeta_y) &=
        \begin{cases}
            \displaystyle\delta_y(r), &\text{ if } y \in \Rt(G),\\
            q(r|y), &\text{ if }y\notin \Rt(G).
        \end{cases} 
    \end{align}
\end{enumerate} 
Then we necessarily have that $\tilde{\Sh}$ agrees with the Shapley value assignment $\Sh$ from \Cref{def:shapley-values}. It then will also have all the properties mentioned in \Cref{thm:properties-shapley-values}. For the converse, note that by \Cref{thm:properties-shapley-values} the Shapley values $\Sh$ satisfies all the above properties.
\begin{proof}
Let $\tilde \Sh$ be an assignment rule satisfying all these properties. We want to show that necessarily: $\tilde \Sh=\Sh$. For this consider the $R$-PDAMG $(G=(V,E),q)$ and $R$-module $A$ and a general value function $v: V \to A$. By \Cref{rem:unanimity-game} we can then always write any value function $v:V \to A$ as follows:
    \begin{align}
        v(x) &= \sum_{y \in V} \underbrace{\zeta_y(x)}_{\in R} \cdot \underbrace{v_G'(y)}_{\in A},
    \end{align}
    with the unanimity value function $\zeta_y$ on $G$ centred at $y \in V$. 
    By $A$-linearity we then can write for $r \in \Rt(G)$:
 \begin{align}
     \tilde{\Sh}^{(G,q)}_r(v) & = \sum_{y \in V} \tilde{\Sh}^{(G,q)}_r(\zeta_y) \cdot v'_G(y).
 \end{align}
For fixed $y \in V$ let $W:=V\sm \lp\Rt(G) \cup \lC y\rC\rp$. Then $W$ consists only of weak elements for $(G,\zeta_y)$. Note that $V^{\sm W} = \Rt(G) \cup \lC y\rC$, which shows that $G^{\sm W}$ is a hierarchically flat DAMG. Further note, that $\zeta_y|_{V \sm W}$ is the unanimity game of $G^{\sm W}$ centred at $y \in V^{\sm W}$. We then get:
\begin{align}
\tilde{\Sh}^{(G,q)}_r(\zeta_y) 
    &\overset{\text{weak elements}}{=} \tilde{\Sh}^{(G^{\sm W},q^{\sm W})}_r(\zeta_y|_{V \sm W}) 
    \overset{\text{flat hierarchy}}{=} q^{\sm W}(r|y) 
    = s^{\sm W}(r|y) 
    \overset{\text{\Cref{thm:proj-stable-prop}}}{=} s(r|y).
\end{align}
 Plugging this back into the above formula gives us:
 \begin{align}
     \tilde{\Sh}^{(G,q)}_r(v) & = \sum_{y \in V} s(r|y) \cdot v'_G(y) 
     \overset{\text{\Cref{thm:properties-shapley-values}}}{=} \Sh^{(G,q)}_r(v).
 \end{align} 
This shows the claim. 
The converse follows from \Cref{thm:properties-shapley-values} as well.
\end{proof}
\end{theorem}

\begin{remark}
    We have seen in \Cref{thm:pdamg-shapley-unique-flat} how the three properties \emph{$A$-linearity}, \emph{weak elements} and \emph{flat hierarchy} uniquely determine the Shapley values for value functions $v$ on $R$-PDAMGs.
    Our goal is to generalize this to DAMGs and $R$-DAMGs, where the projection weight function $q$ is not predefined. 
    The three properties will again guide us in how to do this. We first observe that \emph{$A$-linearity} and \emph{weak elements} do not depend on the projection weight function $q$ for the arguments $v_i$ and the restriction $v|_{V \sm W}$ of $v$, in contrast to the projection $v^{\sm W}$ of $v$, which implicitly would depend on $q$. 
    The \emph{flat hierarchy} property, $\Sh_r(\zeta_y)=q(r|y)$, then provides a direct relationship between the Shapley values $\Sh_r(\zeta_y)$ and the projection weights $q(r|y)$ in the case of simple, intuitively understandable (i.e.\ hierarchically flat) DAMGs.
    In \Cref{app:shap_on_weighted_damg} we will invert this relationship and use the Shapley values $\Sh_r(\zeta_y)$ to impose naturally justified projection weights $q(r|y)$ on hierarchically flat DAMGs $G$ based on the structure of $G$ itself. Together with the other properties this will then uniquely determine the Shapley values and what projection weights $q$ to use, even in the general case of hierarchically more complex $R$-DAMGs.
\end{remark}

\section{Shapley values on weighted directed acyclic multigraphs} \label{app:shap_on_weighted_damg}
\subsection{Edge and root weighted directed acyclic multigraphs \label{sec:weighted_DAMGs}}

\begin{definition}[Edge and root weighted directed acyclic multigraph ($R$-DAMG)]\label{def:weighted-damg}
Let $R$ be a ring and $G=(V,E)$ be a DAMG.
\begin{enumerate}
    \item 
An \emph{edge weight function} on $G$ with values in $R$ is just a map:
\begin{align}
    \varsigma: E &\to R, & e=(x\tuh[e]y) & \mapsto \varsigma(x\tuh[e] y):=\varsigma(e) \in R.
\end{align}
By abuse of the same notation, we abbreviate the (corresponding) \emph{pairwise (edge) weight function}:
\begin{align}
    \varsigma: V \times V & \to R, & \varsigma(x,y) &:= \sum_{e \in E(x,y)} \varsigma(x\tuh[e] y).
\end{align}
Note that $\varsigma(x,y)=0$ if $E(x,y)=\emptyset$. 
\item We call the tuple $(G,\varsigma)$ an \emph{edge weighted directed acyclic multigraph}.
\item We call any map $\tau: \Rt(G) \to R$ a \emph{root weight function} or \emph{root strength function} on $G$.
\item We call the tuple $(G,\tau)$ a \emph{root weighted directed acyclic multigraph}.
\item We call a tuple $(G,\varsigma,\tau)$ consisting of a DAMG $G$, an edge weight function $\varsigma: E \to R$ and a root weight function $\tau: \Rt(G) \to R$ an \emph{edge and root ($R$-)weighted directed acyclic multigraph} or \emph{$R$-DAMG} for short.
\end{enumerate}
\end{definition}

\begin{example}[Standard edge and root weight functions of directed acyclic multigraphs]
\label{eg:damg-std-edge-root-weights}
Let $G=(V,E)$ be a DAMG. 
\begin{enumerate}
    \item The \emph{standard edge weight function} of $G$ is given by the constant one:
\begin{align}
    \varsigma=\one: E &\to R, &\varsigma(e)&:=1,\\
    \varsigma : V \times V &\to R, &\varsigma(x,y) &= |E(x,y)|.
\end{align}
\item The \emph{standard root weight function} of $G$ is given by the constant one:
\begin{align}
    \tau=\one: \Rt(G) &\to R, &\tau(r)&:=1.
\end{align}
\end{enumerate}
\end{example}

\begin{remark}[Default choice of edge and root weight function on DAMGs]
    Consider a DAMG $G=(V,E)$ that does not come with both, edge weight function $\varsigma:E\to R$ and root weight function $\tau:\Rt(G) \to R$. 
    In such a case, the missing function is then, by default, the standard constant one from \Cref{eg:damg-std-edge-root-weights} with the same ring $R$. If both weight functions are missing, we pick $R=\Z$ and $\varsigma=\one$, $\tau=\one$.
    In this way, we can always convert DAMGs $G$, edge weighted DAMGs $(G,\varsigma)$, and root weighted DAMGs $(G,\tau)$ into $R$-DAMGs via: $(G,\one,\one)$, $(G,\varsigma,\one)$ or $(G,\one,\tau)$, respectively.
\end{remark}

\begin{definition}[$R$-DAMG automorphism]
\label{def:r-damg-automorphism}
Let $(G=(V,E),\varsigma,\tau)$ be an $R$-DAMG.
An \emph{$R$-DAMG automorphism of $(G,\varsigma,\tau)$}:
\begin{align}
    \alpha=(\alpha_V,\alpha_E): (G,\varsigma,\tau) &\bij (G,\varsigma,\tau),
\end{align}
consists of two bijective maps:
\begin{align}
    \alpha_V: V&\bij V, & \alpha_E: E &\bij E,
\end{align}
such that for every $e \in E$ and $r \in \Rt(G)$ we have:
\begin{align}
    \tl(\alpha_E(e))&=\alpha_V(\tl(e)), &\hd(\alpha_E(e))&=\alpha_V(\hd(e)), &
    \varsigma(\alpha_E(e)) &= \varsigma(e), & \tau(\alpha_V(r))&=\tau(r).
\end{align}
Note that a \emph{DAMG automorphism of $G$} is just a $\Z$-DAMG automorphism of $(G,\one,\one)$.
\end{definition}

\begin{notation}[Extending the root weight function]
\label{not:root-weight-extension}
Let $G=(V,E)$ be a DAMG, $\varsigma: E \to R$ be an edge weight function and $\tau:\Rt(G) \to R$ be a root weight function.
Then we extend $\tau$ to the whole of $V$ via the following recursion formula:
    \begin{align}
     \tau: V & \to R, &   \tau(y) &:= \begin{cases}
         \tau(r), &\text{ if }y=r \in\Rt(G),\\
         \displaystyle\sum_{z \in \Pa^G(y)} \tau(z) \cdot \varsigma(z,y),&\text{ if } y\notin \Rt(G).
     \end{cases}  \label{eq:str-recursion}
    \end{align}
We call $\tau(y)$ for $y \in V$ the \emph{element weight of $y$} or the \emph{strength of $y$}.
\end{notation}

We now extend the pairwise weight function, which measures the weights coming from edges between a pair of nodes, to total path weights, which aggregate the weights on all directed paths between every pair of nodes.

\begin{definition}[Total path weights of weighted DAMGs]
\label{def:total-weights}
Let $G=(V,E)$ be a DAMG and $\varsigma:E \to R$ be an edge weight function on $G$.
We then define the \emph{total path weight function} $\sigma=\sigma^{(G,\varsigma)}$ of $(G,\varsigma)$ recursively as follows:
\begin{align}
    \sigma=\sigma^{(G,\varsigma)}: V \times V & \to R, &
        \sigma(x,y) &:= \begin{cases}
            0, &\text{ if } y \notin \Desc^G(x), \\
            1, &\text{ if } y =x,\\
            \displaystyle\sum_{z \in \Pa^G(y)} \sigma(x,z) \cdot \varsigma(z,y), &\text{ if } y \in \Desc^G(x)\sm\lC x\rC.
        \end{cases}    \label{eq:total-weights-recursion}
    \end{align}
\end{definition}

\begin{lemma}
\label{lem:path-weight-fucntion-general}
Let $G=(V,E)$ be a DAMG, $\varsigma: E \to R$ be an edge weight function and $\tau:\Rt(G) \to R$ be a root weight function and $\tau: V \to R$ be its extension to $V$.
We then have for all $y \in V$ the following relationship: 
    \begin{align}
        \tau(y) &= \sum_{r \in \Rt(G)} \tau(r) \cdot \sigma(r,y).
    \end{align}
More generally, let $\Xcal \ins V$ be any horizontal subset of $G$, then we have for all $y \in \Desc^G(\Xcal)$:
    \begin{align}
        \tau(y) &= \sum_{x \in \Xcal} \tau(x) \cdot \sigma(x,y).
    \end{align}
\begin{proof}
The first claim follows from the second claim with $\Xcal=\Rt(G)$ and $V=\Desc^G(\Rt(G))$.
Now let $\Xcal \ins V$ be a general horizontal subset of $G$ and $y \in \Desc^G(\Xcal)$. 
First, assume that $y \in \Xcal$.
Then, by the made assumption, we necessarily have for $y \in \Xcal$:
\begin{align}
    \sigma(x,y) &=\delta_y(x).
\end{align}
Then this implies:
\begin{align}
 \sum_{x \in \Xcal} \tau(x) \cdot  \sigma(x,y) &= \sum_{x \in \Xcal} \tau(x) \cdot  \delta_y(x) = \tau(y).
\end{align}
Now consider $y \in \Desc^G(\Xcal)\sm \Xcal$.
Then by recursion we get:
\begin{align}
\sum_{x \in \Xcal} \tau(x) \cdot  \sigma(x,y) 
        &= \sum_{x \in \Xcal} \tau(x) \cdot \sum_{z \in \Pa^G(y)} \sigma(x,z) \cdot \varsigma(z,y) \\
        &= \sum_{z \in \Pa^G(y)} \sum_{x \in \Xcal} \tau(x) \cdot  \sigma(x,z) \cdot \varsigma(z,y) \\
        &= \sum_{z \in \Pa^G(y)}  \tau(z) \cdot \varsigma(z,y) \\
        &= \tau(y).
\end{align}
This shows the claim.
\end{proof}
\end{lemma}

\begin{remark}[Two kinds of edge weights on DAMGs]
    For DAMGs $G=(V,E)$ we have seen that we have two kinds of edge weight functions, in symbols $\varsigma: E \to R$ and $q: E \to R$. 
    
    The latter ones, $q: E \to R$, are used to re-attribute synergy values $w(y)$ to other elements $x \in V$ with the share $q(x|y) \cdot w(y)$. To preserve the total synergy sum we generally consider $q$ normalized: $\sum_{x \in V} q(x|y) =1$.  To distinguish it from other edge weights, we also refer to it as \emph{projection weight function} or \emph{projection kernel} of $G$.

    The former kind of edge weights, $\varsigma: E \to R$, in contrast, can be arbitrary and are more in line with typical weighted graphs. For instance, when we want to count the number of directed paths between two elements $x,y \in V$, then it would make sense to give every edge the weight of one: $\varsigma(e)=1$. To distinguish these kind of edge weights from the projection weights, we also call them \emph{edge strengths} of $G$ and also pair it with a \emph{root strength function} $\tau:\Rt(G)\to R$.

    Despite their different interpretations, most of their properties hold for both types of edge weight functions in the same way.

    One of the main topics of this appendix is how to construct projection weights $q(e)$ from edge strengths $\varsigma(e)$ (and element strengths $\tau(x)$) in a principled way. We will establish especially good properties for the choice of projection weights in \Cref{def:r-damg-induced-proj-weights}.
\end{remark}

\subsection{The projection weights for edge and root weighted directed acyclic multigraphs}

Consider a DAMG $G=(V,E)$ together with an edge weight function $\varsigma:E\to R$, and, possibly a root weight function $\tau:\Rt(G) \to R$.

The general challenge in this setting is now how one constructs from the triple $(G,\varsigma,\tau)$ projection weights $q:E \to R$, which are used to construct the Shapley values $\Sh^{(G,q)}_r(v)$.

A first naive approach would be to directly normalize the edge weights for each non-root $y \in V \sm \Rt(G)$ directly, like in \Cref{eg:edge-uniform-damg}:
\begin{align}
    q: E &\to R, & q(x\tuh[e]y) &:=\frac{\displaystyle\varsigma(x\tuh[e]y)}{\displaystyle\sum_{z \in \Pa^G(y)}\varsigma(z,y)}, \quad \text{ or } \quad q(x\tuh[e]y) :=\frac{\displaystyle\tau(x) \cdot \varsigma(x\tuh[e]y)}{\displaystyle\sum_{z \in \Pa^G(y)}\tau(z) \cdot \varsigma(z,y)}.
\end{align}
However, this normalization procedure does not take the whole hierarchical structure of $G$ into account. 
In \Cref{def:r-damg-induced-proj-weights} we will provide a proper way how to induce projection weights $q(e)$ from the triple $(G,\varsigma,\tau)$. Furthermore, exactly these weights will play a central role in the axiomatic determination of the Shapley values in \Cref{thm:shapley-main}.

\begin{definition}[Induced projection weights of an edge and root weighted DAMG]
\label{def:r-damg-induced-proj-weights}
Let $R$ be a ring, $G=(V,E)$ be a DAMG, $\varsigma:E \to R$ be an edge weight function and $\tau: \Rt(G) \to R$ a root weight function and $\tau:V \to R$ its extension to $V$.
We then define the \emph{induced projection weight function} $q=q_{(G,\varsigma,\tau)}$ of $(G,\varsigma,\tau)$  for $x\tuh[e]y \in E$ and $x,y \in V$ as follows:
\begin{align}
    q_{(G,\varsigma,\tau)}(x\tuh[e]y) &:= \frac{\tau(x)}{\tau(y)}\cdot \varsigma(x\tuh[e]y), &
    q_{(G,\varsigma,\tau)}(x|y) &=\frac{\tau(x)}{\tau(y)}\cdot \varsigma(x,y).
\end{align}
Note that the values of $q$ lie in the smallest ring, $R$-algebra, resp., that contains those fractions\footnote{The process of expanding a ring by certain denominators is called a \emph{localization of the ring} in commutative algebra.}:
\begin{align}
    \Z_{(G,\varsigma,\tau)} &:=\Z_q:= \Z\lB\lC q_{(G,\varsigma,\tau)}(e) \st e \in E \rC\rB, &
   R_{(G,\varsigma,\tau)} &:=R_q:= R\lB\lC q_{(G,\varsigma,\tau)}(e) \st e \in E \rC\rB.
\end{align}
Also note that by the recursion formula from \Cref{not:root-weight-extension} the induced projection weight function $q_{(G,\varsigma,\tau)}$ is indeed normalized:
\begin{align}
    y \in V \sm \Rt(G) & \implies \sum_{x \in V} q_{(G,\varsigma,\tau)}(x|y) = 1.
\end{align}
\end{definition}

\begin{lemma}[Total path weights of the induced projection weights]
\label{lem:str-norm-path-weights}
Let $G=(V,E)$ be a DAMG, $\varsigma:E \to R$ be an edge weight function and $\tau: \Rt(G) \to R$ a root weight function and $\tau: V \to R$ its extension to $V$.
Let $q=q_{(G,\varsigma,\tau)}$ be the induced projection weight function of \Cref{def:r-damg-induced-proj-weights} and $s=s_{(G,\varsigma,\tau)}$ the corresponding total path weight function from \Cref{not:total-weights-q}. 
Also let $\sigma$ be the total path weight function of $\varsigma$ from \Cref{def:total-weights}.
We then have the following equality for all $x,y \in V$:
    \begin{align}
     s_{(G,\varsigma,\tau)}(x|y) &= \frac{\tau(x)}{\tau(y)} \cdot \sigma(x,y).
    \end{align}
\begin{proof}
By recursion we have:
\begin{align}
y \notin \Desc^G(x) \implies && s(x|y) &=0 =\frac{\tau(x)}{\tau(y)} \cdot \sigma(x,y);\\
y = x \implies && s(x|y) &= 1 =\frac{\tau(x)}{\tau(y)} \cdot \sigma(x,y); \\
y \in \Desc^G(x)\sm\lC x\rC \implies &&  s(x|y) &=  \sum_{z \in \Pa^G(y)} s(x|z) \cdot  q(z|y) \\
&&&= \sum_{z \in \Pa^G(y)} \frac{\tau(x)}{\tau(z)}\cdot \sigma(x,z) \cdot  \frac{\tau(z)}{\tau(y)}\cdot \varsigma(z,y) \\
&&&= \frac{\tau(x)}{\tau(y)} \sum_{z \in \Pa^G(y)}\sigma(x,z)\cdot \varsigma(z,y) \\
&&&= \frac{\tau(x)}{\tau(y)} \cdot \sigma(x,y). 
\end{align}
This shows the claim.
\end{proof}
\end{lemma}

\begin{lemma}[Path and element strength under projection]
\label{lem:proj-str}
Let $G=(V,E)$ be a DAMG, $\varsigma:E \to R$ be an edge weight function and $\tau: \Rt(G) \to R$ a root weight function. Let $S \ins V$ be an admissible projection set for $G$. 
Let $\varsigma^{\sm S}$ be the projected edge weight function of $\varsigma$ to $G^{\sm S}$ and $\sigma^{\sm S}$ be the corresponding total path weight function of $\varsigma^{\sm S}$.
Then we have the equality:
\begin{align}
    \sigma^{\sm S}=\sigma|_{V^{\sm S}}: V^{\sm S} \times V^{\sm S} & \to R, & \sigma^{\sm S}(x,y) &= \sigma(x,y).
\end{align}
Furthermore, let $\tau^{\sm S}: V^{\sm S} \to R$ be the extension of the root weight function $\tau|_{\Rt(G^{\sm S})}$ to $V^{\sm S}$ via the recursion in \Cref{eq:str-recursion} on $G^{\sm S}$. Then we have the equality:
\begin{align}
    \tau^{\sm S}=\tau|_{V^{\sm S}} : V^{\sm S} &\to R, & \tau^{\sm S}(x) &= \tau(x).
\end{align}
\begin{proof}
The first claim follows from \Cref{lem:proj-total-weights}, \Cref{thm:proj-stable-prop}. Note that no normalization was required.
Now let $S$ be an admissible projection set and  $T:=S \sm \Rt(G)$ and $U:=S \cap \Rt(G)$. 
First consider $y \in V^{\sm S}$ with $\Pa^{G^{\sm T}}(y) \cap U \neq \emptyset$.
Then $\Pa^{G^{\sm T}}(y) \ins S \cap \Rt(G)$ by assumption on $S$, which implies: $\Pa^{G^{\sm S}}(y) = \emptyset$, so $y \in \Rt(G^{\sm S})$.
By assumption we then have: 
\begin{align}
    \tau^{\sm S}(y)&=\tau|_{\Rt(G^{\sm S})}(y) = \tau(y).
\end{align} 
Now, let $y \notin \Rt(G^{\sm S})$ with $\Pa^{G^{\sm T}}(y) \cap U = \emptyset$. 
By induction we can w.l.o.g.\ assume $S=\lC z \rC$ with $z \notin \Rt(G)$ or $z \notin \Pa^G(y)$.
Then we get by recursion: 
\begin{align}
\tau^{\sm z}(y)
    &= \sum_{x \in \Pa^{G^{\sm z}}(y)} \tau^{\sm z}(x) \cdot \varsigma^{\sm z}(x,y) \\
    &= \sum_{x \in V^{\sm z}} \tau^{\sm z}(x) \cdot \varsigma^{\sm z}(x,y) \\
    &= \sum_{x \in V^{\sm z}} \tau(x) \cdot \lp \varsigma(x,y)+\varsigma(x,z) \cdot \varsigma(z,y) \rp \\
    &= \sum_{x \in V^{\sm z}} \tau(x) \cdot \varsigma(x,y)+\sum_{x \in V^{\sm z}} \tau(x) \cdot \varsigma(x,z) \cdot \varsigma(z,y) + \tau(z) \cdot \overbrace{\varsigma(z,z)}^{=0} \cdot \varsigma(z,y)  \\
    &= \sum_{x \in V^{\sm z}} \tau(x) \cdot \varsigma(x,y)+\underbrace{\sum_{x \in V} \tau(x) \cdot \varsigma(x,z)}_{=\tau(z), \text{ if }z\notin \Rt(G)} \cdot \underbrace{\varsigma(z,y)}_{\substack{=0,\\\text{ if }z \notin \Pa^G(y)}}  \\
    &= \sum_{x \in V^{\sm z}} \tau(x) \cdot \varsigma(x,y)+\tau(z) \cdot \varsigma(z,y)  \\
    &= \sum_{x \in V} \tau(x) \cdot \varsigma(x,y)  \\
    &=\tau(y).
\end{align}
This shows the claim.
\end{proof} 
\end{lemma}

\begin{lemma}[Induced projection weights commute with projections]
\label{lem:proj-r-damg-induced-weights}
Let $G=(V,E)$ be a DAMG, $\varsigma:E \to R$ be an edge weight function and $\tau: \Rt(G) \to R$ a root weight function. Let $S \ins V$ be an admissible projection set for $G$. 
Let $q_{(G,\varsigma,\tau)}$ be the induced projection weight function from \Cref{def:r-damg-induced-proj-weights}.
Then we have the equality:
\begin{align}
    (q_{(G,\varsigma,\tau)})^{\sm S} &= q_{(G^{\sm S},\tau^{\sm S},\varsigma^{\sm S})} : E^{\sm S} \to R_q
\end{align}
\begin{proof}
    W.l.o.g.\ we can assume $S=\lC z\rC$. Then we get for $q:=q_{(G,\varsigma,\tau)}$:
 \begin{align}
     q^{\sm z}(x\tuh[e]y) &= 
     \begin{cases}
         q(x\tuh[e]y), & \text{ if } x\tuh[e]y \in G,\\
         q(x\tuh[e_1]z)\cdot q(z\tuh[e_2]y), & \text{ if }e=e_1e_2,\;x\tuh[e_1]z\tuh[e_2]y \in G,
     \end{cases} \\
     &=
      \begin{cases}
         \frac{\tau(x)}{\tau(y)}\cdot\varsigma(x\tuh[e]y), & \text{ if } x\tuh[e]y \in G,\\
         \frac{\tau(x)}{\tau(z)}\cdot\varsigma(x\tuh[e_1]z)\cdot\frac{\tau(z)}{\tau(y)}\cdot\varsigma(z\tuh[e_2]y), & \text{ if }e=e_1e_2,\; x\tuh[e_1]z\tuh[e_2]y \in G,
     \end{cases} \\
     &= \frac{\tau(x)}{\tau(y)}\cdot\varsigma^{\sm z}(x\tuh[e]y).\\
     &\overset{\ref{lem:proj-str}}{=} \frac{\tau^{\sm z}(x)}{\tau^{\sm z}(y)}\cdot\varsigma^{\sm z}(x\tuh[e]y), \\
      &=q_{(G^{\sm z},\tau^{\sm z},\varsigma^{\sm z})}(x\tuh[e]y).
 \end{align}
 This shows the claim.
\end{proof}
\end{lemma}

\begin{example}[Path uniform weights from constant edge and root weights]
\label{eg:r-damg-path-uniform-weights}
Let $G=(V,E)$ be a DAMG and $\varsigma:E \to R$ and $\tau:\Rt(G) \to R$ be the standard edge and root weight functions given by the constant one: $\varsigma(e):=1$ and $\tau(r):=1$, see \Cref{eg:damg-std-edge-root-weights}.
With this and the notations from \Cref{not:damg} we have for all $x,y \in V$ the identifications:
    \begin{align}
        \varsigma(x,y)&= |E(x,y)|,
        & \sigma(x,y) &= \pi^G(x,y), & \tau(x) &= \pi^G(x), \\
        q(x|y) &=\frac{\pi^G(x) }{\pi^G(y)}\cdot |E(x,y)|, &
        s(x|y) &=\frac{\pi^G(x) }{\pi^G(y)}\cdot \pi^G(x,y).
    \end{align}
\end{example}

\subsection{Characterization of Shapley values on edge and root weighted DAMGs}

\begin{theorem}[Main theorem for Shapley values on $R$-DAMGs]
\label{thm:shapley-main}
    Let $R$ be a ring and $\Sh$ be an assignment rule that assigns to every $R$-DAMG $(G=(V,E),\varsigma,\tau)$, every $R_q$-module\footnote{The localized ring $R_q$ is introduced in \Cref{def:r-damg-induced-proj-weights}; note that if $R$ is a field (e.g.\ $R=\Q$ or $R=\R$) then $R_q=R$.} $A$, every value function $v:V \to A$ and every $r \in \Rt(G)$ a value $\Sh^{(G,\varsigma,\tau)}_r(v) \in A$.
    
Further, assume that we have the following properties:
    \begin{enumerate}
        \item \emph{$A$-Linearity}: For every $R$-DAMG $(G=(V,E),\varsigma,\tau)$, $R_q$-module $A$, finite index set $I$, elements $a_i \in A$ and value functions $v_i: V \to R_q$, for $i \in I$, and root $r \in \Rt(G)$ we have:
        \begin{align}
            \Sh^{(G,\varsigma,\tau)}_r\lp  \sum_{i \in I} v_i \cdot a_i\rp&=  \sum_{i \in I} \Sh^{(G,\varsigma,\tau)}_r(v_i) \cdot a_i \in A.
        \end{align}
        \item \emph{Weak elements}: For every $R$-DAMG $(G=(V,E),\varsigma,\tau)$, $R_q$-module $A$, value function $v: V \to A$ and subset $W \ins V \sm \Rt(G)$ of weak elements for $(G,v)$ and every root $r \in \Rt(G)$ we have:
        \begin{align}
            \Sh^{(G,\varsigma,\tau)}_r(v) &=\Sh^{(G^{\sm W},\varsigma^{\sm W},\tau^{\sm W})}_r(v|_{V\sm W}),
        \end{align}
        where $v|_{V\sm W}$ is the restriction of $v$ to $V \sm W$.
        \item \emph{Flat hierarchy}: For every hierarchically flat $R$-DAMG $(G=(V,E),\varsigma,\tau)$, every $y \in V$, unanimity value function $\zeta_y$ on $G$ centred at $y$ and root $r \in \Rt(G)$ we have the following weighted uniformity:
        \begin{align}
            \Sh^{(G,\varsigma,\tau)}_r(\zeta_y) &=
            \begin{cases}
                \displaystyle\delta_y(r), &\text{ if } y \in \Rt(G),\\
                \frac{\displaystyle\tau(r) \cdot \varsigma(r,y)}{\displaystyle\sum_{u \in \Rt(G)}\tau(u)\cdot\varsigma(u,y)}, &\text{ if }y\notin \Rt(G).
            \end{cases} 
        \end{align}
    \end{enumerate}
    Then $\Sh$ necessarily satisfies the following:
    \begin{enumerate}[resume]
        \item \emph{Induced projection weights}: For every $R$-DAMG $(G=(V,E),\varsigma,\tau)$, $R_q$-module $A$, value function $v:V\to A$ and root $r \in \Rt(G)$ we have: \label{enu:proj-weights-weighted}
    \begin{align}
        \Sh_r^{(G,\varsigma,\tau)}(v) &= \Sh^{(G,q)}_r(v), \label{eq:shapley-R-DAMGs-shapley-PDAMG}
    \end{align}
    where the rhs is the Shapley value on $R_q$-PDAMGs from \Cref{def:shapley-values} for the special case where $q=q_{(G,\varsigma,\tau)}$ is the induced projection weight function for $(G,\varsigma,\tau)$ from \Cref{def:r-damg-induced-proj-weights}.
    \end{enumerate}
In addition, $\Sh$ also satisfies the following:
\begin{enumerate}[resume]
    \item \emph{Explicit formula}: For every $R$-DAMG $(G=(V,E),\varsigma,\tau)$, $R_q$-module $A$, value function $v:V\to A$ and root $r \in \Rt(G)$ we have: \label{enu:formula-weighted}
            \begin{align}
        \Sh_r^{(G,\varsigma,\tau)}(v) &= \sum_{y \in V} \frac{\tau(r)}{\tau(y)}\cdot \sigma(r,y) \cdot v'_G(y), \label{eq:shapley-value-r-damg-final}
    \end{align}
    where $\sigma$ is the total path weight function of $\varsigma$ 
    and $\tau(y) = \sum_{u \in \Rt(G)} \tau(u) \cdot \sigma(u,y)$.
\end{enumerate}
    Furthermore, $\Sh$ then also satisfies the following properties:
\begin{enumerate}[resume]
    \item \emph{Projection}: For every $R$-DAMG $(G=(V,E),\varsigma,\tau)$, $R_q$-module $A$, value function $v: V \to A$, subset $S \ins V \sm \Rt(G)$ and root $r\in \Rt(G)$ we have: \label{enu:projection-weighted}
    \begin{align}
       \Sh^{(G,\varsigma,\tau)}_r(v) &= \Sh^{(G^{\sm S},\varsigma^{\sm S},\tau^{\sm S})}_r(v^{\sm S}),
    \end{align}
    where $v^{\sm S}$ is the projection of $v$ onto $V\sm S$ w.r.t.\ the induced projection weights $q_{(G,\varsigma,\tau)}$.
    \item \emph{Edgeless graph}: For every edgeless $R$-DAMG $(G=(V,\emptyset),\varsigma,\tau)$, $R_q$-module $A$, value function $v: V \to A$ and root $r \in \Rt(G)$ we have: \label{enu:edgeless-weighted}
        \begin{align}
            \Sh^{(G,\varsigma,\tau)}_r(v) &= v(r).
        \end{align}
\end{enumerate}
    Finally, $\Sh$ also satisfies the following classical properties:
\begin{enumerate}[resume]
    \item \emph{$R$-Linearity}: 
    For every $R$-DAMG $(G=(V,E),\varsigma,\tau)$, $R_q$-module $A$, value functions $v_1,v_2: V \to A$, scalars $c_1,c_2 \in R_q$ and root $r\in \Rt(G)$ we have:
    \begin{align}
        \Sh^{(G,\varsigma,\tau)}_r(c_1 \cdot v_1 + c_2 \cdot v_2)&= c_1 \cdot \Sh^{(G,\varsigma,\tau)}_r(v_1) + c_2 \cdot \Sh^{(G,\varsigma,\tau)}_r(v_2).
    \end{align}
    \item \emph{Efficiency}: For every $R$-DAMG $(G=(V,E),\varsigma,\tau)$, $R_q$-module $A$ and value function $v: V \to A$ we have:
    \begin{align}
        \sum_{r \in \Rt(G)} \Sh^{(G,\varsigma,\tau)}_r(v) = \sum_{x \in V} v'_G(x).
    \end{align}
    \item \emph{Null root}: For every $R$-DAMG $(G=(V,E),\varsigma,\tau)$, $R_q$-module $A$, value function $v: V \to A$ and every null root $r\in \Rt(G)$ for $(G,v)$ we have:
        \begin{align}
            \Sh^{(G,\varsigma,\tau)}_r(v)&=0.
        \end{align}
    \item \emph{Symmetry}: For every $R$-DAMG $(G=(V,E),\varsigma,\tau)$, $R_q$-module $A$, value function $v: V \to A$, $R$-DAMG automorphism
    $\alpha: (G,\varsigma,\tau) \bij (G,\varsigma,\tau)$ and root $r\in \Rt(G)$ we have: 
    \begin{align}
        \Sh^{(G,\varsigma,\tau)}_{\alpha(r)}(v)&= \Sh^{(G,\varsigma,\tau)}_r(v^\alpha),
    \end{align}
    where the value function $v^\alpha: V \to A$ is defined on elements $x\in V$ via: $v^\alpha(x):=v(\alpha(x))$.
    
    In particular, for fixed $r \in \Rt(G)$, we have the implication:
    \begin{align}
        \lp \forall y \in \Desc^G(r). \;\;  v_G'(\alpha(y)) = v_G'(y) \rp \implies \Sh^{(G,\varsigma,\tau)}_{\alpha(r)}(v)&= \Sh^{(G,\varsigma,\tau)}_r(v).
    \end{align}
\end{enumerate}
Conversely, if we define $\Sh$ either through A.) property \ref{enu:proj-weights-weighted}, or, B.) property \ref{enu:formula-weighted}, or, C.) the properties \ref{enu:projection-weighted} and \ref{enu:edgeless-weighted}, then $\Sh$ satisfies all the other properties from above.
\begin{proof}
By \Cref{rem:unanimity-game} we can always write any value function $v:V \to A$ as follows:
    \begin{align}
        v(x) &= \sum_{y \in V} \underbrace{\zeta_y(x)}_{\in R_q} \cdot \underbrace{v_G'(y)}_{\in A},
    \end{align}
    with the unanimity value function $\zeta_y$ on $G$ centred at $y \in V$. 
    By $A$-linearity we then can write for $r \in \Rt(G)$:
 \begin{align}
     \Sh^{(G,\varsigma,\tau)}_r(v) & = \sum_{y \in V} \Sh^{(G,\varsigma,\tau)}_r(\zeta_y) \cdot v'_G(y).
 \end{align}
For fixed $y \in V$ let $W:=V\sm \lp\Rt(G) \cup \lC y\rC\rp$. Then $W$ consists only of weak elements for $(G,\zeta_y)$. Note that $V^{\sm W} = \Rt(G) \cup \lC y\rC$, which shows that $G^{\sm W}$ is a hierarchically flat DAMG. Further note, that $\zeta_y|_{V \sm W}$ is the unanimity game of $G^{\sm W}$ centred at $y \in V^{\sm W}$. We then get:
\begin{align}
\Sh^{(G,\varsigma,\tau)}_r(\zeta_y) 
    &\overset{\text{weak elements}}{=} \Sh^{(G^{\sm W},\varsigma^{\sm W},\tau^{\sm W})}_r(\zeta_y|_{V \sm W}) \\
    &\overset{\text{flat hierarchy}}{=} \frac{\displaystyle\tau^{\sm W}(r) \cdot \varsigma^{\sm W}(r,y)}{\displaystyle\sum_{u \in \Rt(G^{\sm W})} \tau^{\sm W}(u) \cdot \varsigma^{\sm W}(u,y)} \\
    &\overset{\text{\Cref{lem:path-weight-fucntion-general}}}{=}
     \frac{\tau^{\sm W}(r)}{\tau^{\sm W}(y)} \cdot \sigma^{\sm W}(r,y) \\
    &\overset{\text{\Cref{lem:proj-str}}}{=} \frac{\tau(r)}{\tau(y)} \cdot \sigma(r,y).
\end{align}
 Plugging this back into the above formula shows the claim about the explicit formula for the Shapley values.

 Let now $\Sh$ be given by the explicit formula. To then see that this agrees with the Shapley value $\Sh^{(G,q)}_r(v)$ from \Cref{def:shapley-values} with induced projection weight function $q=q_{(G,\varsigma,\tau)}$, see \Cref{def:r-damg-induced-proj-weights}, we need to check that for the total path weights $s(r|y)$ of $(G,q)$, see \Cref{not:total-weights-q},  we have the equality:
 \begin{align}
     s(r|y) &= \frac{\tau(r)}{\tau(y)} \cdot \sigma(r,y).
 \end{align}
 However, this was already proven in \Cref{lem:str-norm-path-weights}.
 So $\Sh^{(G,\varsigma,\tau)}=\Sh^{(G,q)}$.

Regarding symmetry, first note that we have for every $R$-DAMG automorphism $\alpha:(G,\varsigma,\tau)\bij (G,\varsigma,\tau)$  and $x\tuh[e]y \in G$:
\begin{align}
   \varsigma(\alpha(e)) &=\varsigma(e), & \tau(\alpha(x)) &= \tau(x), &
   \tau(\alpha(y)) &= \tau(y),
\end{align}
which implies:
\begin{align}
    q_{(G,\varsigma,\tau)}(\alpha(e))&= \frac{\tau(\alpha(x))}{\tau(\alpha(y))}\cdot \varsigma(\alpha(e))  =\frac{\tau(x)}{\tau(y)}\cdot \varsigma(e) = q_{(G,\varsigma,\tau)}(e).
\end{align}
For $q=q_{(G,\varsigma,\tau)}$ this shows that $\alpha$ is also an $R$-PDAMG automorphism $\alpha:(G,q)\bij (G,q)$. The symmetry property then follows from \Cref{thm:properties-shapley-values}.

Further note, that for $S \ins V \sm \Rt(G)$ by \Cref{lem:proj-r-damg-induced-weights} we have the following compatibility relationship:
\begin{align}
  q_{(G^{\sm S},\varsigma^{\sm S},\tau^{\sm S})} &=  (q_{(G,\varsigma,\tau)})^{\sm S}.
\end{align}
With this, all the other properties of $\Sh$ follow from \Cref{thm:properties-shapley-values} with the induced projection weight function $q=q_{(G,\varsigma,\tau)}$ from \Cref{def:r-damg-induced-proj-weights}.
\end{proof}
\end{theorem}

% \begin{remark}[Shapley values with edge and root weights]
% \label{rem:shapley-value-edge-player-str}
% Let $G=(V,E)$ be a DAMG, $\varsigma:E \to R$ be an edge weight function and $\tau: \Rt(G) \to R$ a root weight function. Let $q=q_{(G,\varsigma,\tau)}$ be the induced projection weight function from \Cref{def:r-damg-induced-proj-weights}.
% We can then construct the \emph{Shapley values with edge and root weights} for any value function $v:V \to A$ with values in an $R_q$-module $A$ via the following formula for $r \in \Rt(G)$:
% \begin{align}
%     \Sh^{(G,\varsigma,\tau)}_r(v) &:= \Sh^{(G,q)}_r(v)= \sum_{y \in V} \frac{\tau(r)}{\tau(y)}\cdot \sigma(r,y)  \cdot v_G'(y). \label{eq:shapley-value-edge-player-str}
% \end{align}
% It is then clear that $\Sh^{(G,\varsigma,\tau)}$ satisfies all the properties from \Cref{thm:properties-shapley-values}. 
% Also note that by \Cref{lem:proj-str} and \Cref{lem:proj-r-damg-induced-weights} we have for $S \ins V \sm \Rt(G)$ and $r \in \Rt(G)$:
% \begin{align}
%     \Sh^{(G^{\sm S},\varsigma^{\sm S},\tau^{\sm S})}_r(v^{\sm S}) &= \Sh^{(G^{\sm S},q^{\sm S})}_r(v^{\sm S}).
% \end{align}
% We will uniquely characterize $\Sh^{(G,\varsigma,\tau)}$ axiomatically in \Cref{thm:shapley-main}.
% \end{remark}

\end{document}